\documentclass[iop]{emulateapj}
\usepackage{graphics,enumerate,amsmath, morefloats}
\usepackage[utf8]{inputenc}
\usepackage[T1]{fontenc}
\usepackage{lmodern}

\newcommand{\oiiw}{\mbox{[\ion{O}{2}] $\lambda$3727}}

\newcommand{\niiw}{\mbox{[\ion{N}{2}] $\lambda \lambda$6548,6583}}

\newcommand{\siiw}{\mbox{[\ion{S}{2}] $\lambda \lambda$6716,6731}}

\newcommand{\siiiw}{\mbox{[\ion{S}{3}] $\lambda \lambda$9069,9532}}

\newcommand{\oii}{\mbox{[\ion{O}{2}]}}
\newcommand{\nii}{\mbox{[\ion{N}{2}]}}
\newcommand{\hal}{\mbox{H$\alpha$}}
\newcommand{\sii}{\mbox{[\ion{S}{2}]}}
\newcommand{\hb}{\text{H$\beta$}}
\newcommand{\hd}{\text{H$\delta$}}
\newcommand{\oiii}{\text{[\ion{O}{3}]}}

\newcommand{\caii}{\mbox{\ion{Ca}{2}}}

\newcommand{\HI}{{\sc H\,i}}
\newcommand{\Reff}{{R$_{e}$}}

\shorttitle{SDSS-IV/MaNGA IFS Galaxy Survey}
\shortauthors{Yan et al.}

\begin{document}
\title{SDSS-IV MaNGA IFS Galaxy Survey --- Survey Design, Execution, and Initial Data Quality}
\author{Renbin Yan$^{1}$, Kevin Bundy$^{2}$, David R. Law$^{3}$, Matthew A. Bershady$^{4}$, Brett Andrews$^{5}$, Brian Cherinka$^{6}$, Aleksandar M. Diamond-Stanic$^{4}$, Niv Drory$^{7}$, Nicholas MacDonald$^{8}$, Jos\'e R.~S\'anchez-Gallego$^{8}$, Daniel Thomas$^{9,10}$, David A. Wake$^{11}$, Anne-Marie Weijmans$^{12}$, Kyle B. Westfall$^{9,10}$, Kai Zhang$^{1}$, Alfonso Arag\'on-Salamanca$^{13}$, Francesco Belfiore$^{14,15}$, Dmitry Bizyaev$^{16,17}$, Guillermo A. Blanc$^{18,19,20}$, Michael R. Blanton$^{21}$, Joel Brownstein$^{22}$, Michele Cappellari$^{23}$, Richard D'Souza$^{24}$, Eric Emsellem$^{25,26,27}$, Hai Fu$^{28}$, Patrick Gaulme$^{16}$, Mark T. Graham$^{23}$, Daniel Goddard$^{9,10}$, James E. Gunn$^{29}$, Paul Harding$^{30}$, Amy Jones$^{24}$, Karen Kinemuchi$^{16}$, Cheng Li$^{31,32}$, Hongyu Li$^{33,34}$, Roberto Maiolino$^{14,15}$, Shude Mao$^{32,33,35}$, Claudia Maraston$^{9,10}$, Karen Masters$^{9,10}$, Michael R. Merrifield$^{13}$, Daniel Oravetz$^{16}$, Kaike Pan$^{16}$, John K. Parejko$^{8}$, Sebastian F. Sanchez$^{36}$, David Schlegel$^{37}$, Audrey Simmons$^{16}$, Karun Thanjavur$^{38}$, Jeremy Tinker$^{21}$, Christy Tremonti$^{4}$, Remco van den Bosch$^{39}$, Zheng Zheng$^{33}$}

\affil{$^1$ Department of Physics and Astronomy, University of Kentucky, 505 Rose St., Lexington, KY 40506-0057, USA; yanrenbin@uky.edu}
\affil{$^2$ Kavli IPMU (WPI), UTIAS, The University of Tokyo, Kashiwa, Chiba 277-8583, Japan}
\affil{$^3$ Space Telescope Science Institute, 3700 San Martin Drive, Baltimore, MD 21218, USA}
\affil{$^4$ Department of Astronomy, University of Winsconsin-Madison, 475 N. Charter Street, Madison, WI 53706-1582, USA}
\affil{$^5$ Department of Physics and Astronomy and Pittsburgh Particle Physics, Astrophysics and Cosmology Center (PITT PACC), University of Pittsburgh, 3941 OHara St, Pittsburgh, PA 15260, USA}
\affil{$^6$ Department of Physics and Astronomy, Johns Hopkins University, Bloomberg Center, 3400 N. Charles St, Baltimore, MD 21218, USA}
\affil{$^7$ McDonald Observatory, University of Texas at Austin, 1 University Station, Austin, TX 78712-0259, USA}
\affil{$^8$ Department of Astronomy, Box 351580, University of Washington, Seattle, WA 98195, USA}
\affil{$^9$ Institute of Cosmology and Gravitation, University of Portsmouth, Portsmouth, UK}
\affil{$^{10}$ SEPnet, South East Physics Network ({\tt www.sepnet.ac.uk})}

\affil{$^{11}$ Department of Physical Sciences, The Open University, Milton Keynes, MK7 6AA, UK}
\affil{$^{12}$ School of Physics and Astronomy, University of St Andrews, North Haugh, St Andrews KY16 9SS, UK}
\affil{$^{13}$ School of Physics and Astronomy, University of Nottingham, University Park, Nottingham NG7 2RD, UK}
\affil{$^{14}$ Cavendish Laboratory, University of Cambridge, 19 J. J. Thomson Avenue, Cambridge CB3 0HE, UK}
\affil{$^{15}$ Kavli Institute for Cosmology, University of Cambridge, Madingley Road, CB3 0HA Cambridge, UK}
\affil{$^{16}$ Apache Point Observatory and New Mexico State University, P.O. Box 59, Sunspot, NM 88349, USA}
\affil{$^{17}$ Sternberg Astronomical Institute, Moscow State University, Universitetskij pr. 13, Moscow, Russia}
\affil{$^{18}$ Departamento de Astronomía, Universidad de Chile, Camino el Observatorio 1515, Las Condes, Santiago, Chile}
\affil{$^{19}$ Centro de Astrofísica y Tecnologías Afines (CATA), Camino del Observatorio 1515, Las Condes, Santiago, Chile}
\affil{$^{20}$ Visiting Astronomer, Observatories of the Carnegie Institution for Science, 813 Santa Barbara St, Pasadena, CA, 91101, USA}
\affil{$^{21}$ Center for Cosmology and Particle Physics, Department of Physics, New York University, 4 Washington Place, NY 10003, New York}
\affil{$^{22}$ Department of Physics and Astronomy, University of Utah, 115 S 1400 E, Salt Lake City, UT 84112, USA}
\affil{$^{23}$ Sub-Department of Astrophysics, Department of Physics, University of Oxford, DenysWilkinson Building, Keble Road, Oxford OX1 3RH, UK}
\affil{$^{24}$ Max Plank Institute for Astrophysics, Munich, D-85741, Garching, Germany}
\affil{$^{25}$ European Southern Observatory, Karl-Schwarzschild-Str. 2, D-85748 Garching, Germany}
\affil{$^{26}$ Universit\'e Lyon 1, Observatoire de Lyon, Centre de Recherche Astrophysique de Lyon, 9 avenue Charles Andr\'e, F-69561 Saint Genis Laval Cedex, France}
\affil{$^{27}$ Ecole Normale Sup\'erieure de Lyon, 9 avenue Charles Andr\'e, F-69230 Saint-Genis Laval, France}
\affil{$^{28}$ Department of Physics and Astronomy, 203 Van Allen Hall, The University of Iowa, Iowa City, IA 52242-1479, USA}
\affil{$^{29}$ Department of Astrophysical Sciences, Princeton University, Princeton, NJ 08544, USA}
\affil{$^{30}$ Department of Astronomy, Case Western Reserve University, Cleveland, OH 44106, USA}
\affil{$^{31}$ Shanghai Astronomical Observatory, Nandan Road 80, Shanghai 200030, China}
\affil{$^{32}$ Department of Physics and Tsinghua Center for Astrophysics, Tsinghua University, Beijing 100086, China}
\affil{$^{33}$ National Astronomical Observatories, Chinese Academy of Sciences, 20A Datun Road, Chaoyang District, Beijing 100012, China}
\affil{$^{34}$ University of Chinese Academy of Sciences, Beijing 100049, China}
\affil{$^{35}$ Jodrell Bank Centre for Astrophysics, School of Physics and Astronomy, The University of Manchester, Oxford Road, Manchester M13 9PL, UK}
\affil{$^{36}$ Instituto de Astronomia, Universidad Nacional Autonoma deMexico, A.P. 70-264, 04510 Mexico D.F., Mexico}
\affil{$^{37}$ Physics Division, Lawrence Berkeley National Laboratory, Berkeley, CA 94720-8160, USA}
\affil{$^{38}$ Department of Physics and Astronomy, University of Victoria, Victoria, BC V8P5C2, Canada}
\affil{$^{39}$ Max Planck Institute for Astronomy, K\"onigstuhl 17, D-69117 Heidelberg, Germany}

\begin{abstract}
The MaNGA Survey (Mapping Nearby Galaxies at Apache Point Observatory) is one of three core programs in the Sloan Digital Sky Survey IV. It is obtaining integral field spectroscopy (IFS) for 10K nearby galaxies at a spectral resolution of $R\sim2000$ from $3622-10,354{\rm \AA}$.  The design of the survey is driven by a set of science requirements on the precision of estimates of the following properties: star formation rate surface density, gas metallicity, stellar population age, metallicity, and abundance ratio, and their gradients; stellar and gas kinematics; and enclosed gravitational mass as a function of radius. We describe how these science requirements set the depth of the observations and dictate sample selection. The majority of targeted galaxies are selected to ensure uniform spatial coverage in units of effective radius ($R_e$) while maximizing spatial resolution. About 2/3 of the sample is covered out to $1.5R_e$ (Primary sample), and 1/3 of the sample is covered to $2.5R_e$ (Secondary sample). We describe the survey execution with details that would be useful in the design of similar future surveys. We also present statistics on the achieved data quality, specifically, the point spread function, sampling uniformity, spectral resolution, sky subtraction, and flux calibration. For our Primary sample, the median r-band signal-to-noise ratio is $\sim73$ per $1.4{\rm \AA}$ pixel for spectra stacked between 1--1.5 R$_{e}$. Measurements of various galaxy properties from the first year data show that we are meeting or exceeding the defined requirements for the majority of our science goals. 
\end{abstract}

\section{Introduction}

Large spectroscopic galaxy surveys, such as the Sloan Digital Sky Survey \citep{York00}, and the Two-degree Field Galaxy Redshift Survey \citep{Colless01}, have revolutionized the way we study galaxy evolution. The huge statistical power brought in by targeting a large number of galaxies using the same instrument with excellent calibration enabled huge progress. Not only have these efforts quantified accurately with great precision those trends and scaling relations that were previously known, such as the color-bimodality \citep{Strateva01, BaldryGB04}, the color-density relation \citep{Hogg03, BlantonEH05}, the mass-metallicity relation for gas \citep{TremontiHK04} and stars \citep{Thomas10,Johansson12}, and the Fundamental Plane \citep{Bernardi03}, they have also discovered many new relations and trends, such as the dependence of star formation history on stellar mass \citep{KauffmannHW03}, the star formation rate vs. stellar mass relation \citep{Brinchmann04, Salim07, Wuyts11}, the strong mass dependence of the radio-loud AGN fraction \citep{Best05}, large scale galactic conformity \citep{Kauffmann13}, and many others. They also connected large scale structure studies and galaxy evolution studies thanks to environmental measurements enabled by dense and uniform sampling of complete galaxy samples (see \citealt{BlantonM09} and references therein). 

However, these massive surveys lacked spatial coverage in individual galaxies. The single 3\arcsec\ fibers used by SDSS, for instance, cannot cover most of the light in nearby galaxies. For example, comparing the flux incident on the SDSS 3\arcsec\ fibers with the total flux of all main sample galaxies in SDSS, 80\% of galaxies have less than 36\% of their light covered by the fiber. The spectra provide a lot of information, about both stellar and gaseous components, but they only sample the center of the galaxies and can give a strongly biased picture. Nearly all studies based on SDSS have to take this aperture effect into account in their analysis. Many studies combining spectroscopic information with photometry also need to make corrections, extrapolations, or use simplified assumptions. For example, to obtain the total star formation rate in a galaxy, one either has to apply large aperture corrections to the spectroscopically-derived star formation rate based on the central region \citep{Brinchmann04}, or turn to broadband photometry which suffers more from dust extinction and degeneracies in stellar population modeling \citep{Salim05, Salim07}. Furthermore, a full kinematic description is impossible with single-fiber observations.
Past long-slit surveys are also inefficient at obtaining the spatial information as one only probes a narrow elongated region and the signal-to-noise is poor in galaxy outskirts. 

Integral field spectroscopy (IFS) solves these problems. Several IFS surveys have made great progress in recent years (see \citealt{Cappellari16} for a review). SAURON \citep{Bacon01} and ATLAS${\rm 3D}$ \citep{Cappellari11} surveyed 260 early-type galaxies in the nearby universe using a lenslet array intergral field instrument, SAURON, on the William Herschel Telescope on La Palma. They had a relatively narrow wavelength coverage ($4800-5380{\rm \AA}$) and focused exclusively on early-type galaxies. 
The DiskMass survey \citep{Bershady10} used two fiber-fed Integral Field Unit (IFU), SparsePak on WIYN and PPak on the Calar Alto 3.5m Telescope. It targeted 146 nearly face-on disk galaxies to study stellar and gas kinematics. For the purpose of kinematic measurements, this survey utilized high spectral resolution in three narrow wavelength windows around 515, 660, and 860 nm. 
The VENGA \citep{Blanc13} survey used a fiber-fed integral field spectrograph, VIRUS-P on the 2.7-m Telescope at McDonald Observatory, and targeted 30 nearby spiral galaxies. Recently, the CALIFA survey used the PPak instrument and targeted 600 nearby galaxies selected to sample a wide variety of stellar mass and star formation rate. With the improved sample size and wide wavelength coverage, CALIFA has produced numerous results, such as the universal metallicity gradient among star-forming galaxies \citep{Sanchez14}, the nature of LINER-like galaxies \citep{Kehrig12, Singh13, Gomes16}, the spatially-resolved growth history \citep{Perez13, Sanchez-Blazquez14}, the spatially-resolved stellar mass-metallicity relation \citep{GonzalezDelgado14}, and the resolved star formation main sequence \citep{Cano-Diaz16, GonzalezDelgado16}. However, if one were to do an SDSS-like study of galaxies by binning galaxies by stellar mass, environment, and morphology, one quickly loses sample size for significant statistics \citep[e.g.][]{GonzalezDelgado16}. The main limitation for the sample size is that all these surveys are targeting galaxies one by one and are inefficient at building up a large statistical sample.

To address this issue, two large IFS surveys of the general galaxy population targeting thousands of galaxies are ongoing right now. Both utilize multiple fiber bundles to target multiple galaxies at the same time, enabling much more efficient observing. One of them is the SAMI Galaxy survey \citep{Bryant15, Allen15} using the fiber-fed SAMI instrument \citep{Croom12} on the 3.5m Anglo-Australian Telescope at Siding Spring Observatory. SAMI will eventually target 3400 galaxies and has already produced results on many topics, including the kinematic morphology-density relation \citep{Fogarty14}, outflows and extraplanar gas \citep{Ho14,Ho16}, dynamical scaling relations \citep{Cortese14}, dynamical M/L ratio of disk galaxies \citep{Cecil16}, and aperture corrections for star formation rates \citep{Richards16}. The other large IFS survey is the SDSS-IV/MaNGA galaxy survey operating at the 2.5m Sloan Foundation Telescope. Given the large 3$^\circ$ field of view of the SDSS telescope and sizeable detector real estate, MaNGA uses multiple fiber bundles to target 17 galaxies (and 12 standard stars) at the same time. This allows us to build a 10K galaxy IFS sample with much wider and continuous wavelength coverage than other surveys, enabling powerful statistical studies of the spatially-resolved properties of nearby galaxies.

This paper complements MaNGA's other descriptive publications by providing a complete picture of the survey's design and execution, and an evaluation of the resulting data quality. In Section~\ref{sec:requirements}, we describe the science requirements of our survey, and how they flow down to specific decisions on the sample design and observing strategy. We summarize the hardware in Section~\ref{sec:hardware} and the sample design in Section~\ref{sec:sample}. In Section~\ref{sec:execution}, we describe the execution of the survey, including the observing strategy, setting of the completeness thresholds, choice of the fields, plate design, observing procedure, and the optimization of the instrument focus. In Section~\ref{sec:progress} we dscribe our survey progress and projection. In Section~\ref{sec:quality}, we provide an evaluation of the initial data quality: PSF, sampling uniformity, spectral resolution, sky subtraction accuracy, and flux calibration accuracy. In the end (Section~\ref{sec:verification}), we present a series of tests checking whether we are meeting the science requirements. We summarize in Section~\ref{sec:summary}.

\section{Science Requirements}\label{sec:requirements}

In \cite{Bundy15}, we have described the scientific motivation of the survey. In this paper, we provide the specific requirements that drive the design. \cite{Bundy15} listed four key science questions we aim to address: how galactic disks grow through accretion, how galactic bulges and ellipticals build up, how star formation shuts down by internal and/or external processes, and how mass and angular momentum is distributed in galaxies. In order to put the survey design on a quantitative footing, we need to turn these general questions into specific requirements that dictate design choices and can be verified with initial data. The seven key measurement requirements discussed below flow down from the key science questions.  

In setting these requirements, we would like to be particularly pedantic about distinguishing precision from accuracy. `Precision' of a measurement is set by the random errors, while 'accuracy' of a measurement is set by both random and systematic errors. Below, for direct observables that are independent of model assumptions, such as redshift, radial velocity, velocity dispersion, we can set requirements on both precision and accuracy. But for derived quantities that are model dependent, such as star formation rate, stellar age, gas metallicity, dark matter fraction, etc., we only set requirements on the precision of the measurements, under a specific set of model assumptions. The model-dependence of these quantities means that their accuracies completely depend on how accurate the models are, which are often difficult to assess. Investigating the accuracy of the models are important goals for astronomy, but are beyond the scope of our tasks of designing the survey.

In choosing the threshold values for each requirement, we generally follow the rule that we require the precision to be better than half of the $1\sigma$ scatter in the expected distribution of the quantity. Now we describe the requirements.



%
%
%
%
%



\begin{enumerate}



\item We require the star formation rate surface density to be measured to a {\it precision} of 0.15 dex per spatial resolution element in our target galaxies where star formation rate surface density is higher than 0.01 ${\rm M_\odot yr^{-1} kpc^{-2}}$ and reddening (E(B-V)) is less than 0.5. 

Star formation rate is an essential indicator of galaxy growth. It is required to address the questions of disk growth, bulge growth, and the question of how star formation shuts down. With MaNGA, we measure the surface density of the star formation rate. The main estimator of star formation rate at our disposal is the extinction-corrected \hal\ luminosity, which traces the instantenous star formation rate and is often used as the basis for calibrating other star formation rate indicators \citep{Kennicutt98,Kennicutt12}. The SFR in star-forming galaxies correlates tightly with stellar mass, which is often referred as the star formation main sequence \citep{Brinchmann04, Salim05, NoeskeWF07, Renzini15}. The $1\sigma$ scatter acround this relationship in the local Universe is about 0.3 dex \citep{Renzini15}, which includes both intrinsic scatter and measurement noise. With MaNGA data, we will measure the spatially-resolved version of this relationship --- star formation surface density vs. stellar mass surface density \citep{Cano-Diaz16}. Thus, we require the precision on SFR surface density to be better than half of the scatter in the global relationship.


Of course, such precision can only be sensible above a certain threshold of emission line strength. We set the threshold to the `knee' in the Schmidt-Kennicutt relation where the SFR efficiency changes significantly \citep{Bigiel08}, which is around 0.01 ${\rm M_\odot yr^{-1} kpc^{-2}}$. In addition, extinction can severely affect our ability to measure star formation rate. We thus set a maximum extinction for our requirement, $E(B-V)=0.5$, which is about the median extinction in the centers of star-forming galaxies in the SDSS main sample.




\item Measurements of the gas metallicity gradients in galaxies should have a precision of 0.04 dex per Re.

Gas metallicity gradients can provide crucial insights about the cycling of gas in galaxies. We will measure gas metallicity from strong emission lines. Supernovae explosions and mass loss from asymptotic giant branch stars could return gas enriched with newly-synthesized material. Feedback from star formation can drive gas enriched with heavy elements back into the hot halo which could rain back down later when it cools. How the enriched gas is redistributed in the galaxy disk can be reflected by the radial metallicity gradient. \cite{Fu12} used a radially-resolved semi-analytical model and demonstrated that the gradient is sensitive to the fraction of gas ejected into the hot halo. The higher the fraction of gas ejected into the halo, which eventually redistributes over the disk, the flatter is the metallicity gradient. The models also predict gradients to be a function of stellar mass and bulge-to-disk ratios. Observationally, recent work by \cite{Sanchez14} showed a universal gradient of -0.1 dex/Re with a sigma of 0.09 dex/Re. In order to discern the potential dependence on mass and bulge fraction, we need to measure the gradient to at least a factor of two better than the scatter seen in CALIFA. Therefore, we chose 0.04 dex per Re as our requirement.

On the other hand, the behaviour of the gradient at large radius could directly probe the gas accretion at large radius. Moran et al. (2012) found 10\% of massive disk galaxies show abrupt drops at large radius while Sanchez et al. (2014) found flattening in metallicity at the same place. Resolving this apparent discrepancy requires a large sample size, wide stellar mass sampling, and sufficient precision to resolve the intrinsic scatter which we expect will be no smaller than the scatter at small radii. 


\item Measurements of light-weighted stellar age gradients in {\it star-forming and newly-quenched galaxies} should have a precision of 0.1 dex per Re.

We can measure light-weighted stellar age in {\it star-forming and newly-quenched galaxies} using D4000 and \hd\ absorption indices, following \cite{KauffmannHW03}. This will help address the question about bulge growth and about quenching. The average stellar age measurements are sensitive to the assumed stellar population models with many ingradients including the initial mass function, the stellar spectral library, the template star formation history, and metallicity evolution \citep{Chen10}. Therefore, we set our requirements on the precision of the measurement under a specific set of stellar population models. Recent work on CALIFA disk galaxies shows that gradients range from +0.4 to -0.7 dex/Re, with a $1\sigma$ scatter of 0.2 dex/Re  at fixed stellar masses, which is comparable to the scatter in stellar age itself \citep{GonzalezDelgado15}. We set our requirements to half of the scatter. 

\item Measurements of light-weighted stellar age, metallicity and abundance gradients in {\it quiescent galaxies} should have a precision of 0.1 dex per decade in Re.

In quiescent galaxies (galaxies without recent star formation), the stellar age, metallicity, and abundance can be measured from either absorption line indices or from full spectral fitting. Our precision requirement of 0.1 dex per deacde in Re is chosen because typical age gradients range from -0.1 to 0.4 dex per decade in Re with a $1\sigma$ scatter of about 0.2 dex \citep{Mehlert03,Kuntschner10,Spolaor10} and typical abundance gradients range from -0.5 to +0.2 dex per decade in Re with a scatter of about 0.2 dex \citep{Mehlert03,Spolaor10,Koleva11}.

\item Measurements of the  baryonic specific angular momentum ($\lambda_R$) within 1Re should have an accuracy of 0.05 for $\lambda\sim0.1$ for quiescent galaxies in order to differentiate fast and slow rotators, given the dividing line between the two is roughly around $\lambda=0.1$\citep{Cappellari07, Emsellem07, Emsellem11}. Since the measurement of $\lambda$ is model-independent, this is an accuracy requirement.

\item Measurements of the enclosed gravitating mass within 1.5Re should have an accuracy of 10\% when the kinematics appear to be axissymmetric. 


\item Measurements of the dark matter fraction within 1.5Re or 2.5Re should have a precision of 10\% in bulge-dominated gas-poor galaxies. The measurement becomes more difficult for star-forming galaxies due to the larger uncertainty on stellar mass estimates from stellar population models. Thus, we only set the requirements for bulge-dominated gas-poor galaxies.

This is a precision requirement because the measurement is model-dependent. The assumptions about the dark matter profile, the mass-to-light ratio and its gradient will change the result systematically.

\noindent The last two requirements are not only set because they are interesting on their own, they are also set by the desire to measure the stellar mass-to-light ratio in bulge-dominated, gas-poor galaxies via dynamical modeling to 15-25\% to investigate IMF variations. This constraint is driven by the fact that the maximum M/L difference attributed to the IMF is roughly a factor of two \citep[e.g.][]{Cappellari12, ConroyvD12b}.

\end{enumerate}

In addition, we require measuring the dependence of all of the above physical properties on stellar mass, morphology, and environment. This is one of the major motivations to obtain a sample as large as MaNGA. 



\subsection{Requirements on the depth of the observation} \label{sec:requirements_deriv}

The above requirements flow down to quantitative limits on our survey design. 

\begin{enumerate}
\item 
With the calibration provided by \cite{Kennicutt12}, a SFR surface density of 0.01 ${\rm M_\odot yr^{-1} kpc^{-2}}$ corresponds to an extinction-corrected \hal\ surface brightness (SB) of $1.86\times10^{39} {\rm erg~s^{-1} kpc^{-2}}$.  At a reddening of $E(B-V)=0.5$, the Balmer decrement is 4.85 and the {\it observed} \hal\ luminosity is 31.3\% of the intrinsic \hal\ luminosity, corresponding to an \hal\ surface brightness of $5.83\times10^{38} {\rm erg~s}^{-1} {\rm kpc}^{-2}$.  This is $1.15\times10^{-16}/(1+z)^4 {\rm erg~s}^{-1} {\rm cm}^{-2} {\rm arcsec}^{-2}$ in observed surface brightness. Because of the surface brightness dimming with redshift, this requirement is most stringent at the high-$z$ limit of our sample. At $z\sim0.15$, this corresponds to an \hal\ SB of $6.58\times10^{-17} {\rm erg~s}^{-1} {\rm cm}^{-2} {\rm arcsec}^{-2}$ and an \hb\ SB of $1.36\times10^{-17} {\rm erg~s}^{-1} {\rm cm}^{-2} {\rm arcsec}^{-2}$. 

To convert the measurement precision of SFR surface density to the required signal-to-noise in the spectra, we need to consider how the SFR estimate is made. Both the uncertainty on the raw \hal\ measurement and the uncertainty on the extinction correction derived from the Balmer decrement (\hal/\hb\ ratio) need to be considered. In \cite{Yan16}, we have done a detailed derivation of this dependence. According to Eqn. (12) in that paper, we find the measurement noise on \hb\ dominates the final error budget. At an extinction of $E(B-V)=0.5$, \hal\ is 4.85 times as strong as \hb\ and we assume it has at least 3 times higher S/N than \hb. In order to measure SFR to 0.15 dex or equivalently $(0.15\ln10 =)$ 34.5\% in fractional uncertainty, we need to measure \hal\ to better than 5\% and \hb\ to better than 15\%.  

The uncertainty on line measurement depends not only on the strength of the line, but also on the continuum and sky background level. Any systematic error in the stellar continuum subtraction would also contribute to the error of the line flux. Here, we ignore the latter and only consider the Poisson noise contributed by the stellar continuum, the sky background flux, and the emission line itself. The median sky background in dark time at APO is 22.2 and 21.2 mag arcsec$^{-2}$ in $g$- and $r$-bands, respectively. The outskirts ($r>0.5$\Reff) of nearby galaxies are mostly fainter than this sky background. We place the science requirements in this sky-background-dominated regime. To make conservative estimates, we assume the stellar continuum is as strong as the median sky background. In the very centers of galaxies, the stellar continuum could dominate over the sky background and yield lower S/N for fixed line flux. This would only affect a small fraction ($\sim20\%$ in $g$ and $\sim10\%$ in $r$) of the areas we probe.  

The uncertainty on line measurement depends on the details of the measurement method --- whether it is summed over a wavelength window or fitted with a Gaussian would yield different result. Here, for simplicity, we just require the peak amplitude to be measured to better than 15\% for \hb\ and 5\% for \hal. 

In a 2.5\arcsec\-diameter aperture, assuming a fixed line width of 70 km/s which is roughly the instrumental dispersion, the above line fluxes correspond to a peak amplitude of 21.3 $\mu$Jy and 139 $\mu$Jy for \hb\ and \hal\ at $z=0.15$. 
Thus, the uncertainty from the background and the lines should be less than 3.15 (\hb) and 6.95 (\hal) $\mu$Jy.  
The stellar continuum, assumed to be equal to the median sky background, has a flux density in $g$- and $r$-band of 23.6 and 59.2 $\mu$Jy in a 2.5\arcsec-diameter aperture. The sky-subtracted signal is the sum of the line amplitude and the stellar continuum, which is 44.9$\mu$Jy or 19.8 mag for \hb, and 198 $\mu$Jy or 18.2 magnitude for \hal.
Summarizing, we require the S/N to be greater than $(44.9/3.15=)14.3$ for $g=19.8$ and $(198/6.95=)28.5$ for $r=18.2$ in a 2.5\arcsec-diameter aperture.  




\item In order to measure the gas-phase metallicity gradient to better than 0.04 dex per Re, we need to measure it over several ($>=4$) elliptical annuli and with an accuracy better than 0.05 dex per annulus. We set the outer most annulus to be from 1-1.5 Re with thinner inner annuli as the S/N is higher there. To reach this precision, using the R23 method \citep{KewleyD02} as an example, we find that an Peak-Amplitude-to-Noise ratio of at least 7 are needed on \oii, \hb, and \oiii. This translates to a continuum S/N greater than 10 per pixel near \hb. This requirement is on stacked spectra for an elliptical annulus.

\item For star-forming galaxies and newly quenched systems (younger than 1 Gyr), to measure the mean stellar age to better than 0.1 dex using the 4000\AA\ break and Balmer indices requires a median S/N of greater than 10 per pixel in r-band. This estimate is based on \cite{KauffmannHW03}.

\item For quiescent galaxies, in order to measure the stellar age, metallicity and abundance gradient to better than 0.1 dex per decade in Re, we need to measure this over at least 4 elliptical annuli to a precision of better than 0.12 dex in each annulus. 
Based on our prototype data obtained in January of 2013 using prototype hardware \citep{Bundy15, Wilkinson15}, and an estimation of the accuracy in deriving stellar population properties from SDSS spectra in \cite{Johansson12}, we found that empirically, at a median stacked S/N of greater than 33 per pixel in the r-band continuum, we could measure stellar age to a precision of $\pm0.12 {\rm dex}$, metallicity to a precision of $\pm0.1 {\rm dex}$, and abundance to a precision of $\pm0.1 {\rm dex}$. This is using the method of absorption line indices in the Lick/IDS system. If we use full spectrum fitting, simulations show that we can determine age and metallicity to a precision better than 0.1 dex or better with even lower S/N (C. Conroy, priv. comm.). We went with the more conservative estimate in designing our survey.

\item The $\lambda_R$ measurement depends on the flux profile of the galaxy, the velocity
field, and the velocity dispersion ($\sigma$) field. According to \cite{Emsellem07},
\begin{equation}
\lambda_R = {\sum_{i=1}^{i=N_p} F_i R_i |V_i| \over \sum_{i=1}^{i=N_p} F_i R_i \sqrt{V_i^2+\sigma_i^2}} ,
\end{equation}
where $F_i$, $R_i$, $V_i$, and $\sigma_i$ are the flux, center
distance, velocity, and velocity dispersion in the $i$-th spatial bin. In
order to make a simple estimate, we assume a quiescent galaxy with
uniform $\sigma$ throughout and a flat rotation curve that sharply
decreases to zero in the center. Basically, we are assuming every spatial bin
have the same $\sigma$ and $|V|$. In this case, $\sum_{i=1}^{i=N_p} F_i
R_i$ can be cancelled out in the above equation and it becomes easier
to tie the uncertainty on $\lambda_R$ to the uncertainties on $V$ and
$\sigma$. The relation is
\begin{equation}
\Delta_\lambda = \lambda (1-\lambda^2) \sqrt{\left({\Delta_V \over V}\right)^2+\left({\Delta_\sigma \over \sigma}\right)^2}
\end{equation}
For a slow rotator with $\lambda_R\sim0.1$ and $\sigma=100 {\rm km/s}$, 
to obtain a precision of $\Delta \lambda=0.05$, we need a S/N per bin of at
least 17 per pixel in $r$-band, 
according to the velocity and dispersion uncertainty achieved
in the prototype data. The galaxy also needs to be
reasonably well resolved in order to measure the $V$ and $\sigma$
maps. According to experience, we require approximately 20 independent spatial
bins in order to have a reliable measurement. Therefore, the final
technical requirement is at least 20 spatial bins with S/N per pixel
of at least 17. This requirement can be reached for the majority of
the 1.5\Reff\ sample.

\item 
Enclosed masses will be derived via dynamical modeling \citep{Cappellari08}. The enclosed mass inferred by the models scales as predicted by the virial theorem $M\propto V_{\rm rms}^2= V^2 + \sigma^2$. This implies that the fractional mass error is twice the fractional $V_{\rm rms}$ error. In addition to this random error, there is typically a contribution from various modeling systematic effects of about 6\% (e.g. Fig.9 of \citealt{Cappellari13}). Assuming the random and systematic percent errors add quadratically, with N bins in which we measure $V_{\rm rms}$, the mass error scales with $2/\sqrt{N}$ of the error on $V_{\rm rms}$. In order to measure the mass to 10\% with 4 bins at 1.5\Reff, we need to measure $V_{\rm rms}$ to 11.3\%. 
\begin{equation}
\left({\Delta V_{\rm rms} \over V_{\rm rms}}\right)^2=  {V^2 \over V_{\rm rms}^4} \Delta V^2 + {\sigma ^2 \over V_{\rm rms}^4} \Delta \sigma^2
\end{equation}

From the test-run data, we found the following empirical relationship between S/N and the uncertainties on $V$ and $\sigma$. 
\begin{eqnarray}
\Delta V  &= 75/(S/N)^{0.95} \\
\Delta \sigma &= 101/(S/N)^{0.96}
\end{eqnarray}
For simplicity, we assume $\Delta V$ scales with S/N in the same way as $\Delta \sigma$ to make a conservative estimate. For a system with $V_{\rm rms}$ of 60 km/s (similar to our instrumental resolution), to measure it to 11.3\%, we need a S/N per bin of 16 per pixel in one quarter of an annulus.



\item 
The requirements on the dark matter fraction estimate are more difficult to
derive. Ideally, we
would determine the number of spatial elements needed and the
precision required on the velocity and velocity dispersion in each element,
which can then be turned into a spatial coverage requirement and a
signal-to-noise requirement. However, the real situation is more  
complex.  The precision on the velocity and velocity dispersion
measurement not only depends on the S/N of the spectra, but also
depends on the strength of the absorption features (hence the metallicity
of the galaxy). The number of spatial elements (e.g. Voronoi bins with
a threshold S/N, as described in \citealt{Cappellari03}) one can
construct depends on the surface brightness profile of the
galaxy. Even if these factors are fixed, the precision of the dark
matter fraction we can achieve also depends on the complexity of the
kinematic structure of the galaxy. 

To make progress, we assess whether we can meet our requirement using
simulated galaxies assuming our baseline sample and observing
strategy. Such simulations, in conjunction with the MaNGA prototype data,
show that it is possible to recover dynamical estimates of the dark
matter fraction within 1\Reff\ with anisotropic Jeans mass modelling
(JAM: \citealt{Cappellari08}) to better than 10\% in galaxies with
the second moment of the velocity distribution function larger than 60 {\rm km/s} 
and bundle sizes larger
than 19 fibers (so that their velocity fields are also well-resolved
spatially). 


\end{enumerate}

Among the requirements above, the most stringent requirements are from 
the stellar population gradient measurement on quiescent galaxies and
from the measurement on enclosed gravitating mass. Both of these
come down to roughly the same requirement of a continuum S/N greater
than 33 per pixel in the $r$-band portion of the spectrum stacked across all fibers between 1 and 1.5~\Reff\ and across all exposures. We require the majority of our Primary sample galaxies to satisfy this requirement.  It sets the final exposure time of our observations.

\subsection{Requirements on the Sample}

Besides our requirement on the depth of the observation, we also have requirements
about the mass distribution of galaxies, spatial coverage, number density on the sky, 
and availability of environment information.

We have the following requirements on our sample selection.

\begin{description}
\item [A] The sample selection needs to be simple and reproducible so that one can
easily reproduce the statistical distribution of any galaxy property for a volume-limited
sample down to a certain stellar mass.
\item [B] The sample needs to be representative at all stellar
  masses ($10^9 < M_*/M_\odot < 10^{12}$) in order to have enough statistical power for studies of both high mass and low mass galaxies.
\item [C] The selection shall give more weight to galaxies with rare color-mass combinations
in order to have enough statistical power to sample these rare galaxies or short stages of evolution.
\item [D] The spatial coverage in all target galaxies should be 
  as uniform as possible in units of \Reff. We desire radial coverage to 1.5~\Reff\ along the major axis to allow
  a long baseline for gradient measurements and to cover most of the light. 
  An elliptical aperture with semi-major axis of 1.5~\Reff\ covers 75\% of the light for an exponential disk and 60\% for a de Vaucouleurs profile.

  We also require coverage to 2.5~\Reff\ for 1/3 of the sample to explore the very outer regions of 
  galaxies. 
Large spatial coverage offers the opportunity to probe the gas metallicity gradient in the ourskirts of disks which may connect with accretion of low metallicity or nearly-pristine material from filaments, to probe the stellar population at large radius, to probe beyond the peak of the rotation curve in disks, and to measure the dark matter fraction where it dominates the potential. The benefit of large coverage also comes with substantial risk as the signal-to-noise will be much lower and spatial resolution is sacrificed. 
  


\item [E] The sample should maximize spatial resolution while satisfying the above requirements. Each galaxy
need to be resolved with at least 3 radial bins in order to measure gradients in stellar age, metallicity, etc.
\item [F] The total sample size should be about 10K, with an approximate 2-to-1 split 
between Primary (1.5~\Reff) and Secondary (2.5~\Reff) samples. The justification for the sample
size was described in \cite{Bundy15}. Briefly, this size will allow us to study any galaxy property
as a function of 3 independent variables (e.g. mass, star formation rate, and environment) with 
6 bins in each variable, and provide us with $\sim50$ objects per bin that will yield a $5\sigma$ detection 
in the median difference of that property between bins assuming the intrinsic scatter of that property in
each bin is of similar level to the median difference. 

\item [G] The combined sub-samples need to have a sky surface-density high
  enough to enable efficient allocation of IFU bundles.
\item [H] The majority of the sample needs to have environment
  information available, and the allocation of bundles needs to be
  unbiased w.r.t. to environment.
\item [I] A significant fraction of the sample needs to overlap
  past and near future \HI\ observations. Attention shall also be paid to
  other ancillary data of interest, such as deep imaging in the optical
  and near-IR, and accesibility for follow-up observation from unique facilities in 
  the southern hemisphere (e.g. the Atacama Large Millimeter/submillimeter Array, here after ALMA).
\end{description}

Often the requirements described in the above sections conflict with each other, as the
design of our survey must fit within the realities of a finite budget and limited observing time. First of all, 
our total available observing time is fixed to half of the dark time in six years. Second, given 
the limited budget, we are restricted to the detector real estate provided by the BOSS Spectrographs. 
With fixed total observing time, getting deeper observations inherently conflicts with getting a larger 
sample size. With fixed number of fibers and detector real estate, obtaining higher spatial resolution 
conflicts with the desire of larger spatial coverage. Given the limited number of galaxies in the nearby 
Universe, higher spatial resolution also conflicts with the efficiency of observing them. These factors 
need to be balanced against each other. 

In practice, given the requirements above and the boundary conditions, we jointly optimize the fiber
bundle size distribution and the sample selection (Wake et al. in prep). With the S/N requirements set by the science 
requirements, we were just able to achieve a sample of about 10K galaxies in 6 years at APO.


\subsection{Requirements on Data Quality} \label{sec:imagerequirement}

Here we describe several requirements we impose on the quality of raw data.

\subsubsection{Flux Calibration}
 
As described above, we need to measure SFR and gas phase metallicity to a certain precision. These measurements are based on emission line strengths and line ratios. Flux calibration uncertainty will contribute to the total uncertainty in these measurements. In \cite{Yan16}, we have done detailed derivations to flow down the science requirements to the requirement on flux calibration. Here we briefly summarize. Our requirement on flux calibration is that it does not dominate the uncertainty in the SFR calibration. Specifically, we require the contribution to SFR error from flux calibration alone to be better than 0.05 dex. This results in a requirement of 3.7\% relative calibration error between \hal\ and \hb\ and an 8.1\% absolute calibration error on \hal.  From the gas-phase metallicity requirement, in order to measure O/H to better than 0.04 dex, using \nii/\oii\ as the metallicity calibrator, we found the relative calibration between these two lines needs to be better than 7\%. 

\subsubsection{Quality of the reconstructed PSF}
MaNGA is both a spectroscopy survey and an imaging survey as we are effectively obtaining an image at every wavelength. Therefore, besides requirements on the depth of the observation, we also need to set requirements on the image quality. 
Because variations in the PSF could change the enclosed flux within a fixed aperture, wavelength-dependent PSF variation (e.g. caused by differential atmosphere refraction) would lead to errors in flux calibration. Our requirement on the relative calibration between \nii\ and \oii\ to be better than 7\% therefore translates to requirements on the effective PSF and the uniformity of observing depth. As derived by \cite{Law15}, we have the following requirements, for the entire wavelength range covered by the BOSS spectrograph: 
\begin{description}
\item [A.] The FWHM of the reconstructed PSF in the final data cube cannot vary by more than 20\% across a bundle.
\item [B.] The reconstructed PSF must have an axis ratio ($b/a$) greater than 0.85.
\item [C.] The effective exposure time cannot vary by more than 15\% across a bundle.
\end{description}

These requirements not only place constraints on how we conduct the observations, but also on how we process the data to recontruct the image at each wavelength to a regular grid. The choices we make on the observing strategy and the image reconstruction algorithm have to be driven by the hardware setup and the seeing condition at the observatory. We will discuss the observing strategy below in Section~\ref{sec:observingstrategy}. And we refer the readers to \cite{Law16} who discussed the choices we made in the image reconstruction algorithm and compared our approach to the approaches adopted the CALIFA survey as described by \cite{Sanchez12} and by the SAMI survey as described by \cite{Sharp15}.

\subsubsection{Sky Subtraction}

Our spectra cover many interesting features in the near-IR. The \caii\ triplet, Na I at 0.82$\mu m$, and FeH Wing-Ford band are sensitive to the inital mass function \citep{Cenarro03, ConroyvD12a}. The \siiiw\ emission lines, combined with \sii\ and \hal, make excellent estimators for ionization parameter and gas metallicity \citep{KewleyDS01}. However, these wavelengths are also dominated by forests of sky emission lines. Thus, in order to take advantage of these features, we need to achieve high quality sky subtraction. We require our sky subtraction to be Poisson-limited, especially in the $8,000-10,000{\rm \AA}$ range. The Poisson limit is required so that we can stack many spectra in the ourskirts of a galaxy (or from many galaxies) to achieve the stated S/N in earlier sections, and to analyze weak features in the near-IR.

\section{Hardware} \label{sec:hardware}

Details of the instrument configuration are described by \cite{Drory15}. We summarize them briefly here. We use the Baryon Oscillation Spectroscopic Survey (BOSS) spectrographs \citep{Smee13} on the Sloan Foundation Telescope \citep{Gunn06}. We modified the fiber feed system used by BOSS so that we can use fiber bundles to achieve integral field spectroscopy. The detector real estate limits the number of fibers we could accomodate in each spectrograph. Bigger bundles result in fewer per field. As described in detail by Wake et al. (2016), we jointly optimized the bundle sizes with our target selection and arrived at the 17 science IFU bundles per plate with five different sizes: 2 19-fiber bundles, 4 37-fiber bundles, 4 61-fiber bundles, 2 91-fiber bundles, and 5 127-fiber bundles. All bundles are hexagonal in shape. Each fiber covers a 2\arcsec\ diameter aperture on the sky with a 2.5\arcsec\ center-to-center spacing between fibers, yielding a 56\% fill factor. The largest bundle has a long axis of 32 arcsec and a short axis of 28 arcsec. We define the effective radius of the bundle to be the radius of a circle that has the same area as the bundle. Table~\ref{tab:bundles} lists the salient features of the bundles. This bundle size distribution roughly matches the apparent size distribution of our targets. Each plate also contains 12 7-fiber mini-bundles and 92 single sky fibers. The sky fibers are always associated with the IFU bundles as detailed in Table~\ref{tab:bundles}, and are always plugged within 14 arcminute to the IFU bundles. In total, we have 1423 fibers that feed the two spectrographs, with 709 fibers in Spectrograph 1 and 714 fibers in Spectrograph 2.  

\begin{table*}
\begin{center}
\caption{Fiber Bundle Configuration}
\begin{tabular}{c|c|c|c|c|c}
\hline\hline
Bundle Size & Purpose  & Number of bundles & Number of & Long axis diameter & Effective radius  \\
(fibers) &  &  per cartridge & Sky Fibers &  (arcsec) & (arcsec) \\\hline
7  & Flux. Cal. & 12 & 1  & 7.0 &  5.45  \\
19 & Science & 2 &  2  & 12.0 & 7.73\\
37 & Science & 4 &  2  & 17.0 & 7.73 \\
61 & Science & 4 &  4  & 22.0 & 10.00\\
91 & Science & 2   & 6  & 27.0 & 12.27 \\
127  & Science &  5 &  8 & 32.0 & 14.54  \\
\hline\hline
\end{tabular}
\label{tab:bundles}
\end{center}
\end{table*}

We have built six sets of these fiber assemblies and installed them in each of six identical cartridges. A cartridge is a removable cylindrical box that holds the fiber assemblies and provides the interface between the focal plane of the telescope and the input to the spectrograph. Aluminum plates are mounted on one side of the cartridge; fibers and fiber bundles get plugged into holes on the plate; at the output of the cartridge, all fibers are aligned in one of two pseudo-slits which feed the two spectrographs. On the pseduo-slits, the 1423 fibers are organized into 44 groups mounted on `v-groove' blocks with small gaps in between. Depending on the bundle size, the fibers in each IFU bundle get assigned to one to four such v-groove block(s). Sky fibers associated with an IFU bundle are assigned to the same v-groove block(s) as the fibers in the bundle, and they are always positioned on the edge of the block so that there is minimal contamination of the sky spectra by galaxy light.

When the cartridge is mounted to the telescope, the plate is located on the focal plane of the telescope and the pseudo slits are inserted into the spectrographs. The light from the fibers is collimated, then split by a dichroic beam splitter into a blue camera and a red camera. The blue camera covers from $\sim3630{\rm \AA}$ to $\sim6300{\rm \AA}$, and the red camera covers from $\sim5900{\rm \AA}$ to $\sim10,300{\rm \AA}$. Finally they are dispersed by a grism and recorded on CCDs of 4K by 4K. Each CCD is readout by 4 amplifiers. The spectral resolution which depends smoothly on wavelength is around $R\sim2000$ (see Section~\ref{sec:specres}).

\section{Sample Design} \label{sec:sample}
Full details of the sample design and tiling are given in Wake et al. in prep. Here we give an overview and provide some insight on how the science requirements drive the sample design. 

Given the science requirements, we desire a sample of nearby galaxies with predetermined redshifts, with which we can estimate the stellar mass or the absolute magnitude and the environment around each galaxy. The effective radius should also be reliably measured so that we can use it to define a sample with uniform spatial coverage. The NASA-Sloan Atlas (NSA) catalog\footnote{http://www.nsatlas.org} provides an ideal basis from which to select our targets. It is based on SDSS imaging with improved background subtraction and deblending \citep{Blanton11}, and is much more complete than SDSS photometry catalogs especially for galaxies brighter than $r_{\rm AB}$ of 16 (our final sample ranges roughly between $r_{\rm AB}$ of 13 and 17). We used a newer version (v1\_0\_1) of the NSA catalog than what is available on the NSA website. This version will also be released as part of Data Release 13 (DR13) of SDSS. 

From the NSA, we select a sample with a roughly flat stellar mass distribution. The derivation of stellar mass is model dependent. Therefore, instead of stellar mass, we use the absolute i-band magnitude ($M_i$) as it has less dependency on model assumptions and is more easily reproducible. Thus, we build our sample to have a flat distribution in absolute $i$-band magnitude. The absolute magnitudes for the sample are derived using the software package {\it kcorrect} (v$4\_2$) \citep{BlantonR07}. 

\begin{figure}
\begin{center}
\includegraphics[width=0.5\textwidth]{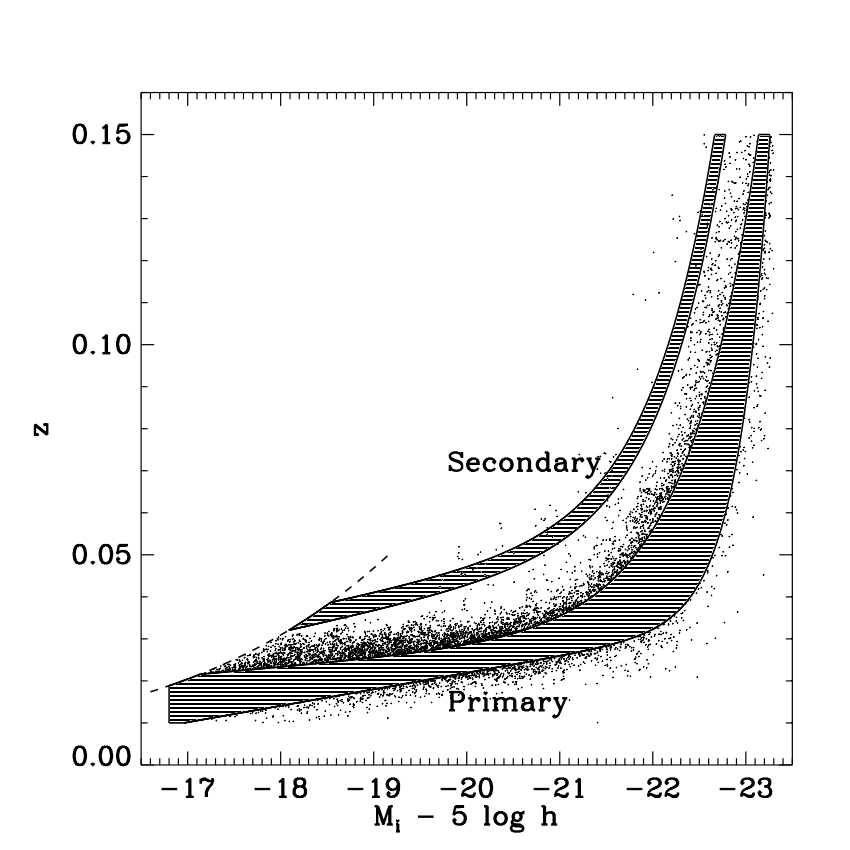}
\caption{This figure shows the luminosity-dependent redshift cuts we use to select the sample for MaNGA. The two shaded bands indicate the Primary sample (the lower band) and the Secondary sample (the upper band). The dashed curve indicates the completeness limit applied at faint magnitudes. Both the Primary and the seondary samples are strictly volume-limited at each $M_i$. The dots represent the Color-Enhanced supplement, which is built to oversample the under-populated regions in color-magnitude space. See text and Wake et al. for more details.}
\label{fig:samplecuts}
\end{center}
\end{figure}

We would like to maximize spatial resolution while ensuring the majority of the sample is covered to a certain radius. The latter constraint means we need to set a minimum redshift so that the angular size of galaxies can fit within our largest bundles. Because the sizes of galaxies increase with stellar mass or luminosity, the minimum redshift has to increase with brightening $M_i$. The maximum redshift is then set accordingly so that we have roughly the same number of galaxies in each absolute magnitude bin. The number per $M_i$ bin is set by the appropriate total number density of targets on the sky to ensure high completeness and high efficiency in targeting. In order to not bias the intrinsic sampling of galaxies, we conduct a volume-limited selection for each absolute magnitude. Our final luminosity-dependent redshift cuts are shown in Figure~\ref{fig:samplecuts} and will be provided by Wake et al. (in prep).



MaNGA's main galaxy sample is composed of three components: Primary sample, Secondary sample, and the Color-Enhanced supplement. Both the Primary sample and the Color-Enhanced supplment aim to cover 1.5~\Reff; the Secondary sample is designed to reach 2.5~\Reff. To balance the potential science opportunity provided by the Secondary sample and the risk associated with it, we decided to allocate 1/3 of the bundle resources to the Secondary sample. The other 2/3 will be spent on the Primary and Color-Enhanced samples. As they have the same spatial coverage goal, the Primary sample and the Color-Enhanced supplement are collectively termed the `Primary+' sample. 

The Primary sample is a volume-limited sample at each absolute i-band magnitude, as is the Secondary sample. The Color-Enhanced supplment is designed to supplement the Primary sample by providing denser coverage in parts of the color-magnitude diagram with fewer galaxies. For example, green valley galaxies, massive blue galaxies, or the least massive red galaxies are rare in a volume-limited flat-stellar-mass sample. 
The mechanism to sample these rarer galaxies is to expand the redshift limits specifically for the relevant color-magnitude bins. The total number of the Color-Enhanced supplment is set to be 1/3 of the Primary sample. For details of how the supplment is defined, see Wake et al. (in prep).



\begin{figure}
\begin{center}
\includegraphics[width=0.5\textwidth]{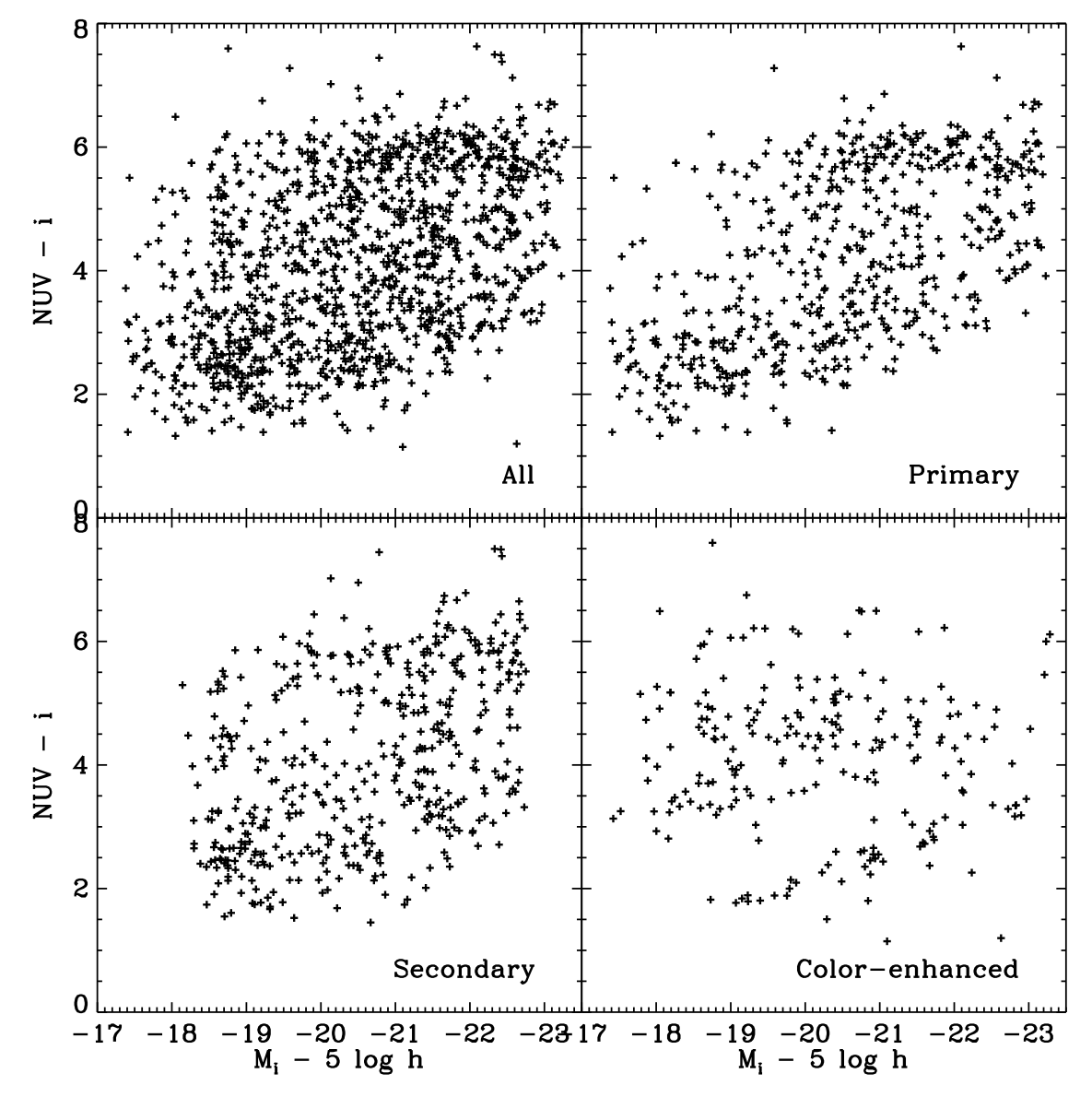}
\caption{Color-magnitude distribution of galaxies observed in our first year of observations, which will be released in Data Release 13. Note the color and magnitudes here are k-corrected to restframe and absolute magnitude, but are not corrected for internal extinction. The upper left panel shows all targets observed; the upper right panel shows the Primay sample; the lower left panel shows the Secondary sample; and the bottom right panel shows the Color-enhanced sample.}
\label{fig:cmd_dr13}
\end{center}
\end{figure}

In Figure~\ref{fig:cmd_dr13}, we show the color-magnitude distribution of galaxies observed in our first year of observations. The color and magnitudes here are k-corrected to restframe color and absolute magnitudes. However, there is no correction for internal extinction. Consequently there are trends in the distribution of color-magnitude that correlate with inclination, particularly for blue galaxies. This effect could be significant for the Color-Enhanced supplement as it selects preferentially face-on systems at blue colors and edge-on systems at red colors.

\begin{figure}
\begin{center}
\includegraphics[width=0.5\textwidth]{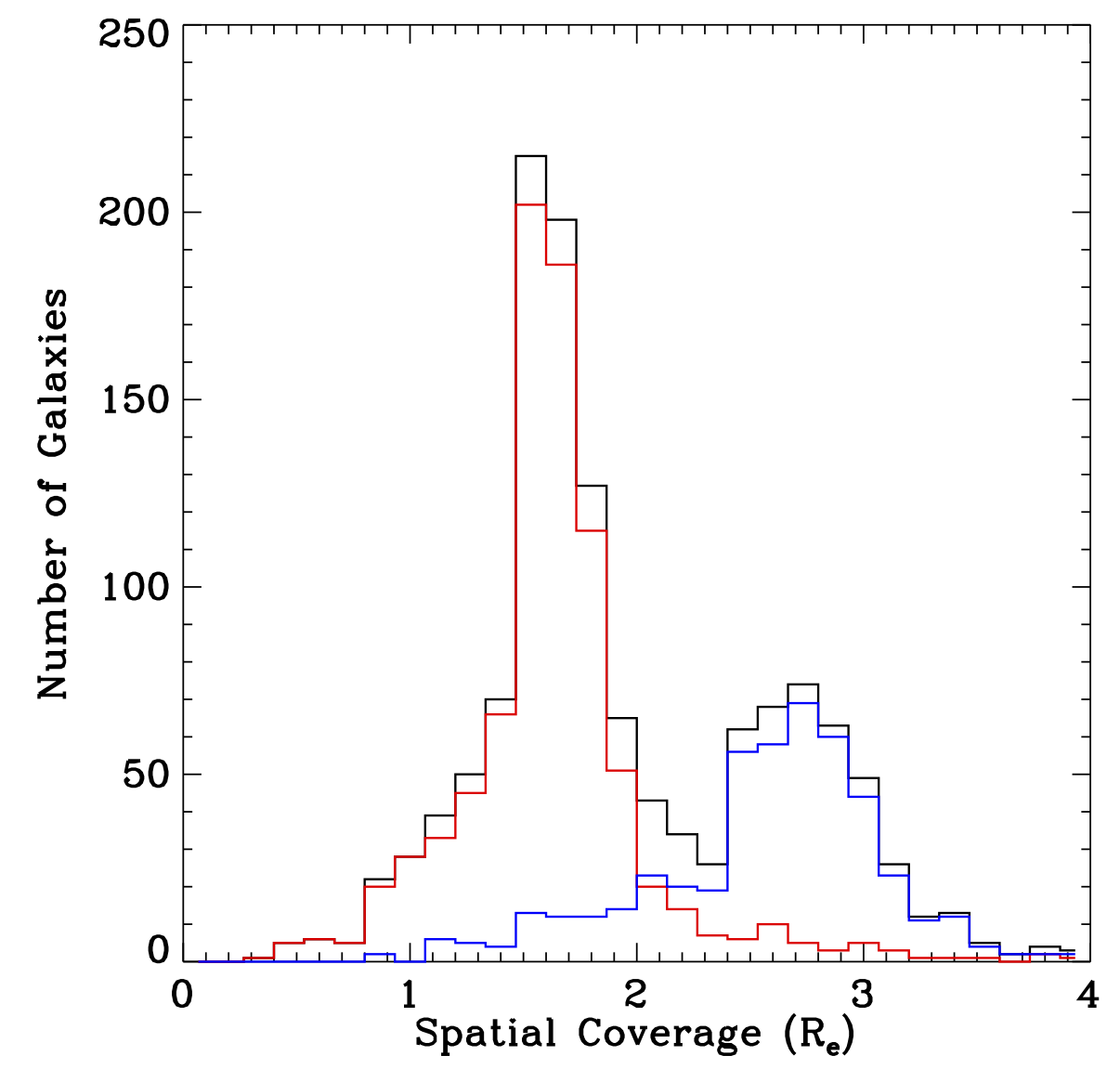}
\caption{Spatial coverage distribution of galaxies observed in our first year of observations. The red histogram shows that for the Primary+ sample (Primary and Color-Enhanced). The blue histogram shows that for the Secondary sample. The black histogram is the total distribution.}
\label{fig:spatialcoverage}
\end{center}
\end{figure}

We would like to warn the users of the MaNGA sample that the sample is not a simple volume-limited sample with a single luminosity cut. Just taking the sample as a whole does not yield a representative sample of the galaxy population at any redshfit. For almost all science topics, one needs to carefully weight each galaxy by the inverse of its selection probability to reconstruct what a volume-limited sample would be like. In Wake et al. (in prep.), we will provide the detailed prescription about how to reconstruct a volume-limited sample.


We allocate the targets to a large number of 7 square degree circular tiles that cover the entire SDSS DR7 footprint. The tiles are allowed to overlap to achieve roughly equal sampling completeness in different environments. Overall, we achieve a high allocation efficiency, with less than 2\% of all fiber bundles unallocated.
The tiles that do not have enough target galaxies were eventually filled with ancillary targets or galaxies outside our selection cuts.  In total, there are about 1800 tiles and we expect to only observe about 575 to 600 of them, given the limited observing time. The choice of which tiles to observe will be discussed below in Section \ref{sec:fieldchoice}. 


There are a large number of ancillary science programs that could be done with the unique capability of the MaNGA instrument. In summer 2014, we had a call for ancillary proposals within the SDSS-IV collaboration. We received a large number of requests. Some of these could be used as fillers on plates with available bundles, others target rare objects that would not be observed unless we replace main sample targets with them. Given the large demand, we decided to dedicate 5\% of fiber bundles, in addition to the unallocated bundles, to ancillary target observations. In the end, 95.7\% of all bundles are assigned to the main sample targets (45.1\% Primary, 15.0\% Color-Enhanced, and 35.7\% Secondary) and 5.1\% are assigned to ancillary program targets. There is a 2.3\% overlap between these two categories: some high value ancillary targets also belong to our main sample. After including the ancillary targets, the total allocation efficiency is 98.5\%. When designing plates, the 1.5\% of unallocated bundles are used to repeat some already-observed targets or assigned to randomly-selected filler galaxies. We also achieve relatively high completeness in targeting: 85\% of primary sample and 66\% of secondary sample galaxies within our redshift cuts are targeted.



Once targets are assigned tiles, we allocate the different sized IFUs according to the spatial coverage goals of the galaxies in that tile. 
Although the IFU size distribution is designed to match the size distribution of our targets, the matching is not perfect for each tile. When they do not match, we try to assign larger bundles to galaxies with a larger intended coverage. Figure~\ref{fig:spatialcoverage} shows the final spatial coverage distribution of galaxies observed in the first year (DR13). Here the spatial coverage is defined by the ratio of a bundle's effective radius to the major axis~\Reff\ of the target galaxy. 69\% of the Primary+ sample are covered beyond 1.5~\Reff, and 66\% of the Secondary sample are covered to larger than 2.5~\Reff.

\section{Design and Execution of the Observations} \label{sec:execution}

In this section, we describe the operation of the survey in detail. We will first summarize how we set our observing strategy and plate completeness thresholds, then describe how we chose the fields to observe, how we design the plates, and finally how we conduct the observations. 


\subsection{Observing Strategy} \label{sec:observingstrategy}

We first describe our observing strategy as it sets basic requirements for the operation. 

Our fibers are 2\arcsec\ in size, while the typical seeing at Apache Point Observatory is 1.5\arcsec, and the fiber-convolved point spread function (FCPSF) is about 2.5\arcsec, full-width at half-maximum (FWHM). The center-to-center spacing between fibers in our fiber bundle is 2.5\arcsec. With a single pointing for exposures, we will not sufficiently sample the FCPSF. Thus, we need to conduct dithered observations. We employ a 3-point dither pattern which forms an equilateral triangle with 1.44\arcsec\ on a side. This still does not Nyquist-sample the FCPSF. In order to achieve Nyquist-sampling, we would need to execute a 9-point dither, but that is not practical given the scheduling complexity. Therefore, we adopt the 3-point dither as a compromise. The dither pattern is illustrated in Figure 6 of \cite{Law15}.

The MaNGA survey benefits from two unique capabilities provided by the Sloan Telescope and the BOSS spectrograph: the wide field of view and the wide wavelength coverage. However, these same features also bring complications. 
First, the large field of view results in a large variation in atmosphere refraction within the field, which compresses the field in the parallactic angle direction. Second, the wide wavelength coverage of the spectrograph results in a large chromatic differential atmosphere refraction (DAR). Away from zenith, the extreme blue and extreme red wavelengths are both offset from the guiding wavelength. 

Combining both effects, the exact dither pattern is a complicated function of the wavelength, the position of the field on the sky, and the time sequence of the dither pattern. Sometimes the two effects cancel while at other times they add, making the dither more irregular. In order to maintain sampling uniformity, we have set constraints on where and how to execute the dither sequence. \cite{Law15} did extensive simulations to study this effect. We came up with a metric termed the `Uniformity Statistic' ($\Omega$) which is the maximum offset among all locations on a plate and across all wavelengths between the first and last exposures in a standard 3-point dither sequence due to both the uncorrected portion of the atmosphere refraction and the chromatic differential refraction. Generally speaking, $\Omega$ increases with the time separation between dither moves, and with increasing airmass. 
Given our simulation, in order to satisfy the image quality requirements set in Section~\ref{sec:imagerequirement}, $\Omega$ must be less than 0.4\arcsec\ for all wavelengths and all positions on a plate. With this requirement, we set the following guidelines for our observing strategy, which are justified in detail by \cite{Law15}.

\begin{enumerate}
\item Our exposures are set to 15 minutes each in order to maximize signal-to-noise and not be overwhelmed by cosmic rays in the red detectors. 
\item All exposures of a plate have to be taken within a predefined window centered on Meridian (referred as `the visibility window'), with the length of the window set by the Declination of the plate.
\item Each set of the three dithers must be taken within 1 hour of each other in hour angle.
\item All exposures in a set must have individual seeing measures better than 2.5\arcsec (FWHM), a min-to-max variation within the set less than 0.8\arcsec, and a set-averaged seeing better than 2.0\arcsec.
\item All exposures in a set need to have $(S/N)^2$ within a factor of 2 from each other in all four cameras.
\end{enumerate}

The visibility window is set by the declination of the field. The implemented visibility window length is different from the ideal case given by \cite{Law15}. The latter dictates very short windows for fields on the Celestial Equator or at ${\rm Dec}<0$. This makes scheduling difficult at these declinations. Therefore, we relax it slightly for practical reasons. In the simulations presented by \cite{Law15}, we assumed the time span of the three exposures (from the beginning of the first exposure to the end of the last exposure) is one hour, which allows for some moderate delays between exposure for unforseen situations. In reality, most of the time the exposures are executed consecutively without additional delays. Therefore, we assume a shorter window of only 48 minutes (three 15 minute exposures plus two readout/flushing overheads). This results in a much longer visibility window for low declinations. 

The contraints that sets MaNGA's visiblity window limit is quite different from that of single-fiber spectroscopy, such as the BOSS and eBOSS surveys. For single-fiber spectroscopy,  one is mostly concerned with the field differential refraction but not the chromatic differential refraction. And one can mitigate the AR effect by drilling the plate according to the planned observing time \citep{Dawson13}. This helps the single-fiber surveys to cover time when the galactic plane is transiting. However, for MaNGA, the rotation of the chromatic DAR vector relative to our plate is what dominates the uniformity statistic at large hour angles and high airmass. The rotation cannot be addressed by the plate design. Therefore, the mitigation strategy used by BOSS/eBOSS to cover galactic plane transits is not applicable for MaNGA.

As a lesson for future fiber-based IFU surveys, if funding allows, an atmospheric dispersion corrector would be a valuable asset that would alleviate these problems.

\subsection{Plate Completeness Thresholds} \label{sec:completeness_thresholds}


In this section, we discuss how the science requirements in Section~\ref{sec:requirements} drive the plate completeness criteria in our observations.



As stated at the end of Section~\ref{sec:requirements_deriv}, our science requirements can be achieved if we can obtain a S/N of 33 per pixel in the r-band continuum in the stacked spectra between 1 and 1.5\Reff. To achieve this in more than 75\% of our target galaxies, we need a typical exposure time of 2.25 hours (3 sets of 3 15-minute dithers) per plate in good conditions in sky areas with low galactic extinction. 
With this typical exposure time and depth, considering the good weather fraction at APO and all associated overhead, we will be able to observe $\sim10K$ galaxies in 6 years using half of the dark time\footnote{Dark time is defined to be the time when the illumination fraction of the moon is less than 35\%, or when the moon is below horizon and has an illumination fraction betweeen 35\% and 75\%.}. 

Doubling the S/N would result in a sample size smaller by a factor of 4. A sample size of 2.5K galaxies would be too small to meet our requirements. For example, if we want to study the dependence of a galaxy property on multiple variables (e.g., mass, color, environment), adopting several bins in each variable, we would quickly find too few galaxies per bin. A sample of 2.5K also does not represent a significant advance from the sample currently available from surveys like CALIFA (667 galaxies) or SAMI (3400 galaxies). Going for 10K will enable much more additional discovery space, especially in terms of statistical signficance and rare galaxies. In addition, going for higher S/N does not necessarily lead to smaller uncertainties in the measurements of physical properties. In the above discussion of science requirements, we have ignored the systematics associated with calibration and modeling, such as the uncertainty in SFR calibration, the systematics of the strong-line methods for measuring gas metallicity, the systematics associated with the stellar spectra libraries for stellar population modeling, and the uncertainty due to simplified star formation histories. At higher S/N, these systematical errors could dominate over the random noise and prevent us from taking advantage of the deeper observations. On the kinematics analyses, the uncertainties on the angular momentum, the enclosed gravitating mass, and dark matter fraction are also limited by spatial resolution \citep{LiH16} or uncertainties in inclination determination. 

In the other direction, we could consider shorten the exposure times to a minimum, which would be 45 minutes (1 set of 3 dithers). This would triple the sample size with a decrease of 40\% in S/N. With this option, the precision of many of our measurements would be comparable to the currently-observed scatter in scaling relationships. This would prevent us from making progress in identifying the cause of the scatter.  

Therefore, we consider our current choice of the S/N threshold and the sample size of 10K the right balance to maximize the science return given the fixed amount of observing time. 

\bigskip
Below, we describe several details in setting the completeness threshold and the lessons we learned in this process. 

First, we cannot require every galaxy to reach equal depth as their sizes and surface brightnesses vary significantly. It is more practical to set a threshold on the total accumulated $(S/N)^2$ for a fixed reference fiber magnitude (or the equivalent surface brightness), to ensure the majority of them ($>75\%$) meet the science requirement. 

Second, the reference fiber magnitude should be galactic-extinction-corrected. This would yield roughly equal depth regardless of foreground galactic extinction. 

Third, although our science requirement is set on the r-band continuum, in reality, we do not track the $r$-band S/N in real time. This is because the BOSS spectrographs split the light into the blue and red cameras. The $r$-band is split by the dichroic beam splitter. Therefore, during observations, it is much easier to track the $g$-band and $i$-band, which are covered completely by the blue and red cameras, respectively. 
The correspondence between the S/N in $g$-band (or $i$-band) and that in $r$-band (for fixed reference magnitudes) changes with the level of galactic extinction and the sky background. The actual S/N thresholds have to be set through trial and error. 

Our final completeness criteria for each plate are: the total accumulated (S/N)$^2$ among all exposures in complete sets must be above 20 pixel$^{-1}$ fiber$^{-1}$ in the $g$-band for an extinction-corrected fiber magnitude of $g=22$ and above 36 pixel$^{-1}$ fiber$^{-1}$ in the $i$-band for an extinction-corrected fiber magnitude of $i=21$. We obtain as many complete dither sets until we meet these criteria.

\bigskip
We also learned a few lessons. 

First, for projections of the survey speed, it is important to have an accurate of prediction of how S/N per exposure depends on seeing, airmass, and extinction in order to predict how many exposures are needed to reach the S/N thresholds. A single-fiber survey and an IFS survey using the same telescope and instrument can have very different dependences of S/N on these factors. At the beginning of our survey, we did not have enough data in hand to fully assess how our S/N depends on these factors. Therefore, we adopted the S/N relationships used by the SDSS-III/BOSS team, which led us to set too high a S/N threshold. We thus exposed for too long initially on many plates and fell behind schedule. We corrected the problem in April 2015. 

The reason for this difference is that the S/N obtained by single-fiber surveys such as BOSS is very sensitive to seeing, as they are targeting centers of galaxies where the surface brightness profile peaks. But for an IFU survey targeting outskirts of galaxies where the surface brightness is much flatter, the S/N is almost independent of seeing. These different setups also yield different S/N dependencies on airmass due to fiber alignment issues. Our actual S/N dependence on airmass and extinction will be presented in Section~\ref{sec:snrelation} with Figures~\ref{fig:sn2_vs_airmass},\ref{fig:sn2_vs_extinction} and Equations~\ref{eqn:sn_equation_g},\ref{eqn:sn_equation_i}. 

Second, the requirement we set on the spatial sampling uniformity brings a cost to the observing efficiency. Because we dither and the 3 consecutive dithered exposures need to be obtained within 1 hour of each other, sometimes the planned dither sequence can get interrupted by bad weather or bad seeing conditions, leading to a fraction of good exposures that cannot be combined into sets, which we refer as `orphaned exposures'. Sometimes, these `orphaned exposures' can be patched on subsequent nights with exposures with similar S/N and seeing conditions into complete dither sets. Other times they are left behind and can account for 8-10\% of the total number of exposures. We do not include these exposures in the total (S/N)$^2$ calculation or the reconstruction of the data cube.  We trade a bit of observing efficiency for better data quality. These orphaned exposures could still be useful for certain science applications that require greater depth but less image quality.

\subsection{Field Planning} 



\subsubsection{Overall field choices} \label{sec:fieldchoice} 
Given our observing strategy, the visibility window of each tile, the plate completeness thresholds, and the allocated time on the telescope, we can decide which tiles to observe.  We consider the following factors in choosing the tiles. First, the tiles chosen need to cover sky regions available during a specific observing period. Second, we would like to overlap with several imaging surveys or surveys conducted in other wavebands to maximize the scientific return of the MaNGA data. Third, we would like to maximize signal-to-noise ratio obtained per hour. The second and the third considerations do not necessarily agree with each other. We balance the two needs while maintaining the total number of galaxies and obtaining sufficient overlap with other surveys, with the following three considerations.

First, understanding the star formation, chemical evolution, and baryon cycles in galaxies requires knowledge about their gas content. Direct measurements of gas require HI observations at 21 cm\ and CO observations with sub-millimeter telescopes. We therefore choose to overlap significantly with the Arecibo Legacy Fast ALFA (ALFALFA) survey \citep{Giovanelli05}, the planned Apertif Med-Deep fields\citep{Verheijen09}, and low declination regions accessible from ALMA.

Second, deep near-infrared imaging can provide an independent measure of the stellar mass distribution in galaxies, which we can compare with the mass distribution measured from stellar population synthesis modeling of our spectra and that measured from kinematics, or combine to provide joint constraints. Therefore, we choose to overlap as much as possible with the United Kingdom Infrared Telescope (UKIRT) Infrared Deep Sky Survey (UKIDSS) survey footprint \citep{Lawrence07}. 

Third, deeper and higher resolution optical imaging than SDSS would help us identify internal structures in galaxies, such as small bars, spiral arms, etc., which will help us understand trends discovered in our data. At the same time, deep imaging could provide halo mass estimates from weak graviational lensing, which is independent of the other mass measurements we have. Therefore, we choose to overlap with the Hyper Suprime-Came (HSC) Survey deep imaging fields\footnote{http://www.naoj.org/Projects/HSC/index.html}. In the Southern Galactic Cap, we have 3 stripes of SDSS legacy data. Here, we choose to prioritize Stripe 82\ in order to take advantage of the many auxiliary datasets available there. These low declination fields will provide a substantial sample for follow-up observation from ALMA and European Southern Observatory's Very Large Telescopes (VLT). 

We developed a simulation tool to optimally choose the tiles for the available time, on a night by night basis. This simulation tool can both be used to simulate the whole survey footprint or to decide which plate to drill for a particular drill run. 


To simulate the whole survey footprint, we assume 42.75\% of the nights from now until summer 2020 are clear. This is the average good weather fraction after excluding open-dome time that is cloudy or has bad seeing, and after accounting for inefficiency caused by suboptimal conditions\footnote{Over the past one and a half years, we opened the enclosure in 50\% of the time. Among the open-dome time, about 10\% (5\% of the total) have seeing poorer than 2.5\arcsec\ or has very low transparency that would not generate useful exposures. Among the 45\% of the time with good seeing and not-so-bad transparency, good exposures are generated that we could potentially use in complete sets. Among all exposures that make up complete sets, a small fraction produces less than typical S/N, due to either sub-optimal transparency or sky brightness. The last factor increases the total number of sets needed on our plates by 5\%.  To account for these factors, in our survey simulation, we use a weather factor of 42.75\% ($50\% \times 90\% \times 95\% = 42.75\%$) and assume all these time was typical clear conditions.}. Then for each night with scheduled observing time, the simulation tool tries to find the best tile to be observed. We use an observing efficiency of 75\% to account for overhead due to cartridge changes, acquistion, focusing, calibration, and readouts. We use the following logic in determining the best tile. 
Every tile has a visibility window that is determined by the declination of the plate. At the beginning of the observing period, we simulate all tiles that are visible. 
We predict the signal-to-noise obtained in each exposure using the empirical relationship obtained from our first season of observations, which is a function of airmass and galactic extinction. We simulate observing all tiles until they are complete (accumulated signal-to-noise higher than the completeness threshold),  until their visibility window expires, or until the observing time block finishes, whichever comes earlier. Among the plates that complete, we pick the one that requires the shortest time, with signal-to-noise ratio as the tiebreaker (higher signal-to-noise ratio is preferred). If none of the plates complete, we pick the plate with the highest completeness fraction. The simulation then continues with the time allocation. The plates that are partially-complete are stored so that they can be used on a later night. The signal-to-noise obtained in each exposure can be predicted according to empirical relations we obtained, which we describe in Section~\ref{sec:snrelation}. The fields are chosen from the tiled list described in Section~\ref{sec:sample}. 

\begin{figure*}
\begin{center}
\includegraphics[width=1.0\textwidth]{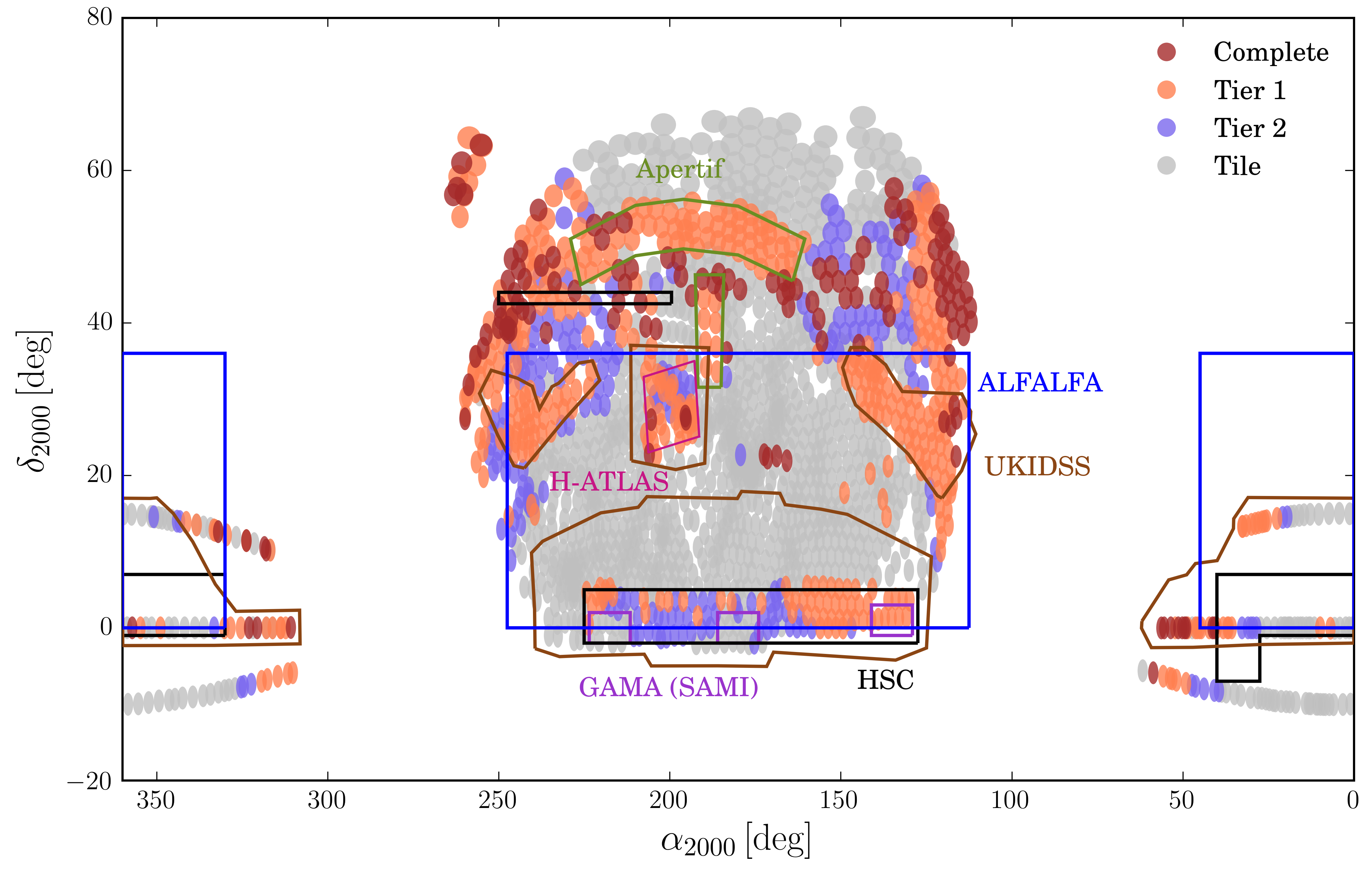}
\caption{The existing and planned footprint of MaNGA on the sky. The grey circles illustrate our tiling of the SDSS Legacy survey footprint with 3 degree diameter tiles. The red circles denote tiles we have already observed. The orange circles denote our planned footprint under the assumption that the average good weather fraction is 50\% at all local sidereal times (LSTs). This is our Tier 1 planned fields and they have high likelihood to be observed by summer 2020. Blue circles denote the additional fields we will possibly observe (Tier 2) if the good weather fraction is 75\% at all LSTs, which are less likely to be observed than Tier 1. The historical good weather fraction at Apache Point Observatory is 50\%. For an up-to-date projection of the high-likelihood tile coordinates, see http://www.sdss.org/surveys/manga/forecast/ .}
\label{fig:surveyfootprint}
\end{center}
\end{figure*}

To prioritize tiles with auxillary data available, we separate all tiles into two different categories: those inside our favored regions and those outside our favored regions. We need to exhaust those inside before picking plates outside the favored regions. 
Figure~\ref{fig:surveyfootprint} shows which tiles we will likely cover in the end. We will have significant overlap with several auxillary surveys in a number of places. Most of the tiles that do not overlap with any other auxiliary data are from already-observed fields. The fields on the edges of the NGC and SGC have to be observed becasue we need to make use of the allocated time when the galactic plane is transiting (around 5hr and 19hr LST). Most of the already-observed fields were observed prior to the field optimization decisions. 

\subsubsection{Monthly Plate Drilling Planning}

Spectroscopic observations using the Sloan Telescope needs plugplate to be prepared in advance. Therefore, at least 5 weeks before the actual observations, we choose the fields to be observed and prepare the plates. Our actual planning for MaNGA usually follows a longer lead time to protect against potential problems in plate design, drilling, and shipping. During the planning phase, given the observing schedule over the next few months, we run a simulation to decide which tiles to observe. This uses essentially the same simulation tool described above, but prioritizes plates according to the following considerations. Plates that are already are on the mountain are prioritized first, followed by those currently in the process of drilling or delivering, then followed by tiles we have not yet drilled. In this simulation, we also assumes 100\% good weather so that we will have enough plates in case of excellent weather.


\subsubsection{Daily Plugging Request}

At the mountain, after each night of observation, we have to decide which plates to observe on the next night so that they can be plugged during the day. This again requires the simulation tool mentioned above. In this case, we only consider plates that are on the mountain and have already been marked. We give highest priority to incomplete plates that are already plugged. Quite often, some partially-completed plates may have individual dithers that do not form a complete set. When simulating the next night, if a plate with incomplete sets can have those sets patched, its priority will get significantly boosted. This minimizes the chance of leaving those incomplete sets behind. Because we have strict requirements on the seeing uniformity and S/N uniformity among dithers in a set, when there is a partly cloudy night or a night with strongly variable seeing, our continuous dither sequence can be broken up into a number of incomplete sets. The patching of dither sets requires similar conditions (seeing, transparency, sky brightness) and similar hour angle of observations (see Section~\ref{sec:observingstrategy}. We have built a Web interface (codenamed `Petunia') to display incomplete set information for each plate and aid realtime plate choice decisions by the SDSS observers. 




\subsection{Plate Design}

Our target selection process allocates targets from all samples, including ancillary programs, to specific tiles and assigns them bundles of a specific sizes. 
Before plates are drilled, the individual targets on those plates are visually inspected to ensure that their photometry measurements are reliable and the automatically-determined centers are accurate. Through visual inspection, we identify those few percent of targets with problems. For galaxies with incorrect centers, we visually determine the correct centers; for galaxies with bad photometry, often due to a nearby saturated star, we reject them and replace them with other galaxies satisfying our selection cuts, including ancillary targets. If there are replacements, the bundle allocation algorithm is rerun so that the allocated bundle size for all galaxies on that plate is optimized. In the end, if we still have bundles left that are unallocated, we place those bundles with galaxies in the NSA catalog but outside the MaNGA selection cuts, with preferences given to those with the largest \Reff.

The target selection only allocates the size of the bundle, but not the placement of the bundle. Since we have multiple bundles for each size, it does not matter which galaxy goes to which bundle as long as the bundle is of the right size. Therefore, among all galaxies assigned to a given sized bundle, we assign them to physical bundles according to their closeness to each bundle's anchoring point in the cartridge to minimize the chance having a bundle's fiber cable stretched across the plate.

\subsubsection{Selection of standard stars}
For flux calibration, we target 12 standard stars on each plate simultaneously with the galaxy targets. These serve as standard stars. We select those with colors similar to the late-F type main sequence stars. F stars are chosen here because they are relatively bright and very common in the Milky Way, and have relatively smooth spectra. Hotter main sequence stars would have even smoother spectra but they are too rare at high galactic latitude. White dwarfs are also good standards but they are too faint to be found in large numbers within the appropriate magnitude range. F stars also have relatively flat spectral energy distribution in $f_\nu$ yielding comparable S/N in the blue and red portions of the spectral coverage. The cuts we use to select the F stars are the following: 

\begin{align}
m_{dist} = [((u-g) - 0.82)^2 + ((g-r)-0.30)^2 \nonumber \\ 
+ ((r-i)-0.09)^2+((i-z)-0.02)^2]^{1/2}.
\end{align}

We select those stars with an observed magnitude (PSFMAG\footnote{The PSFMAG we use are the same as given by SDSS Data Release 12. See http://www.sdss.org/dr12/algorithms/magnitudes/\#mag\_psf}) range between 14.5 and 17.2\ in the $g$-band. If we cannot find enough stars for a plate, we raise the faint limit to 17.7, or 18.2 if necessary. For late-F stars, this magnitude range ensures that they are beyond most of the galactic dust for the galactic latitude we observe ($|b|>19$). Thus, we can safely use the dust map of \cite{SchlegelFD98} to extinction-correct the colors when selecting them and apply the extinction correction to the model spectra when deriving the calibration vector. 

We pick standard stars that are widely spread across each plate so that we sample a range of different airmasses. We allocate each IFU bundle to the standard star nearest its anchor block for ease of plugging.

\subsubsection{Selection of sky positions}
As described in \cite{Drory15}, we designed the sky fibers to be located near each fiber bundle to ensure the sky subtraction is both close to the target on the sky and on the CCD. Different sized bundles have different numbers of sky fibers. The 19-, 37-, 61-, 91-, and 127-fiber bundles have 2,2,4,6,and 8 sky fibers, respectively. For the mini-bundles assigned to the standard star bundles, each has one sky fiber associated with it. The sky fibers are physically connected with their respective IFU bundle and they split off 280\ mm above the ferrule. They have a roaming radius of 14 arcminutes around their associated IFU target. The sky positions are picked from a catalog of sky locations that contain no detections in the SDSS imaging survey. This is produced by the photometric pipeline as described by \cite{Stoughton02}.


After the plates are designed, the design files are sent to the drill shop at the University of Washington. Plates are drilled, measured, and then shipped to APO.



\subsection{Observing preparation and procedure}
At the mountain, when plates are received, they are hand-marked according to the design to facilitate plugging. 

Different from SDSS-III/BOSS and SDSS-IV/eBOSS, the plugging of the MaNGA IFUs are deterministic. Each IFU has a designated hole. Therefore, each IFU hole is marked with its ferrule plugging ID (FRLPLUG) which runs from 20 to 37 for the science IFUs, and from 51-62 for the standard star IFUs. We started from 20 to avoid the confusion with the 16 guide fibers which are also numbered. The associated sky fiber holes are also identified in the marking, but do not require deterministic plugging.

The plates for a given night are mounted into different cartridges before the fibers are plugged into the plate. The other ends of the fibers go to one of two pseudoslits on the side of the cartridge. When a cartridge gets mounted to the telescope, the plate is on the focal plane, while the two pseudoslits are inserted into the two BOSS spectrographs, in which the light is collimated, split into blue and red channels, dispersed, refocused, and recorded by the CCDs in the blue and red cameras separately.

After the plate is plugged, they are mapped by shining a laser beam down each fiber on the pseduo-slit end and stepping the laser along the slit so that each fiber gets illuminated sequentially. On the plate side, we cover the plate with a diffusion screen and record a video of the fiber illumination sequence. The video is then analyzed to find the correspondence between the fibers and the holes in which they are plugged. Since MaNGA IFU bundles are deterministically plugged, this step is only used to confirm the bundles are plugged correctly and are passing light, and to map the sky fibers which are plugged arbitrarily within their own group. The resolution of the video camera is not sufficiently high to distinguish individual fibers within a bundle. The mapping of fibers within a bundle is also deterministic. They were built according to a specific mapping configuration for each sized bundle \citep[see][]{Drory15}. We verified the mapping within bundles on the teststand at University of Wisconsin. The information about fiber mapping in each bundle is recorded in a central metadata repository used by the data reduction pipeline \citep{Law16}. 

When the plate is bolted down inside the cartridge, they are slightly bent to mimic the shape of the focal plane at 5300\AA. The curvature and height of the plate is measured using a profiling bar which indexes to 8 known radial locations on the cartridge.  The bar consist of 5 digital linear micrometers placed at 5 radial positions on the plate.  In total 40 measurements across each plate are taken and stored in the plate database. If the plate curvature does not meet a specified tolerance range, the central pin is adjusted until all numbers are within specifications.



During night-time operations, the on-site observers mount the pre-plugged cartridge onto the telescope. Before observations, the spectrographs need to be focused. The focus can be adjusted at three places: the position of the collimator which are pushed by three pistons, the red camera and the blue camera which can be adjusted by turning their respective focus ring. The red camera focus ring is usually held fixed. The measurement of the best focus is done by taking two arc lamp exposures with half of the collimated beam blocked by a Hartmann screen. The Hartmann screen can block either the left or right half of the beam. Two exposures are taken, one with the left half blocked and the other with the right half blocked. If the spectrograph is out of focus, one will detect a shift in the line centroid between these two Hartmann images. The shift can then be used to compute the movement needed to achieve best focus, based on data obtained from a focus sweep. 


This focusing step is done everytime a new cartridge is mounted, but in a slightly different way for afternoon preparation compared to night observations. During afternoon checkout, we use the shift to determine the necessary collimator movement to get the red camera into focus. Then we retake another set of hartmann exposures to determine the best blue camera ring adjustment needed to get the blue camera into focus. The blue camera focus ring needs to be moved manually, which can take several minutes to finish. During the night observations, to reduce the overhead, only the collimator is moved. We compute the movement needed to balance the amount of defocus between the blue and the red cameras to reach an acceptable compromise. 

The above description would have been accurate if the temperature does not change between afternoon checkout and night observations. In reality, the temperature change can significantly alter the relative focus between the blue and red cameras. Thus, during afternoon checkout, observers intentionally set the cameras out of focus in the right direction by an amount that is empirically determined, so that the relative focus between red and blue cameras stay within tolerance throughout the night. 

After focusing, we take a 4 second arc lamp exposure and a 30-second (before MJD 57325, Oct 29, 2015) or 25-second (on and after that date) flat field for calibration. We reduced the flat from 30 seconds to 25 seconds integration because a region of the red camera in Spectrograph 1, which was replaced in July 2014, becomes non-linear in one quadrant above 35,000 ADUs. All 30-second flats taken before the switch are still in the linear regime. We reduced the exposure time to build a buffer zone for potential variations in flat lamp brightness. After calibration frames, we open the flat field pedals of the telescope, and acquire the field using first the two acquisition guide stars, then all the other 14 guide stars. We then apply the first dither offset and start the exposure sequence. We take as many dithered exposures as necessary to complete a plate or until the end of its visibility window. We also take another set of arc and flat after the science exposures if we have stayed on the same field for more than an hour.


\subsubsection{Focus Optimization for MaNGA}

\begin{figure}
\begin{center}
\includegraphics[width=0.225\textwidth]{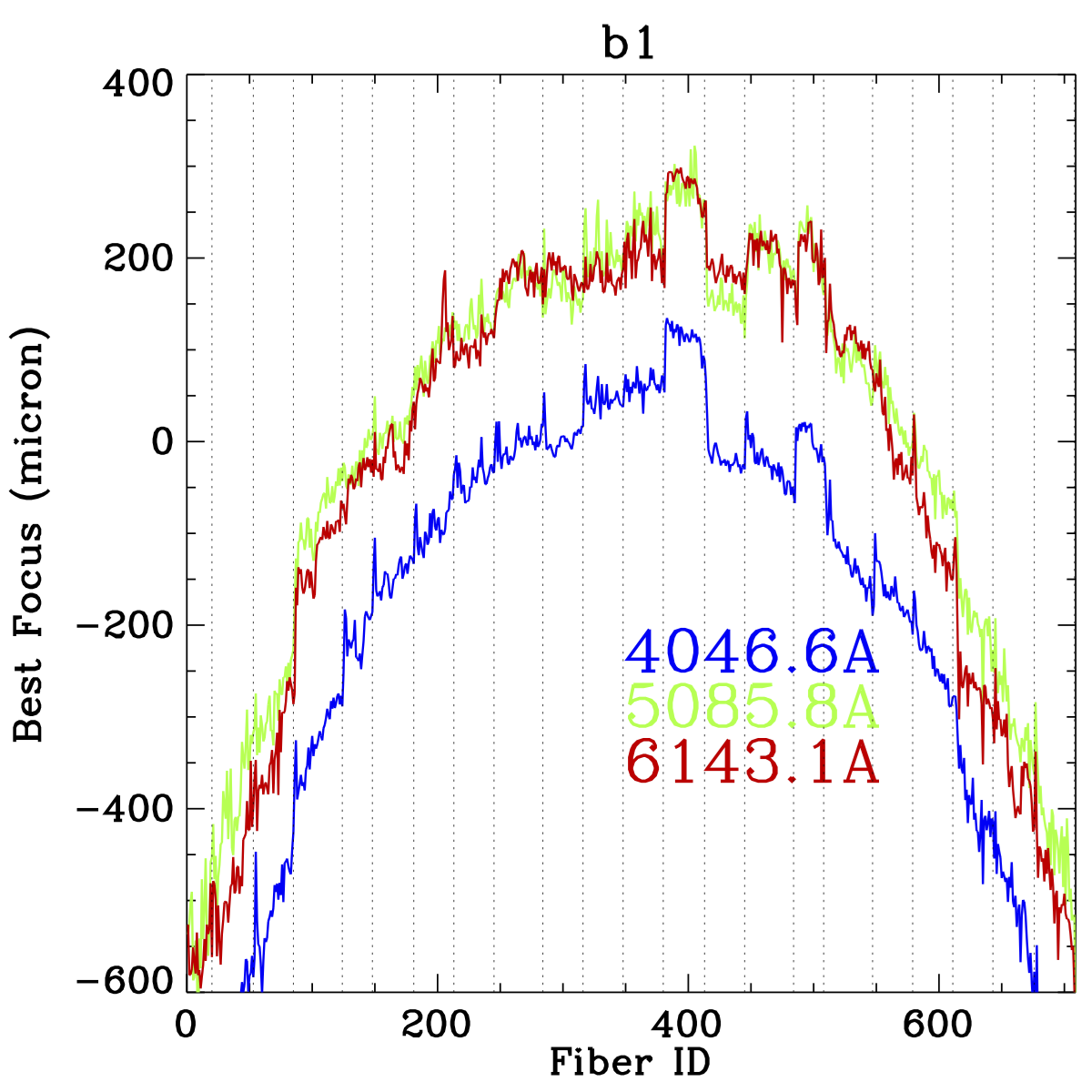}
\includegraphics[width=0.225\textwidth]{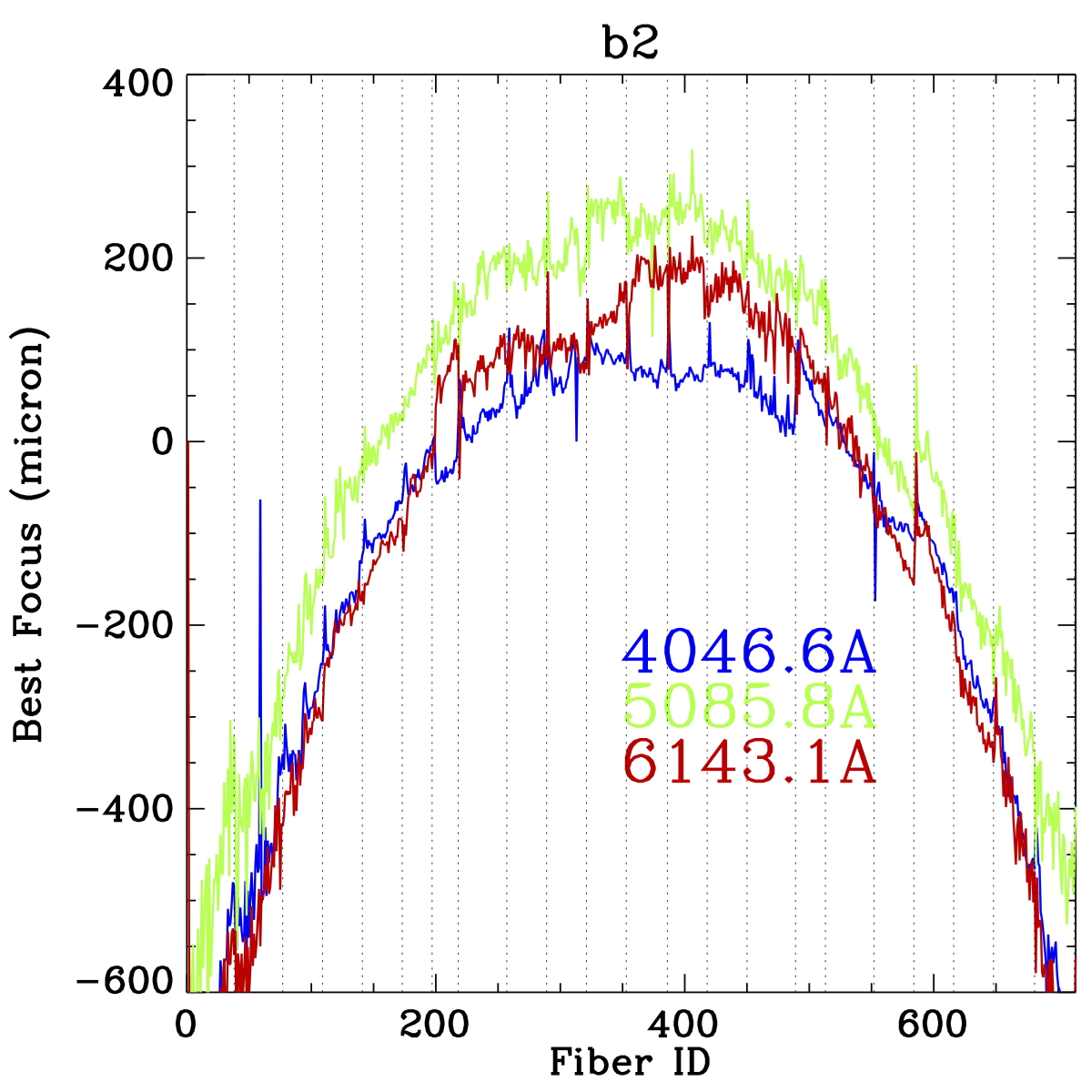}
\includegraphics[width=0.225\textwidth]{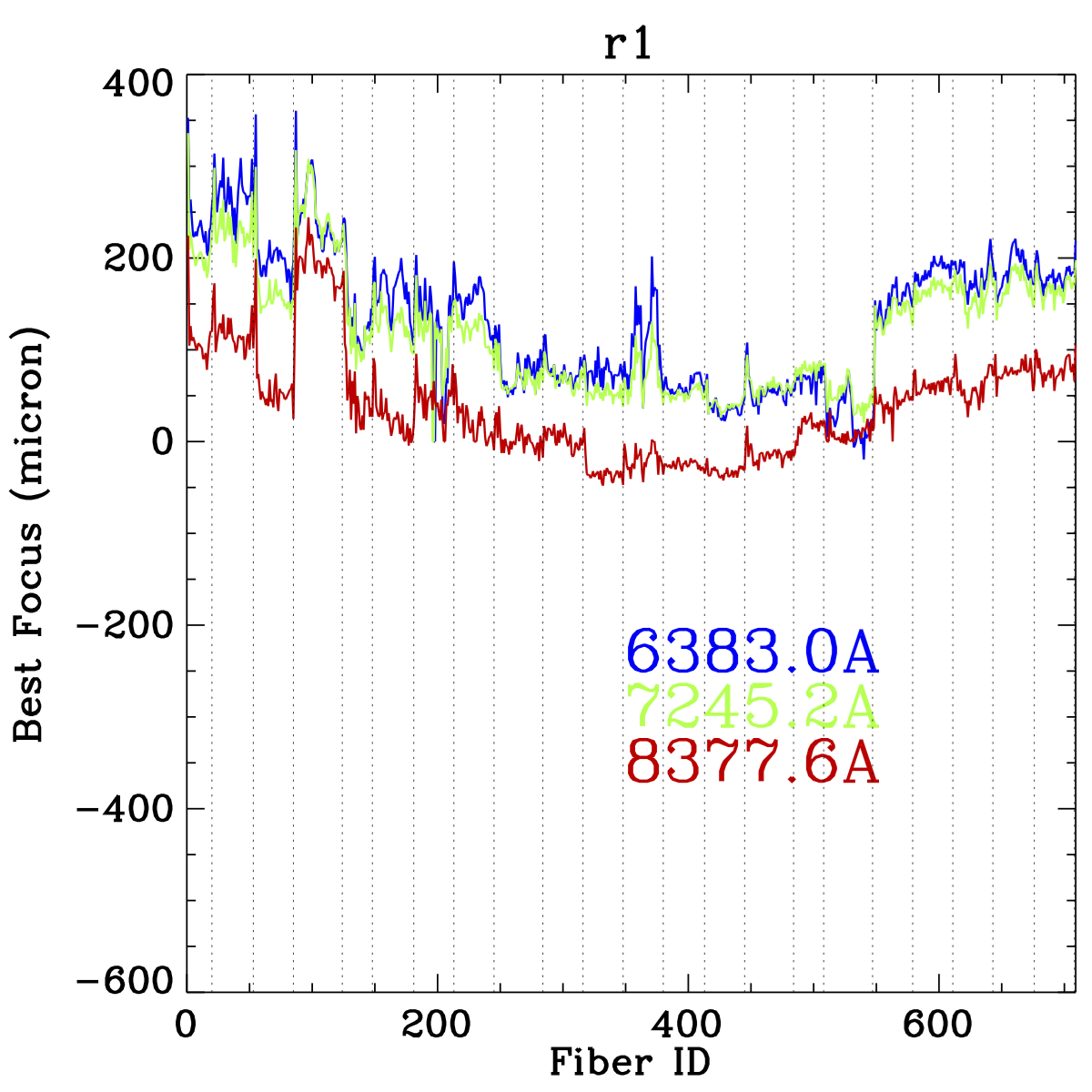}
\includegraphics[width=0.225\textwidth]{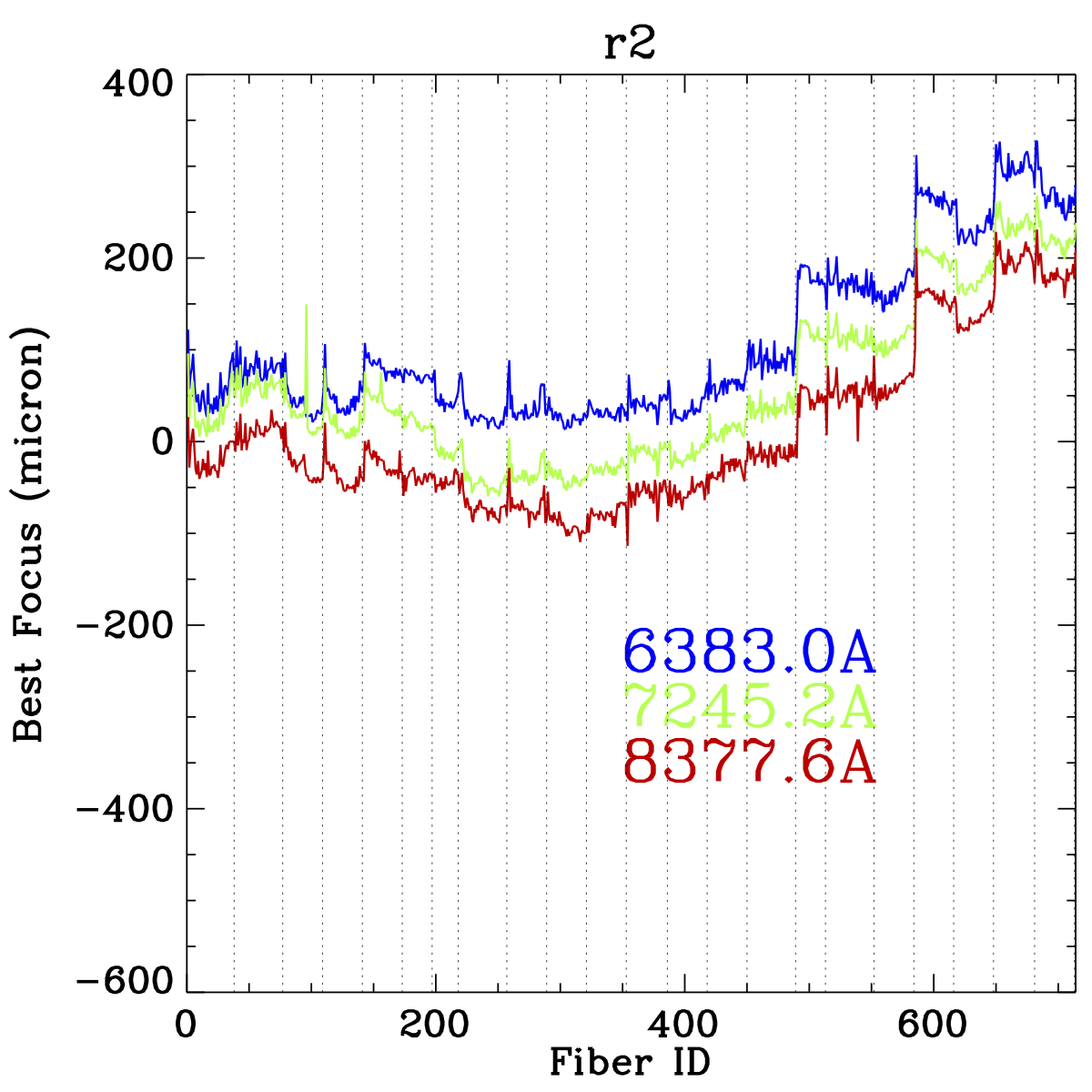}
\caption{The best focus (expressed as the relative collimator position to produce the sharpest LSF) as a function of fiberid for the blue and red camera at a few different wavelengths. The thin dotted vertical lines mark the boundaries between v-groove blocks.}
\label{fig:focalplaneshape}
\end{center}
\end{figure}
\begin{figure}
\begin{center}
\includegraphics[width=0.5\textwidth]{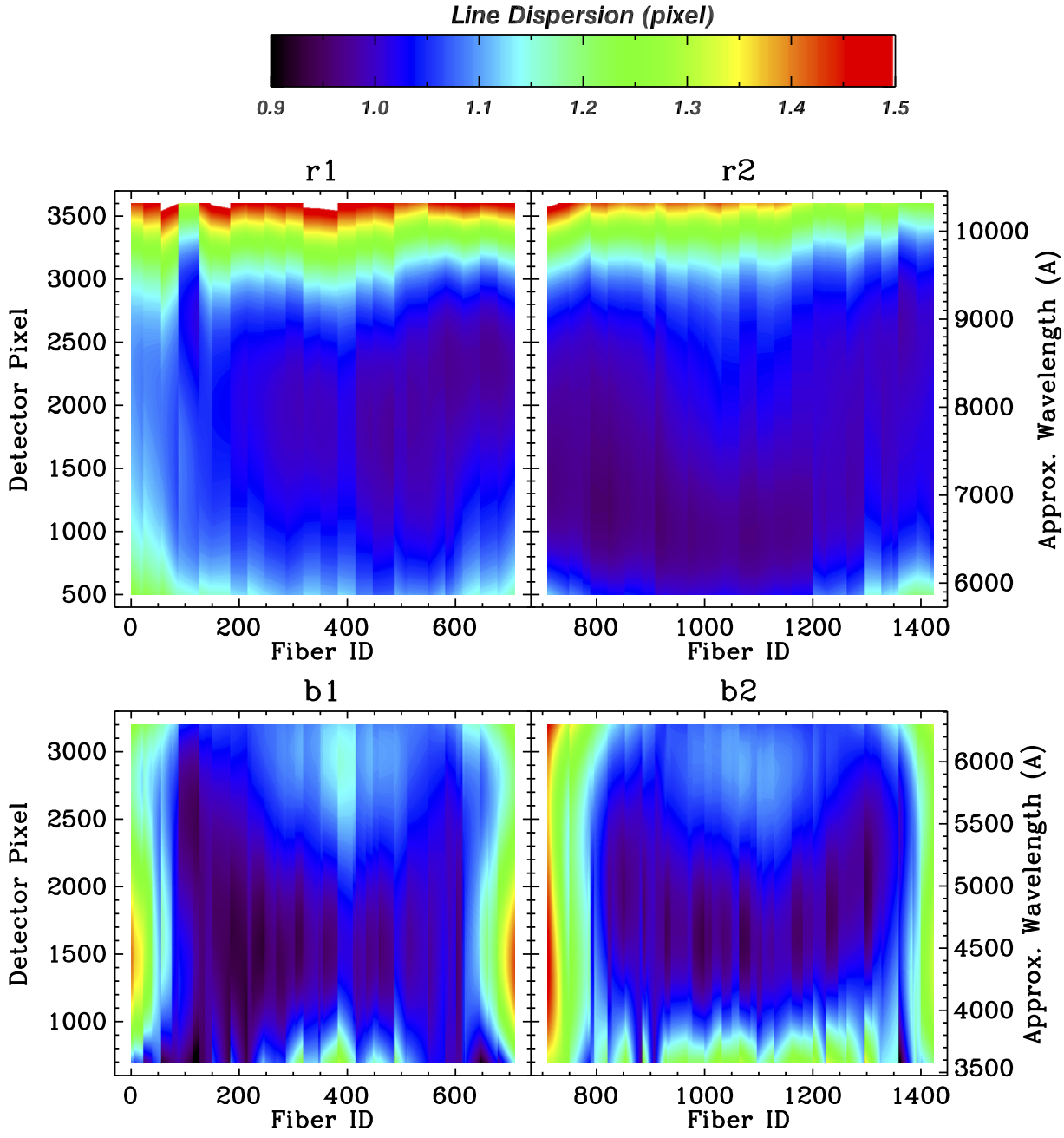}
\caption{Maps of arc line dispersion on the detector for the 4 cameras. The X-axis is the spatial direction along the slit. The Y-axis is the dispersion direction.}
\label{fig:dispersionmap}
\end{center}
\end{figure}

The focal plane of the spectrograph (relative to the surface of the CCD) at the position of the CCD is not flat. It is curved in both the dispersion direction and the spatial direction. Because the focus rings of the cameras are much more difficult to move than the collimator, we use the collimator moves to indirectly probe the shape of the focal plane. We have conducted focus sweeps by moving the collimator from one side of the best focus to the other, taken a flat and an arc at each step. From these focus sweeps, we empirically determine the best focus for each location on the CCD, described by the position of the collimator that yields the sharpest arc line for that location. Figure~\ref{fig:focalplaneshape} shows the best focus as a function of fiber ID for 3 wavelengths on each of the 4 cameras. We pick 3 arc lines in each camera located at the bottom, middle, and top of the CCD. From this figure, we see that the blue cameras have a much more curved focal plane than the red cameras. This causes larger resolution variations among fibers in the blue cameras. When the fibers near the center of the slit are in best focus, the edge fibers are significantly out of focus. In SDSS-III/BOSS, best focus is defined by the central fibers of the slit. This leads to relatively poor resolution in the edge fibers and steep resolution gradients within the fiber blocks at the edges. To mitigate this, we changed the compromise point to the position when fibers located at roughly 1/4 and 3/4 the length of the slits are at their best focus. This is a better compromise as it minimizes the resolution variation among fibers. Figure~\ref{fig:dispersionmap} shows the result. The central fibers now have slightly worse resolution, but the dispersion in line width among fibers is significantly reduced while the same median line width is maintained. 

The focal planes in the red cameras are much flatter. As shown in Figure~\ref{fig:dispersionmap}, we have much less variation in resolution in the red cameras along the spatial direction. 
The dispersion of the red camera degrades significantly towards the red end of the spectrum, probably because these photons are absorbed at a significant depth within the thick chip. We adjusted the best focus for the red camera slightly so that the red end can be in a slightly better focus.

In the proto-type observation conducted in January 2013, we discovered that the red camera in Spectrograph 1 had significant coma at the red end of its wavelength coverage. We studied the history of the coma and found it had been in existence since summer shutdown of 2010. The cause turns out to be that the CCD was not placed at the correction position. This was corrected during the summer shutdown in 2014. There is still a small amount of coma left but the quality returned to the same level as the beginning of SDSS-III. Our sky subtraction quality meets specification after this correction.


\subsection{Dithered observations with guider offset}\label{sec:dithering}
MaNGA observation is dithered by executing guider offsets. 
The guider system uses 16 coherent imaging fiber bundles centered on 16 guide stars in the field. These imaging bundles are plugged in the same physical plate but they are directed to a separate guider camera. The guider camera monitors the positions of the stars inside these 16 imaging bundles, taking a 15 second  exposure every 30 seconds. The images are analyzed to find the offset of the stars relative to their expected positions. The software solves for the optimal axes, rotation, and scale offsets to correct the pointing of the telescope. This keeps all the science fibers pointed at their respective targets. The resulting pointing stability of the telescope is 0.12\arcsec, which is the RMS offset reported by the guider software. 

To dither, we simply modify the location of the expected guide star position. Rather than requiring all the stars be located at the center of the guider fiber bundle, we require them to be located 0.83\arcsec\ from the center in the direction of the dither. This results in slightly worse pointing stability. The reason is that the guider bundle's rotational orientation affects the expected guide star position with the bundle. The angle can be measured in the lab to better than 10 deg. But there is a +/-3.5 deg uncertainty when guide fibers are plugged into the holes from mechanical tolerances. A poor knowledge of the actual angle can propagate to an error in the expected position of the guide star. Another error source is poor knowledge about the exact rotation center in the guider bundle. This is determined from the flat field image of the guider output. In a flat field image, each guider bundle shows a small circle. The center of the circle is set to be the rotational center. However, given mechanical error of the centering of the guider bundle within its ferrule and the plughole tolerance, the actual rotation center can be slightly offset from the center of the flat field circle. Additionally, the position and scale of the guider CCD could potentially change with time and flexure relative to the guider bundle output block on the side of the cartridge. We do not know the level of impact of each of these factors. However, we can give an empirical assessment on our guiding accuracy from our data because we are doing integral field spectroscopy. Comparing our data with imaging reveals exactly how the galaxies are positioned relative to our fiber bundles. We present this comparison in Section~\ref{sec:omega}.



\subsection{Quicklook Verification of Data Quality}

\begin{figure}
\begin{center}
\includegraphics[width=0.45\textwidth]{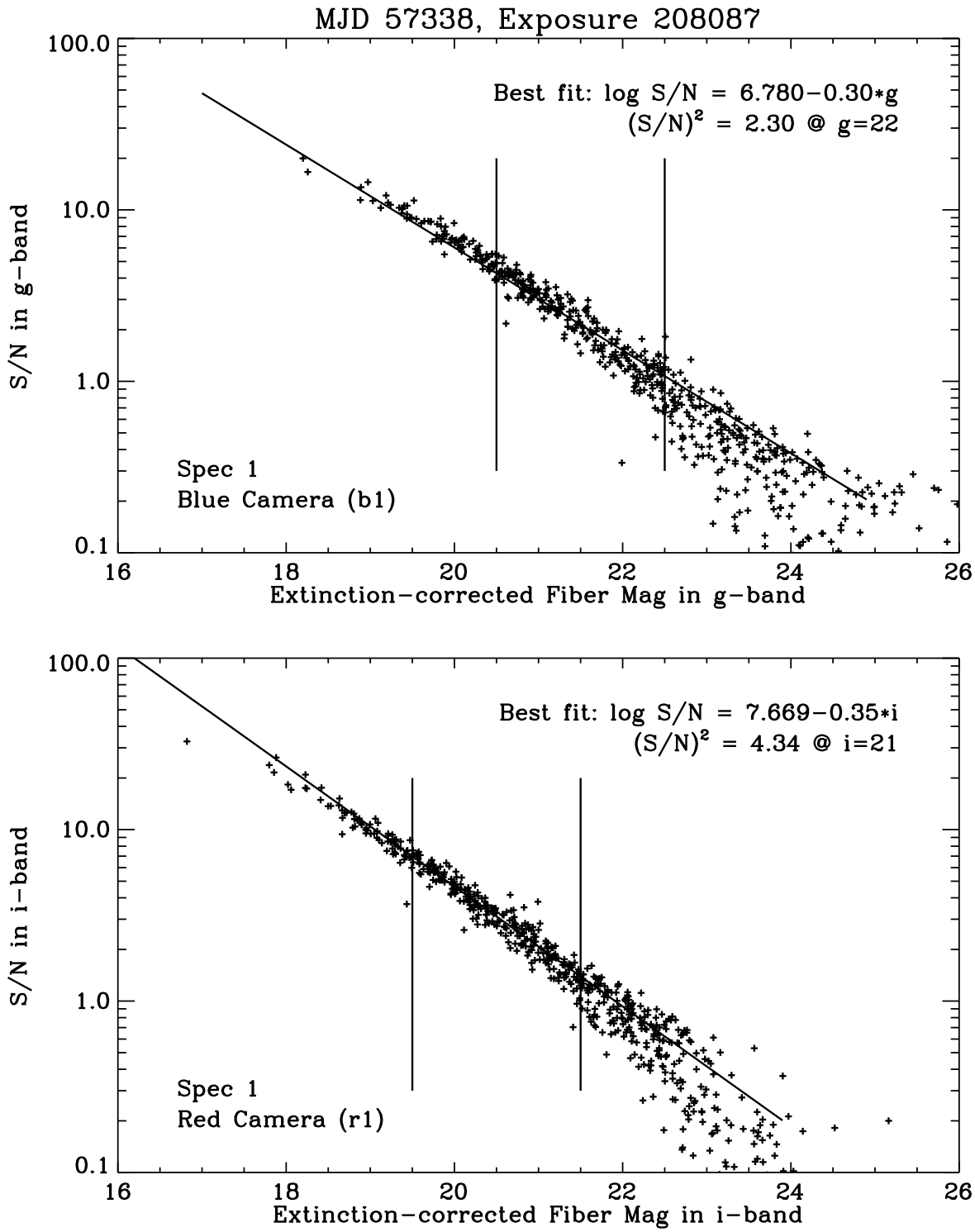}
\caption{Onsite quick reduction results showing the Signal-to-Noise ratio for all fibers in Spectrograph 1 targeting galaxies as a function of the galactic-extinction-corrected fiber magnitudes computed from SDSS images. The vertical lines indicate the range over which we fit the linear relationship with a fixed slope. The top panel is for the blue camera for which the S/N is measured in $g$-band and fit as a function of $g$ magnitude.  The bottom panel is for the red camera for which the evaluation is done in $i$-band.}
\label{fig:logsn_vs_mag}
\end{center}
\end{figure}

As mentioned above, we have a quick-reduction pipeline running at APO called `DOS'. As soon as each exposure finishes reading out, DOS does a quick-and-dirty reduction of the data. It outputs the signal-to-noise obtained for each fiber in the blue and red cameras. It then plots them as a function of expected fiber magnitudes based on the SDSS imaging. It does a linear fit with fixed slopes between the $(S/N)^2$ and the fiber magnitudes for magnitudes between 20.5 and 22.5\ in the $g$-band, and between 19.5 and 21.5\ in the $i$-band. It then outputs the best-fit (S/N) for $g=22$ and $i=21$ as the reference (S/N) for this exposure. As we have two spectrographs and each has two cameras, we have four $(S/N)^2$ values output for each exposure. Figure~\ref{fig:logsn_vs_mag} shows examples. 

The plate completeness judgment is defined by the total accumulated $(S/N)^2$ at the two reference magnitudes in the blue ($g=22$) and red ($i=21$), averaged between the two spectrographs. We require the average accumulated $(S/N)^2$ in the blue to be greater than 20 pixel$^{-1}$ fiber$^{-1}$ and the $(S/N)^2$ in the red to be greater than 36 pixel$^{-1}$ fiber$^{-1}$ for a plate to be considered complete.


\subsection{Data Reduction}

After each night of operations, the data are transfered from APO to the Science Archive Server at the University of Utah, where they are processed by the MaNGA Data Reduction Pipeline (DRP, \citealt{Law16}). The DRP processes the raw data files and produces sky-subtracted, flux-calibrated, coadded data cubes for each galaxy. There are two parts of the DRP: a 2d stage and a 3d stage. The 2d stage processes the raw frames to produce one sky-subtracted and flux-calibrated 1-d spectrum for each fiber in each exposure. The final spectra are sampled on a common wavelength grid. We provide two kinds of wavelength sampling: evenly-spaced in logarithmic wavelength and evenly-spaced in linear wavelength. The 3d stage of the DRP combines all the 1-d spectra associated with each target from all exposures to produce the final data cube for that target, with astrometry adjustments made to each exposure by comparing the data against SDSS images. Besides the final data cube, the 3d stage also makes available all the 1-d spectra per fiber, per exposure associated with each target in a row-stacked form. This is referred to as the `Row Stacked Spectra' (RSS). The data reduction pipeline is described in detail by \cite{Law16}.

\section{Signal-to-Noise Accumulation Speed and Survey Progress} \label{sec:progress}

\subsection{S/N as a function of fiber magnitude}


Here we present the S/N ratio obtained as a function of the fiber magnitude (or surface brightness). 
This differs from our original expectations based on archival data from the BOSS survey, as explained below.

\begin{figure*}
\begin{center}
\includegraphics[width=\textwidth]{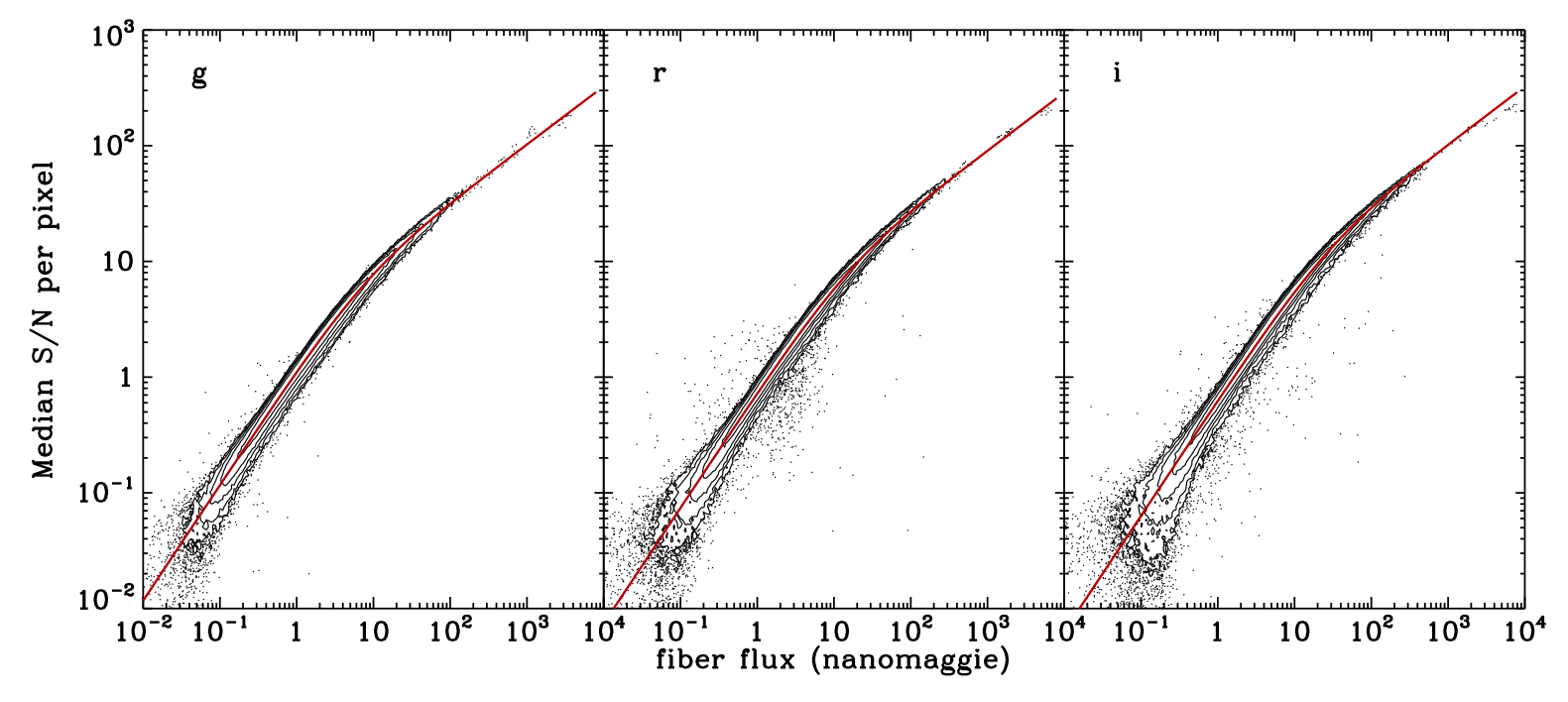}
\caption{S/N ratio per pixel in g, r and i bands as a function of the synthetic fiber flux computed from the flux-calibrated spectra.}
\label{fig:sn_synthemag}
\end{center}
\end{figure*}

For each 1-d spectrum we obtain per exposure, we compute the median S/N per pixel ($\Delta \log \lambda = 10^{-4}$) in each of the four optical bands (g, r, i, and z) that are completely covered by the spectra. The wavelength windows used for the median calculation are: g: $4400-5500\AA$, r: $5601-6749\AA$, i: $6910-8500\AA$, z: $8284-9281\AA$. We derive an empirical relation between the median S/N per pixel and the flux-calibrated synthetic fiber magnitudes, with the latter calculated from flux-calibrated spectra. These are shown in Fig.~\ref{fig:sn_synthemag}. We fit these relationships with the following functional form: 
\begin{equation}
S/N = {a F \over \sqrt{F+b}} \label{eqn:actual_sn_equation} {\rm .}
\end{equation}
Here, $F$ is the synthetic fiber flux in unit of nanomaggies\footnote{https://www.sdss3.org/dr8/algorithms/magnitudes.php}; $a$ and $b$ are fitting parameters. We chose this form because it captures the ingredients of an actual S/N estimation based on Poisson statistics and it provides a very good match to the data, as shown in Figure~\ref{fig:sn_synthemag}. The resulting fitting parameters for the $g$, $r$, and $i$ bands are given in Table.~\ref{tab:snprediction}. The relationships are much tighter than what we get when using fibermagnitudes computed from imaging as there are no added uncertainties from astrometry error, PSF mismatch, or photometry calibration. These formulae can be useful for S/N predictions for future IFU observations with the SDSS Telescope and BOSS spectrographs. 

This set of coefficients can be used to compute the typical S/N we obtain in the data cube for a certain apparent surface brightness. First, convert surface brightness to 2\arcsec\ diameter fiber flux in unit of nanomaggies. Then use Equation~\ref{eqn:actual_sn_equation} and coefficients in Table~\ref{tab:snprediction} to get the S/N per pixel per exposure. Finally, multiply by the square root of the number of exposure (typically 9 for sky regions with low galactic extinction) and the square root of the fiber filling facotr (0.56). The result is the S/N per pixel one would obtain in the final data cube in a 2\arcsec\ diameter aperture. The pixel size is $\Delta \lambda = 10^{-4} \lambda \ln 10$. Typically, for an apparent $r$-band SB of 22.5 mag arcsec$^{-2}$, the final S/N in the data cube over a 2\arcsec-diameter aperture is about 5.1 .

\begin{table}
\caption{Parameters in Equation~\ref{eqn:actual_sn_equation} which relates our S/N with flux in fibers.}
\begin{tabular}{l|r|r}
\hline\hline
Band & $a$ & $b$ \\\hline
$g$ & 3.41710 & 7.65072 \\ 
$r$ & 2.89589 & 12.9510 \\
$i$ & 3.23293 & 23.4808 \\
\hline\hline
\end{tabular}
\label{tab:snprediction}
\end{table}

\subsection{Exposure S/N Dependence on Airmass and Galactic Extinction} \label{sec:snrelation}



\begin{figure}
\begin{center}
\includegraphics[width=0.5\textwidth]{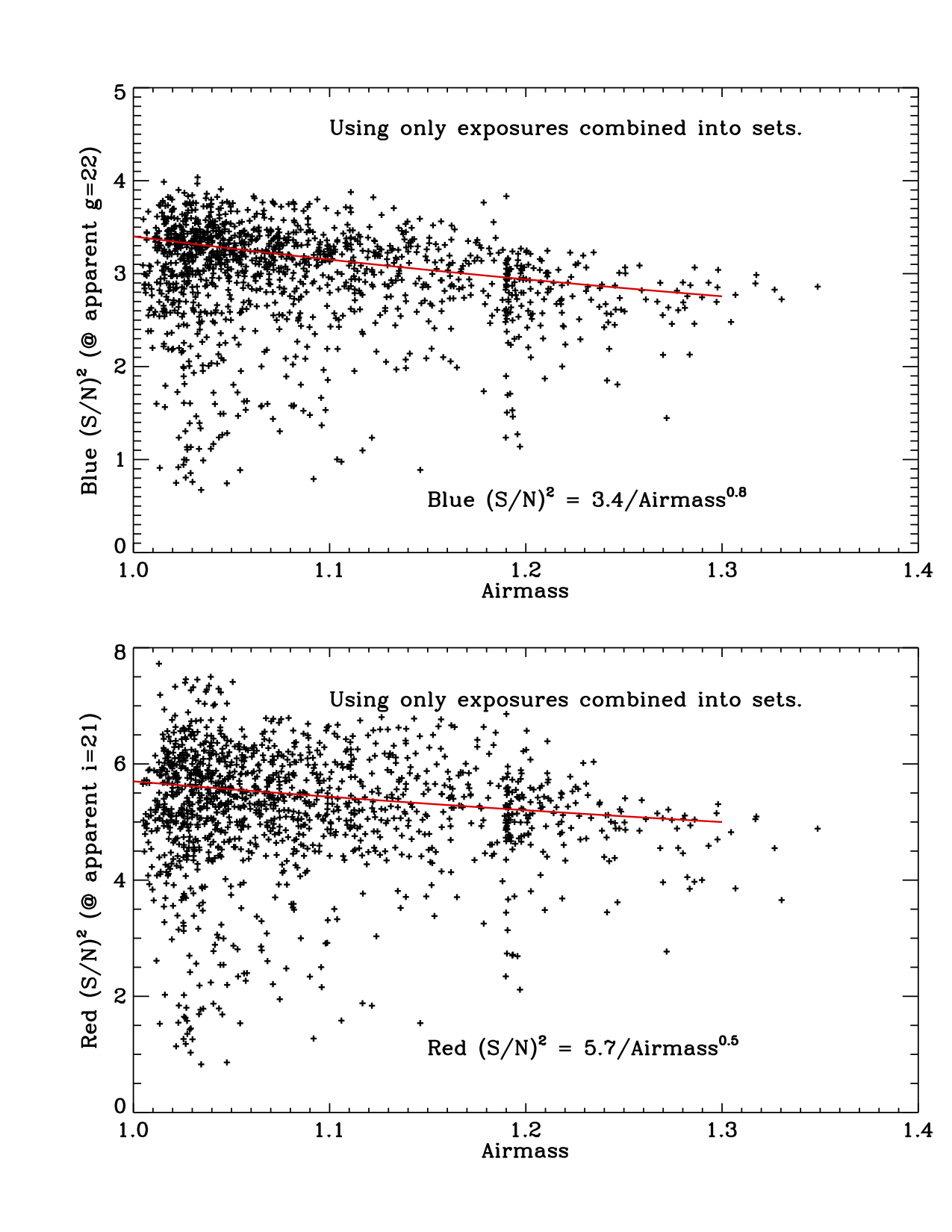}
\caption{$({\rm S/N})^2$ per exposure at fixed apparent magnitude (before galactic extinction correction) as a function of airmass. The large scatter is due to variations in sky background and transparency. The reference magnitudes are g=22\ in the blue and $i=21$ in the red. The lines represent an empirical relationship at typical good conditions. Only exposures that make complete sets are included, as the exposures with orphaned dithers tend to be biased to lower S/N.}
\label{fig:sn2_vs_airmass}
\end{center}
\end{figure}

Given the S/N estimates made by our data reduction pipeline, we can empirically determine the dependence of our S/N on airmass and extinction.
Our S/N depends weakly on airmass. There are several factors that degrade the S/N when we go to higher airmass. First, there is more atmospheric extinction. Second, there is a brighter sky background. For single-fiber spectroscopy like BOSS/eBOSS, S/N is also degraded at high airmass due to both larger field differential refraction and chromatic field differential refraction. Because MaNGA is using fiber bundles to cover galaxies, DAR simply shifts flux from one fiber to the next and does not affect our S/N. 
Therefore, we have a weaker S/N dependence on airmass than BOSS/eBOSS. Figure.~\ref{fig:sn2_vs_airmass} shows how our S/N depends on airmass. Note, here we are using the S/N at fixed apparent magnitude {\it} before correcting for galactic extinction, because we want to separate the dependence on airmass from the dependence on galactic extinction. This is derived from fitting the $\log {\rm S/N}$ vs.~apparent magnitude using the same fixed slope relation as used in DOS, except that we do not apply galactic-extinction. 

The reason for the large scatter in Figure~\ref{fig:sn2_vs_airmass} is variations in transparency and sky brightness, with the moon and clouds both contributing to the sky brightness variation. 

\begin{figure}
\begin{center}
\includegraphics[width=0.48\textwidth]{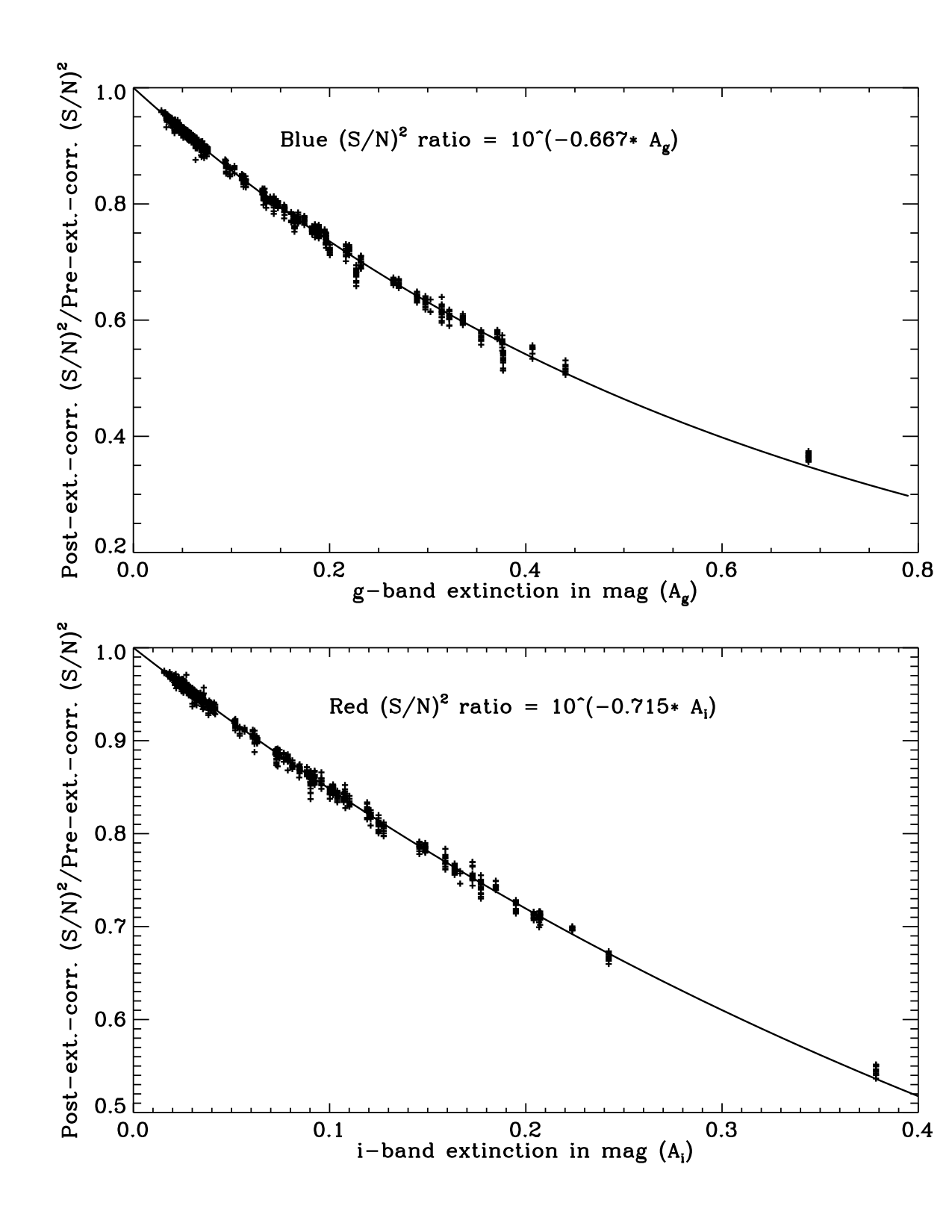}
\caption{Impact of the galactic extinction correction on obtained $(S/N)^2$ for the $g$-band (upper panel) and the $i$-band (lower panel). The ratio of the corrected versus uncorrected $(S/N)^2$ values is plotted against the strength of the correction in magnitudes.
The lines represent empirical relationships that were designed to fit the high extinction regime where the corrections are most important.}
\label{fig:sn2_vs_extinction}
\end{center}
\end{figure}

We estimate our S/N at galactic-extinction-corrected magnitudes. Therefore, our S/N also depends on galactic extinction. Extinction simply shifts our apparent magnitude in the $\log {\rm S/N}$ vs. magnitude plots. Therefore, the effect of extinction should be directly related to the slope of the relation. In Figure~\ref{fig:sn2_vs_extinction}, we show the ratio of the $(S/N)^2$ derived from galactic-extinction-corrected magnitudes to those derived based on uncorrected magnitudes, and plot them against the median galactic extinction of all target galaxies on a plate, for all exposures. The coefficients in the exponent should be twice the slope used in Figure~\ref{fig:logsn_vs_mag}. The best-fit relation is slightly different. This is because for plates with significant extinction, the fixed reference magnitudes correspond to fainter apparent magnitudes where the slopes are slightly steeper. BOSS assumed $(S/N)^2$ scales with $10^{-0.8A}$ which is appropriate for the background dominated regime. The reference magnitudes BOSS and MaNGA use are significant compared to the sky backgrounds leading to a shallower slope.

Given the empirical dependence of S/N on airmass and extinction, we arrived at the following equations for our observing speed.
\begin{eqnarray}
{\rm (S/N)}_{\rm Blue}^2 &= {3.4\over{\rm Airmass}^{0.8}}\times 10^{-0.667A_g}\label{eqn:sn_equation_g}\\
{\rm (S/N)}_{\rm Red}^2 &= {5.7\over{\rm Airmass}^{0.5}} \times10^{-0.715A_i}\label{eqn:sn_equation_i} {\rm .}
\end{eqnarray}
Here $A_g$ and $A_i$ are the galactic extinction in $g$- and $i$-bands, respectively.


One thing worth noting is that our S/N has no dependence on seeing. Single-fiber spectroscopy targeting centers of galaxies, such as BOSS and eBOSS, has a very strong dependence on seeing, because the surface brightness profiles of galaxies peak at the center and are relatively shallow on the outskirts.



\subsection{S/N prediction and expected survey speed}

Given the above empirical relationship between S/N and fiber magnitudes, we can estimate the stacked S/N obtained in the outskirts of our target galaxies and check this against our science requirements.


We estimate the S/N from surface photometry in an outer elliptical annulus. The annulus is set to have the axis ratio and the orientation given by the NSA catalog. The semi-major axis range of the annulus is set to 1--1.5Re or 2/3 -- 1 of the effective radius of the fiber bundle hexagon, whichever is smaller. 
For the smallest bundle we have, for a circular galaxy that has 1.5Re greater or equal to the effective radius of the bundle (5.45\arcsec), this corresponds to the outer ring of fibers. 

Using the images provided by the NASA-Sloan Atlas, we convolve the images to the seeing of the observation before computing the 2\arcsec\ aperture photometry at the positions of all fibers in a bundle that is centered on the galaxy. We use the resulting fiber flux to estimate the S/N obtained in the defined elliptical annulus. Many fibers lie across boundaries of the annulus, for which we only count the flux in proportion to the fraction of the fiber area that lies within the annulus. 

\begin{figure*}
\begin{center}
\includegraphics[width=0.9\textwidth]{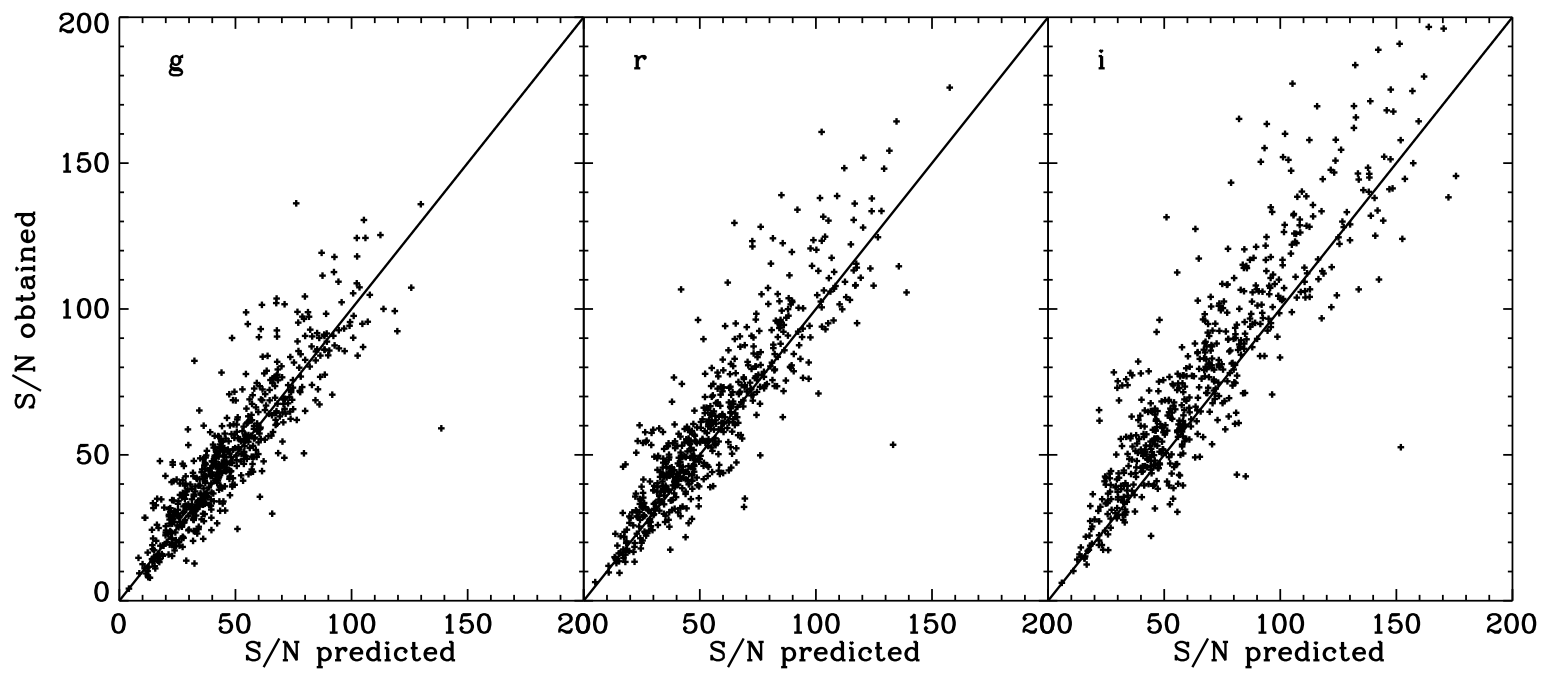}
\caption{Measured S/N compared to predicted S/N based on photometry in the $g-$, $r-$ and $i$-bands, for an elliptical annulus between 1 and 1.5 effective radii of the galaxy or between 2/3 and 1 effective radius of the bundle, whichever is smaller. We display only galaxies that were observed in the first year with the updated completeness threshold. The solid line marks the 1:1 relation. The actual data have slightly higher S/N than our predictions in $i$-band.}
\label{fig:sn_pred_vs_data}
\end{center}
\end{figure*}

We compared this S/N prediction with the stacked RSS spectra of galaxies. We first multiply the flux from each fiber by the fraction of the fiber's area within the annulus, then add them together. The variance is propagated accordingly. We then compute the median S/N in four wavelength windows corresponding to $griz$ bands. Figure~\ref{fig:sn_pred_vs_data} shows the comparison between the predicted S/N based on previous photometry and that from the stacked RSS spectra. The data are very consitent with our predictions in the $g$- and $r$-bands, and show slightly better S/N in the $i$-band.

\subsection{Final S/N Distribution and Projection}

\begin{figure}
\begin{center}
\includegraphics[width=0.45\textwidth]{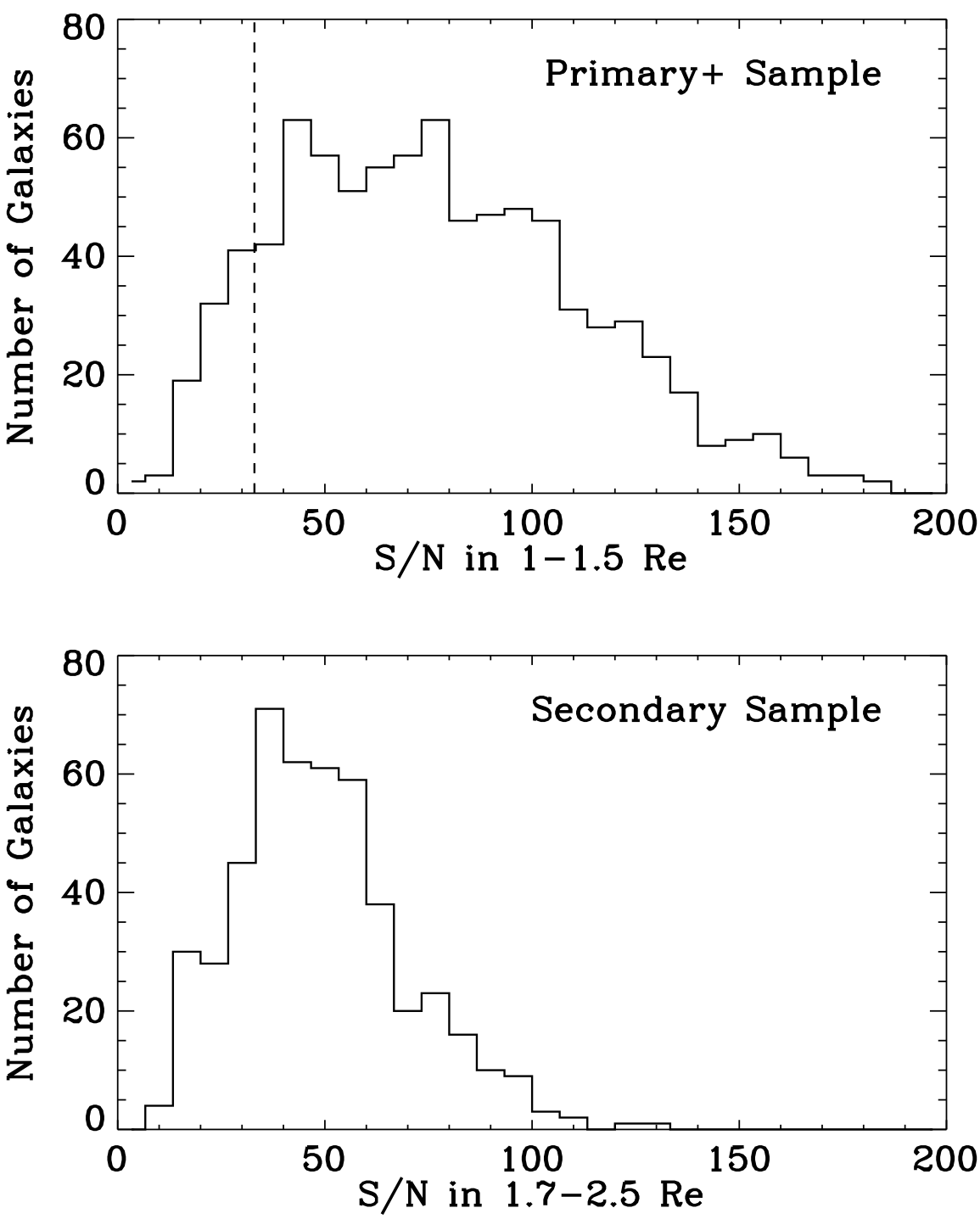}
\caption{Upper panel: distribution of the stacked S/N per 1.4\AA\ pixel in $r$-band in the outer tertile of all Primary+ galaxies observed in the first year. This is the S/N in the stacked spectra in an elliptical annulus between 1 and 1.5 effective radii of the galaxy or between 2/3 and 1 effective radius of the bundle, whichever is smaller. The vertical dashed line indicate the requirements. About Lower panel: distribution of stacked S/N in the outer tertile (1.7-2.5\Reff) of all Secondary galaxies.} 
\label{fig:actualsndistribution}
\end{center}
\end{figure}

Top panel of Figure~\ref{fig:actualsndistribution} shows the actual S/N distributions in stacked spectra in an elliptical annulus between 1 and 1.5 \Reff\ of the galaxy (or between 2/3 and 1 effective radius of the bundle, if the galaxy is not covered to 1.5\Reff) for the Primary+ sample. Our science requirement is to have a S/N greater than 33 for more than 75\% of the sample. In the first year of observation, 89\% of the Primary+ sample reach this S/N. The bottom panel shows that for the Secondary sample for 1.7-2.5 \Reff, or between 2/3 and 1 effective bundle radius, whichever is smaller. About 78\% of the Secondary sample reach an r-band S/N per pixel greater than 33\ in this ellipical annulus.

\begin{figure}
\begin{center}
\includegraphics[width=0.5\textwidth]{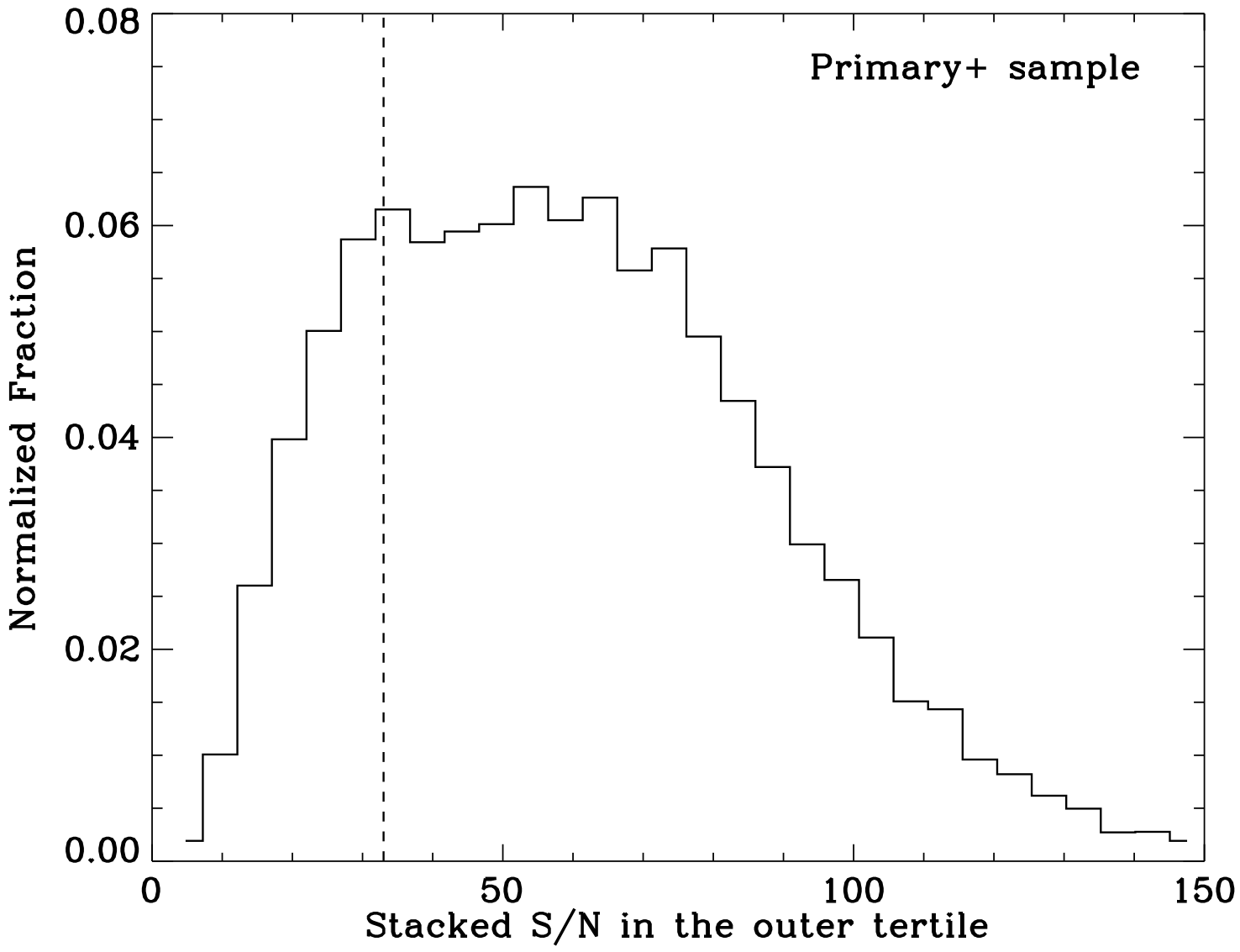}
\caption{Projected final S/N distribution in stacked spectra in the outer 1-1.5\Reff\ among the Primary+ sample we expect from 6 years of observation. The vertical dashed line marks the S/N threshold required for at least 75\% of the galaxies in the Primary+ sample.}
\label{fig:projectedsndistri}
\end{center}
\end{figure}

Figure~\ref{fig:projectedsndistri} shows the expected final S/N distribution in our sample in 6 years. Among the Primary+ sample, 80\% will have a stacked S/N per pixel in r-band in the outer tertile greater than 33, meeting the science requirements. 

\subsection{Projection for the Number of Galaxies}

\begin{figure}
\begin{center}
\includegraphics[width=0.5\textwidth]{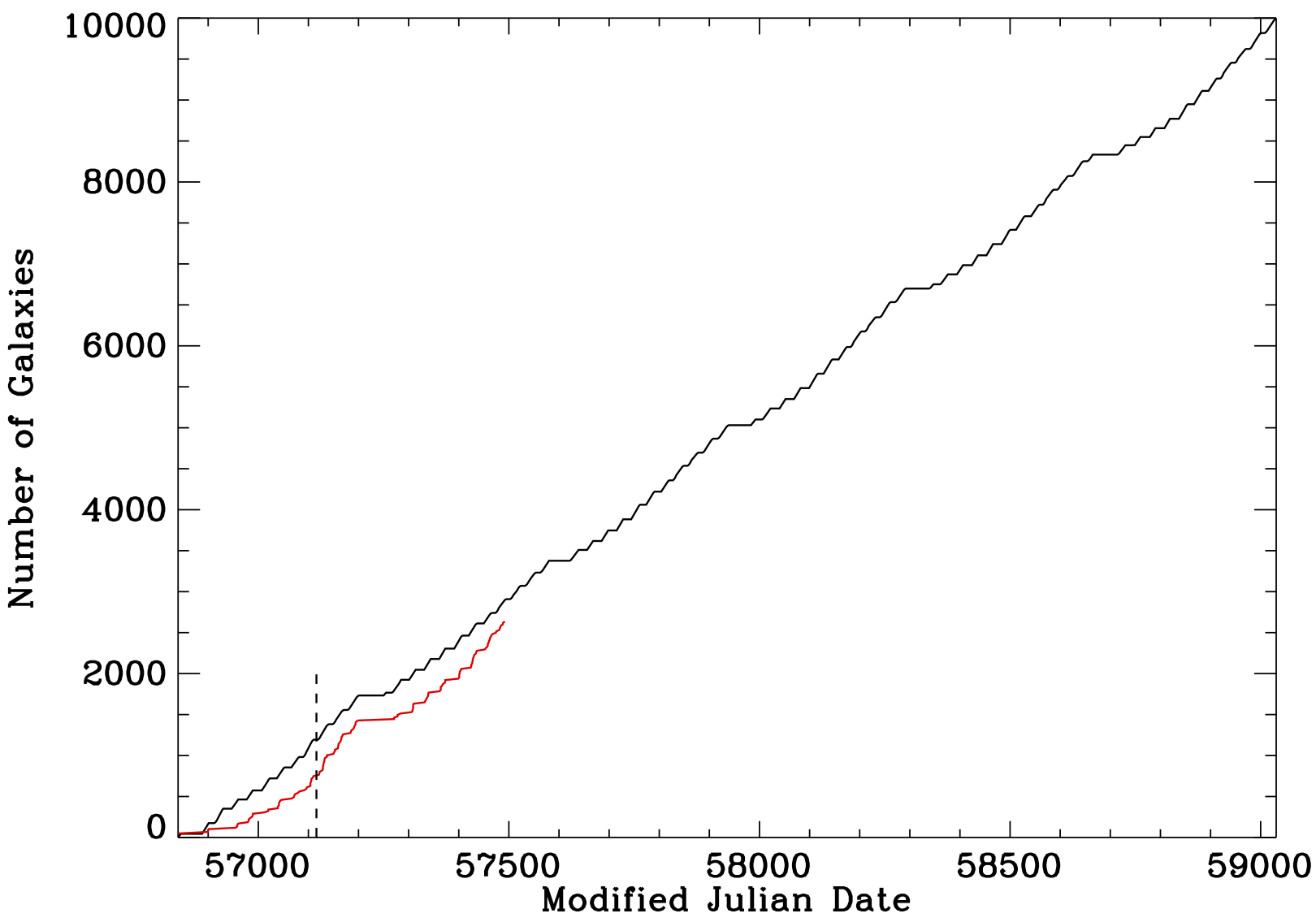}
\caption{Current progress towards the final goal of 10K galaxies. The black line shows the expected number of plates as a function of date and the red line shows the actual number of plates observed. MaNGA is slightly behind schedule because of bad weather and over-exposing of plates in the first season. The dashed line marks the date when exposure times were corrected. Since then, we have been on track with expectations and have made up a fraction of the shortfall.}
\label{fig:surveyprogress}
\end{center}
\end{figure}

As of Apr 18, 2016, we have completed observations of $\sim156$ plates, including two commissiong plates observed in March 2014. These plates contain more than 2550 unique targets. Figure~\ref{fig:surveyprogress} shows our progress with time compared with expectations.  We are slightly behind schedule because of the over-exposing of plates in the first season (see discussion in Section~\ref{sec:completeness_thresholds}). Since then, we are progressing as expected and have made up a fraction of the shortfall.

Given the remaining time we have before Summer of 2020, we expect to finish $\sim575-600$ plates, yielding a final sample of $\sim10K$ galaxies. The planned footprint is presented in Figure~\ref{fig:surveyfootprint}.


%
%


%


\section{Data Quality Assessment} \label{sec:quality}

In this section, we provide critical information about the data and an assessment of data quality. In a companion paper, \cite{Law16}, we describe the data reduction pipeline in detail and also provide additional analysis of data quality. Law et al. provide examples of the reduction quality on individual plates. Here we present the overall statistics of quality metrics for the dataset observed in the first year.

\subsection{Example Spectra from Data Cube}

\begin{figure*}
\begin{center}
\includegraphics[width=\textwidth, viewport=0 120 1020 640,clip]{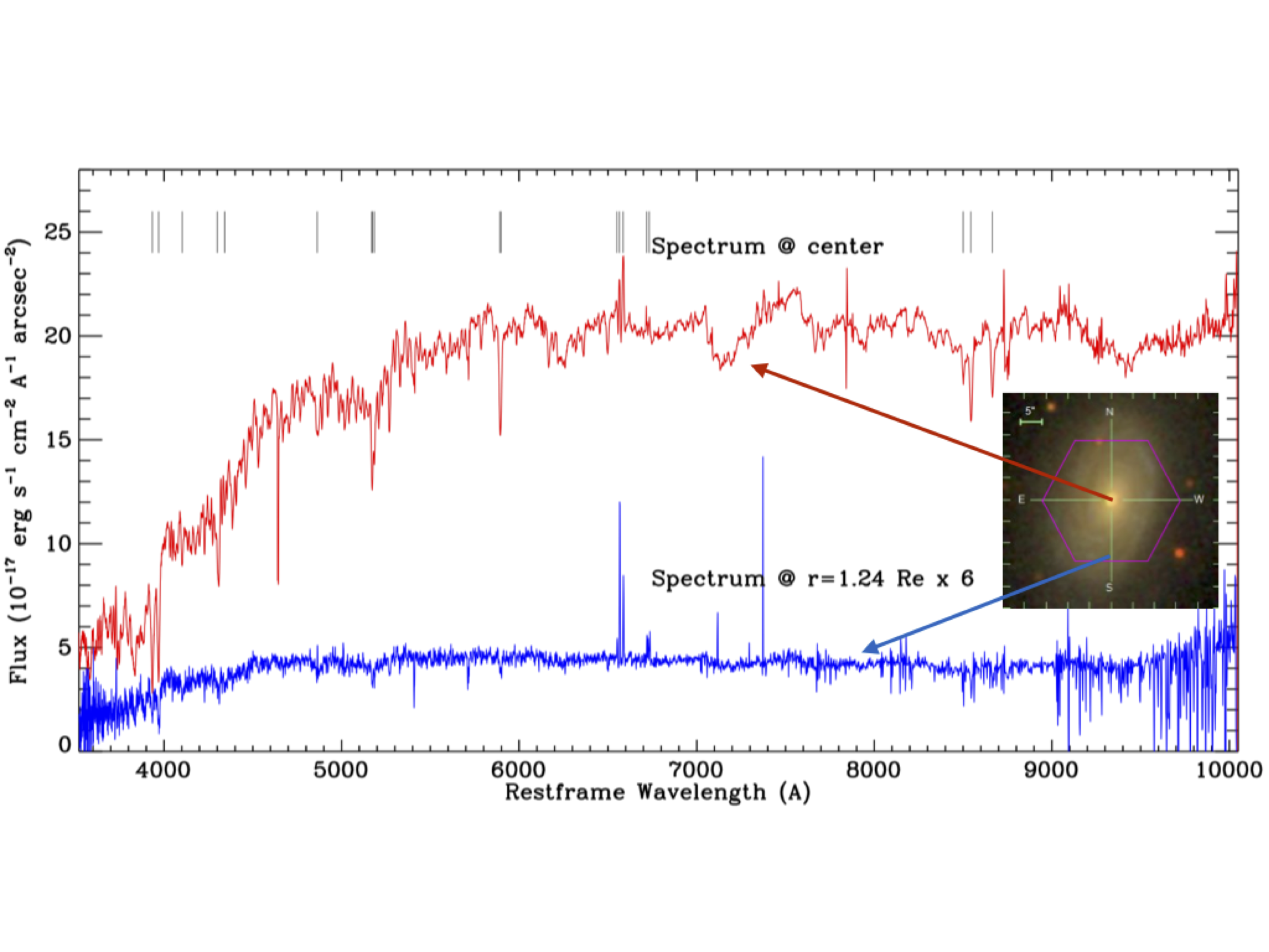}
\caption{Example spectra from a typical MaNGA data cube. This galaxy was observed with IFU 12704 on plate 8138. The inset shows the SDSS color image with the hexagonal IFU footprint overlayed. The top spectrum is from the central spaxel; the bottom spectrum is from 1.24 Re away from the center and is multiplied by a factor of 6 for easier comparison with the central spectrum. {\it No smoothing has been applied.} Even the outer spectrum, which is fainter by a factor of 30\ in r-band flux, has sufficient S/N to clearly detect numerous spectral features, which are marked with the short lines on top. The sharp spikes in the near-IR, particularly in the bottom spectrum, are due to sky subtraction residuals. Note that the two spectra have very different shapes and feature emission lines with significantly different strengths relative to the continuum.}
\label{fig:examplespec}
\end{center}
\end{figure*}

Before we present the metrics, we illustrate the data quality obtained by MaNGA with two typical spectra from a spiral galaxy in Figure~\ref{fig:examplespec}. One spectrum is from the central $0.5\times0.5$ sq. arcsec spaxel in this galaxy, and the other is from an edge spaxel that is 13\arcsec\ away from the center. The central spectrum has a S/N per pixel of 200-250\ in the $r$-band and the edge spectrum has a S/N per pixel of 20-30. {\it No smoothing has been applied to either of the spectra.} One could easily identify multiple spectra features in both spectra, such as CaII H \& K, G-band, H$\beta$, MgI b, NaI D, \hal, \niiw\, \siiw, and CaII triplet. The two spectra have very different $D_n(4000)$ indicating different stellar population ages.  They also display very different emission line ratios in \nii/\hal, indicating different ionization mechanisms are at play at these two locations. 

\subsection{Resulting PSF on the Focal Plane from seeing and guiding}

The point spread function (PSF) is a critical element in the analysis of IFS data. Here, by PSF, we mean the intrinsic light profile of a point source on the focal plane of the telescope. Because we are doing imaging spectroscopy, the knowledge about the PSF is critical for most of the analysis of the data. For an individual exposure, flux incident on a fiber-bundle is spatially undersampled by the fibers and is not completely covered due to gaps between fibers. Thus, the IFU bundles {\it alone} cannot provide an accurate measurement of the PSF shape, but can provide a refinement of the scale of the PSF if the shape is known. 
The guider images can provide such an initial measurement of the PSF shape. For a more detailed description of the guider system, see Section~\ref{sec:dithering}.

The science exposures are 15 minutes long, during which the guider system monitors the position of the guide stars relative to the position of the guide fibers. As described in Section~\ref{sec:dithering}, the guiding is not perfect and introduces some smearing of the integrated PSF over the 15-minute exposure. 

\begin{figure}
\begin{center}
\includegraphics[width=0.5\textwidth]{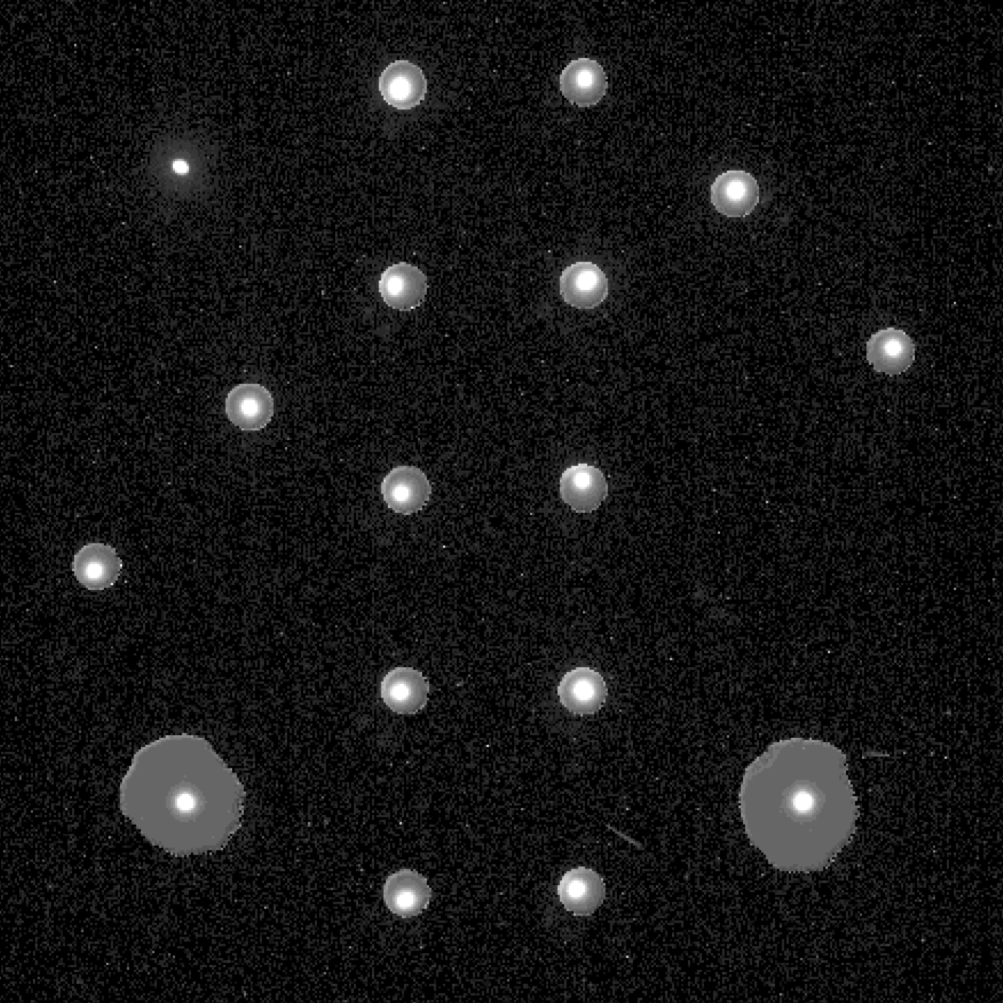}
\caption{An example stacked guider image made from coadding 37 individual flat-fielded guider frames taken during a 15 minute exposure. The 16 guide fibers are positioned in this particular configuration on the fiber output block which is imaged by the guide camera. The bright dot in the upper left is a Tritium spot which can be used to check the focus of the camera. The two larger guide fibers near the bottom left and bottom right are acquisition fibers. They provide a larger area for us to measure the sky background.}
\label{fig:cogimg_example}
\end{center}
\end{figure}

To measure the integrated PSF, we bias-subtract and flat-field all the guider images and then stack the guider images taken during a science exposure. Figure~\ref{fig:cogimg_example} shows a stacked image made from 37 individual guider frames taken during a 15 minute exposure. This gives the time-integrated PSF. The left panel of Figure~\ref{fig:seeing_vs_reconpsf} shows the distribution of seeing FWHM observed for all exposures by the guider during the first year of observations. The median seeing is 1.50\arcsec\ and the range is 1-2.5\arcsec. 

\begin{figure}
\begin{center}
\includegraphics[width=0.45\textwidth]{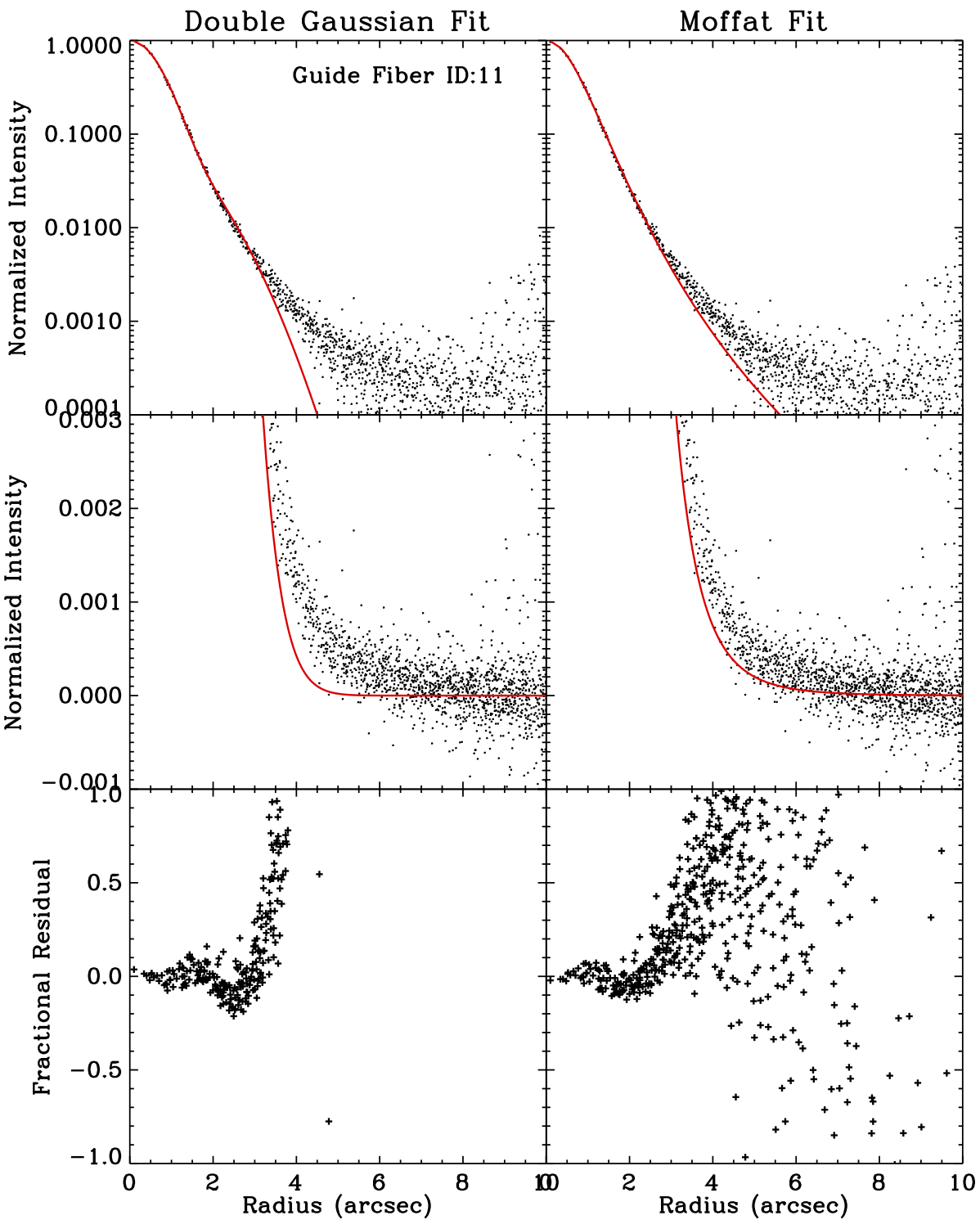}
\caption{The spatial profile of an example guide star image fitted by a double gaussian model (left) and a Moffat model (right). The y-axis of the top 4 panels are the flux ratio relative to the peak flux of the best fit model. 
The top panels show the whole curve in log units in which negative points are excluded. 
The middle panels show the fit in linear units zoomed in around the low flux outskirts. 
The bottom panels show the fractional residual relative to the model. An additional tail beyond 3\arcsec\ that cannot be adequately accounted for by either the double gaussian model or the Moffat model. The tail contains about 3\% of the total flux.}
\label{fig:guiderPSF}
\end{center}
\end{figure}

In our data reduction pipeline, we model the focal-plane PSF seen by the fiber bundles with a double Gaussian function: 
\begin{equation}
F(r) = A_1 exp(-{r^2 \over 2\sigma_1^2}) + A_2 exp(-{r^2 \over 2\sigma_2^2}) .
\label{eqn:doublegau}
\end{equation}
We prefer to model the PSF as a double gaussian rather than a Moffat because Gaussians are much faster to integrate over an aperture\footnote{
We need the computation to be fast because we need to compute many fiber-convolved PSF with different sizes during the flux calibration step\citep{Yan16}.The integration of the Moffat function does not have an analytic formula and we have to integrate numerically which is very slow. Integration of the Guassian function is easy to compute using the Error Function and no numerical integration is required. Therefore, they differ by orders of magnitude in computation time.} than Moffat functions and Moffat functions provide only moderate improvements. 
Figure~\ref{fig:guiderPSF} shows a typical PSF measured by the guider with a 1.5\arcsec\ FWHM. A double Gaussian provides an adequate description of the central region but misses the extended tail beyond 3\arcsec. A Moffat function only does slightly better and cannot fit the extended tail either.
The extended tail contains about 3\% of the total flux. If one is concerned with features in MaNGA galaxies that involve large surface brightness contrasts, this issue may be important. 

The PSF seen by the guider may be slightly different from that seen by the fiber bundles. This is because the guider system modifies the PSF in two ways. First, cross-talk between individual fiber strands in an imaging fiber can smear the PSF. Second, small focus offsets in the imaging camera could also modify the PSF. 

To address the impact of these effects, we refine the scales of the PSF, using mini-bundle observations of standard stars, assuming the shape of the PSF is the same as seen by the guider. For flux calibration purposes, we observe 12 F-type subdwarfs on each plate simultaneously with the science targets. By fitting for the flux ratios among fibers in the mini-bundle, we find that on average the PSF seen by the IFU bundles is 90\% in width compared to the guider PSF. 



\subsubsection{What PSF should I use?}

First, we would like to make a distinction between the `PSF' and the `fiber-convoled PSF'. By `PSF', we mean the intrinsic light profile produced by a point source on the focal plane of the telescope. By `fiber-convolved PSF', we mean this PSF convolved with a Top-hat 2\arcsec-diameter fiber aperture. When using the fiber bundles to observe a star, the flux in each fiber is equal to a sampling off this fiber-convolved PSF at the center position of that fiber.

What PSF should be used in the analysis of the MaNGA data depends on what data products are used in the analysis, the data cubes or the RSS files. 

If one uses the datacubes, one should use the `reconstructed PSF' stored along with the data cube. The data cube is produced in the 3d stage of the DRP using an image reconstruction algorithm, the Modfied Shepard method, to produce a final image on a regular grid of $0.5\times0.5$\arcsec\ square spaxels. The input to this algorithm includes the fiber spectra and the relative positions of all fibers in a bundle at all wavelengths, from all exposures. Suppose one observes a star, the fiber flux is equal to a sampling of a `fiber-convolved PSF'. Therefore, to obtain a corresponding PSF to the data cube, we process the `fiber-convolved PSFs' of all contributing exposures through the same image reconstruction algorithm, applying exactly the same offsets as applied to all exposures (see \citealt{Law16} for details). This produces a 'reconstructed PSF'. To use this PSF, one should not do any more integration with either the 0.5\arcsec\ spaxel or with the fiber aperture, as they are already included. 

\begin{figure*}
\begin{center}
\includegraphics[width=0.33\textwidth]{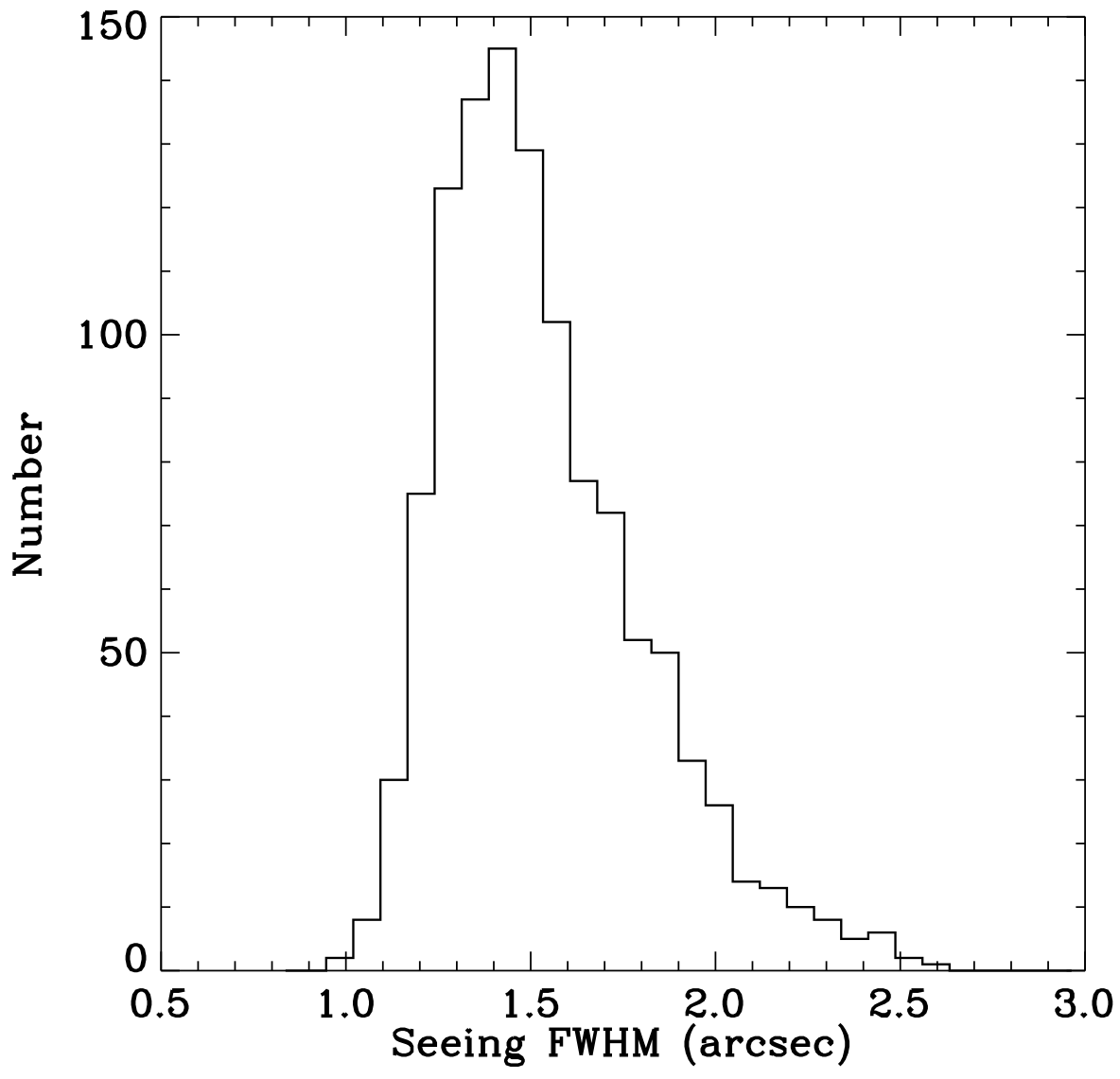}
\includegraphics[width=0.33\textwidth]{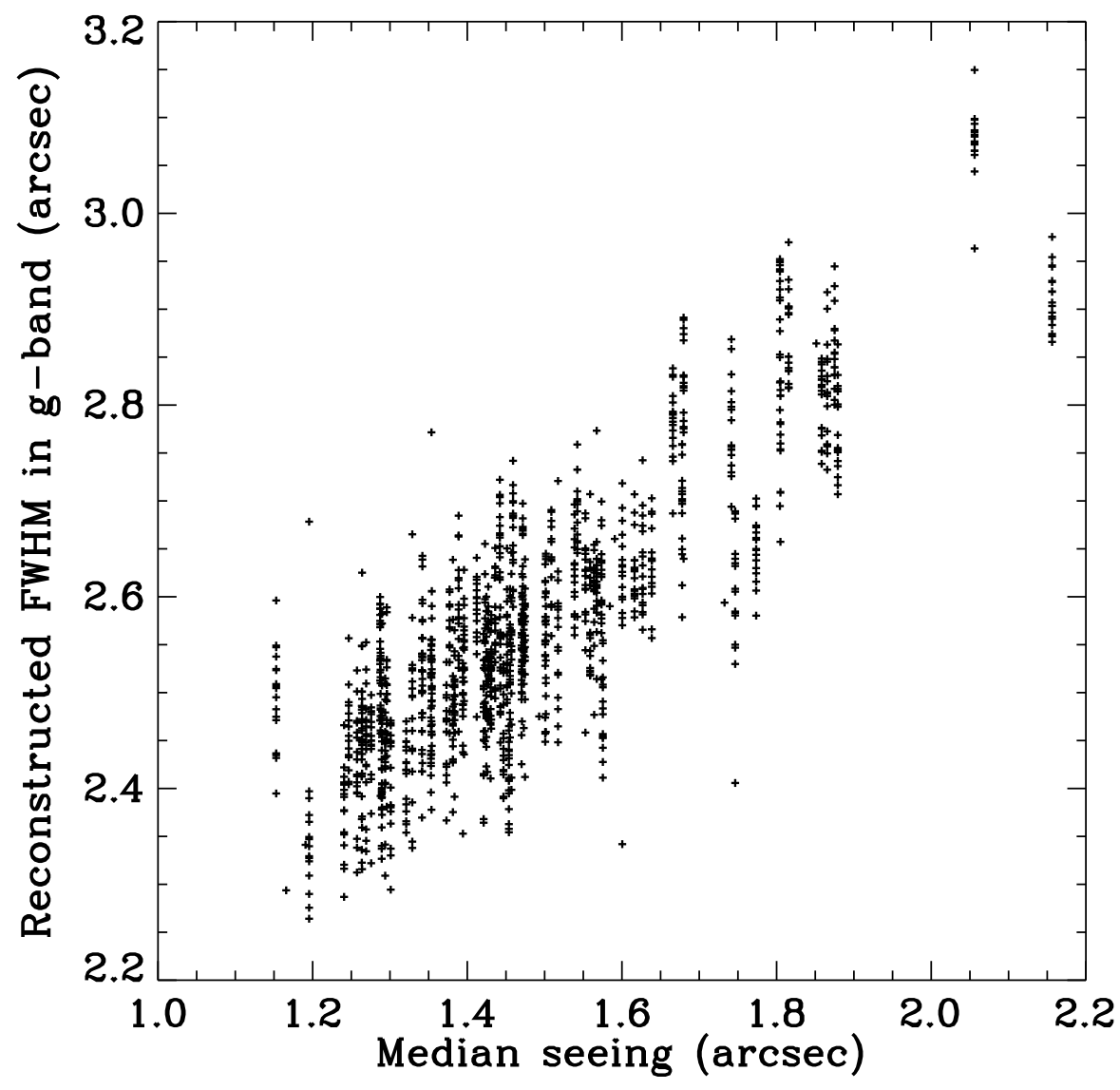}
\includegraphics[width=0.33\textwidth]{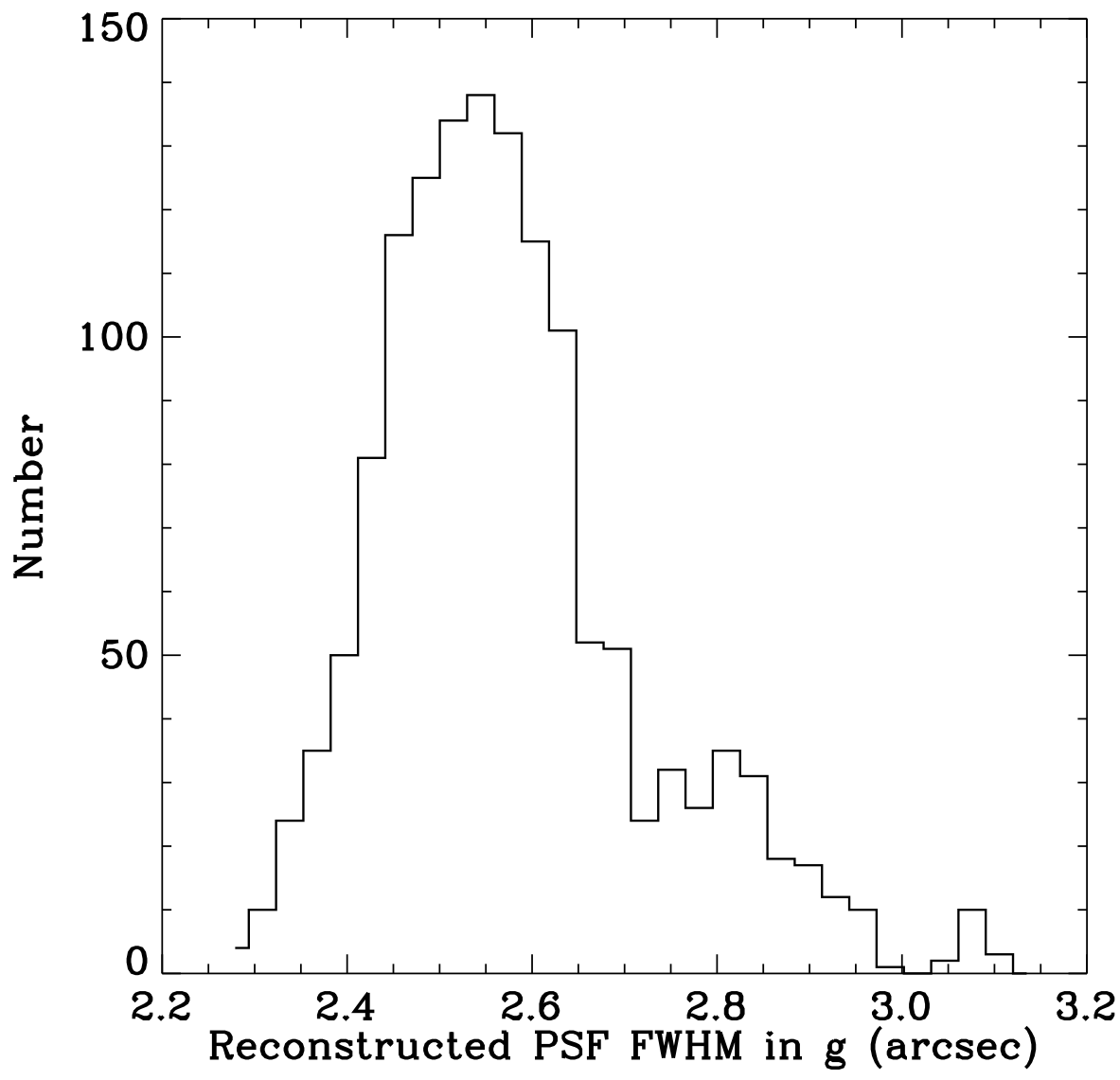}
\caption{Left: The seeing distribution of all MaNGA science exposures taken in the first year of operation. Middle: Median intrinsic per-exposure seeing FWHM vs. FWHM of the reconstructed PSF in the data cube in $g$-band for all galaxies completed in the first year. Right: Distribution of the FWHM of the reconstructed PSF in $g$-band. }
\label{fig:seeing_vs_reconpsf}
\end{center}
\end{figure*}

In the middle panel of Figure~\ref{fig:seeing_vs_reconpsf}, we show that the FWHM of the reconstructed PSF is well correlated with the median seeing of all the exposures going into a data cube. The right panel of this figure shows the final FWHM distribution of this `reconstructed PSF' among all galaxies completed in the first year. The median FWHM is 2.54\arcsec with a tail extending to 2.8\arcsec. This is the final angular resolution of the data cube.





If one uses the RSS files, which store the spectra per fiber per exposure, one would need to compute the PSF associated with each exposure. In this case, one should start with the guider image, measure the PSF (intrinsic light profile on focal plane modeled as double Gaussian or Moffat function), convolve with the 2\arcsec\ fiber aperture, and scale down by 10\% in width to get the per exposure `fiber-convolved PSF'. Note the PSF varies across the focal plane and varies with wavelength. Thus, one needs to adjust it according to the procedure as described by \cite{Yan16} before the fiber aperture convolution. If one uses the RSS file or applies forward modeling to fit the data, this is the PSF to use.

\subsection{Sampling Uniformity from the Actual Dithers} \label{sec:omega}

\begin{figure}
\begin{center}
\includegraphics[width=0.5\textwidth]{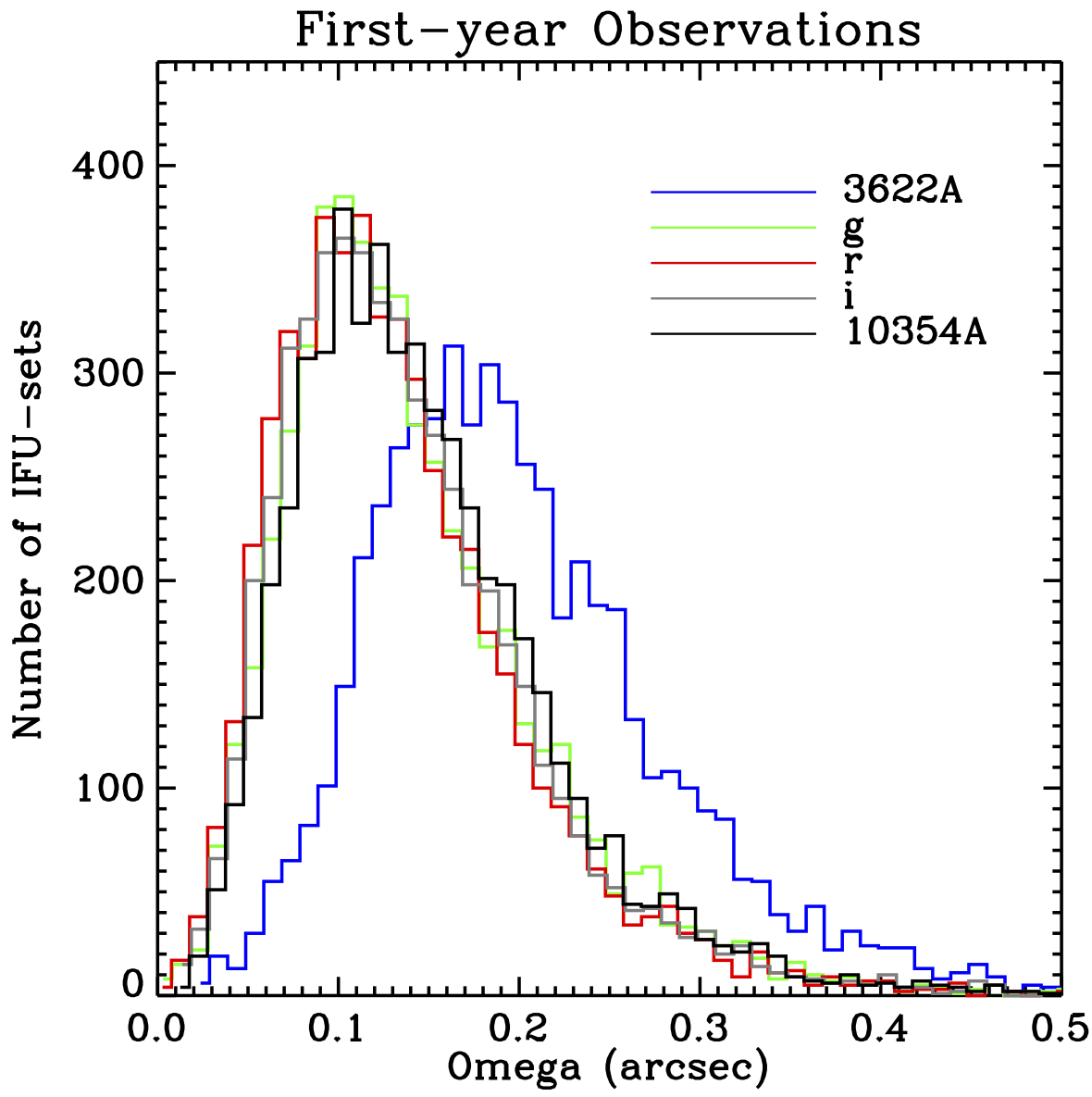}
\caption{Distribution of the maximum $\Omega$ offsets (see Section~\ref{sec:observingstrategy}) among dithers in a set for all IFUs targeting galaxies in the first year of observation. The different histograms show the offsets at five wavelengths: including the effective wavelengths of the $g$, $r$, and $i$ bands, and the bluest and reddest wavelengths we cover. The offset is largest at the bluest wavelengths due to the large chromatic differential refraction in the blue.} 
\label{fig:galaxies_omega}
\end{center}
\end{figure}

Here we evaluate our dithering accuracy. In Section~\ref{sec:observingstrategy}, we defined the dithering Uniformity Statistic ($\Omega$) to be the maximum offset between dithers in a set. There are two parts contributing to this offset: one is due to atmospheric refraction and the optical distorion of the telescope, the other is due to imperfect guiding. The former is theoretically predictable, but the latter is not. We can measure the offset due to imperfect guiding by matching the data to SDSS imaging, as mentioned above. This is done in the extended astrometry module of the data reduction pipeline \citep{Law16}. Combining both components, we directly measure $\Omega$ for all of our galaxies. Figure~\ref{fig:galaxies_omega} shows the Omega distribution for all galaxies observed in the first year. The offset is largest at the bluest wavelength we cover. At 3622\AA, 98.6\% of all galaxies have $\Omega < 0.4\arcsec$. By meeting this requirements, we ensure a high degree of spatially uniform sampling as a function of wavelength across the full wavelength range.

\subsection{Spectral Resolution} \label{sec:specres}

\begin{figure}
\begin{center}
\includegraphics[width=0.5\textwidth]{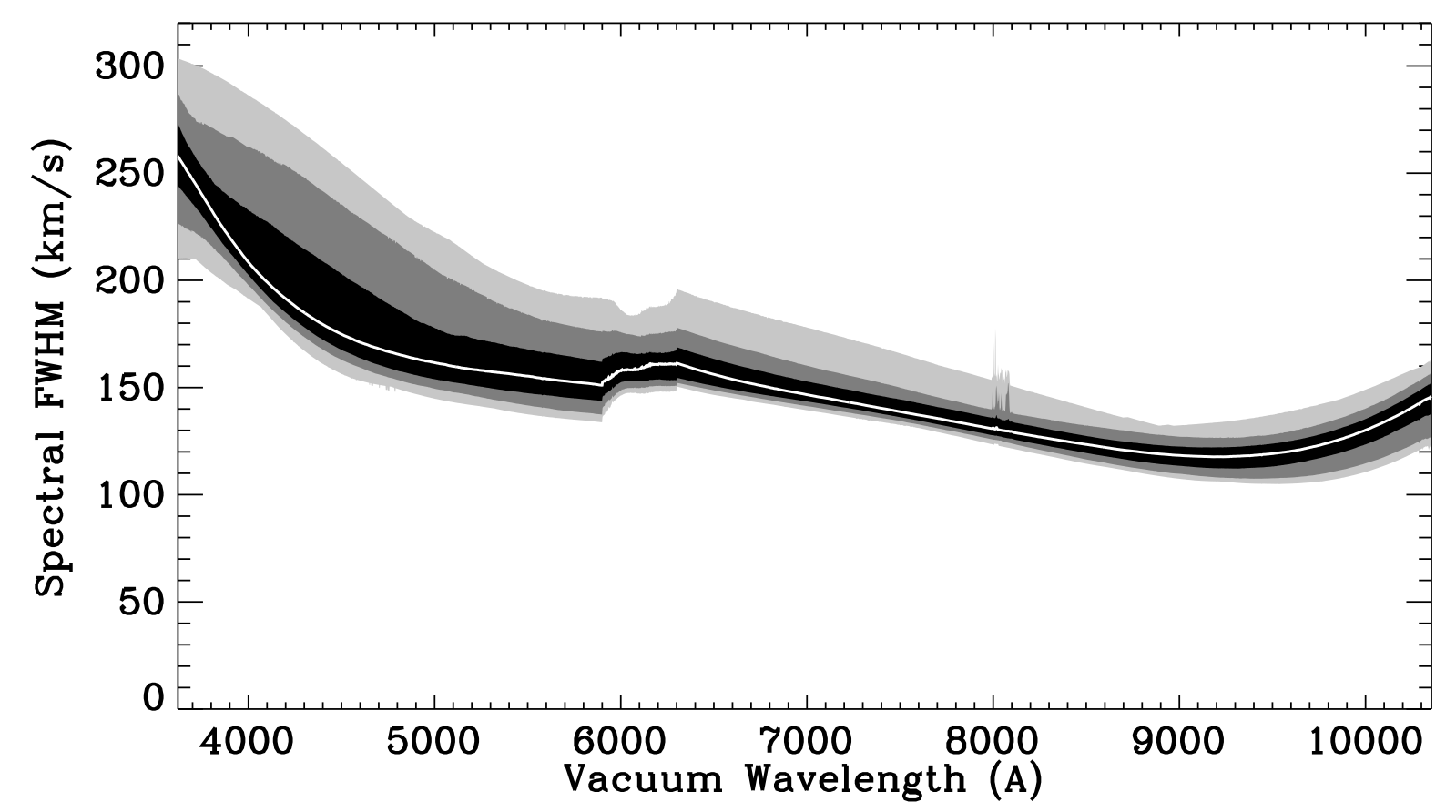}
\caption{The white curve shows the median instrumental resolution expressed in FWHM in velocity units as a function of wavelength among all 1390 galaxies to be released in DR13. The black region indicates the 15.85- to 84.15-percentiles of the distribution at each wavelength, which is roughly $\pm1\sigma$ around the median. The dark gray zone indicates the 2.5- to 97.5-percentiles, and the light gray zone indicates the 0.15- to 99.85-percentiles of the spectral dispersion. The region between 5900\AA\ and 6300\AA\ is where the blue cameras and red cameras overlap in wavelength. The dispersion here is averaged between the two cameras. The feature around 8000\AA\ is due to the middle three rows of the red detector having slightly different pixel widths. The velocity FWHM presented here {\it does not} does not include the 10\% broadening described in the text, i.e., the true values are about 10\% larger than those shown here.}
\label{fig:specres}
\end{center}
\end{figure}

\begin{figure}
\begin{center}
\includegraphics[width=0.5\textwidth]{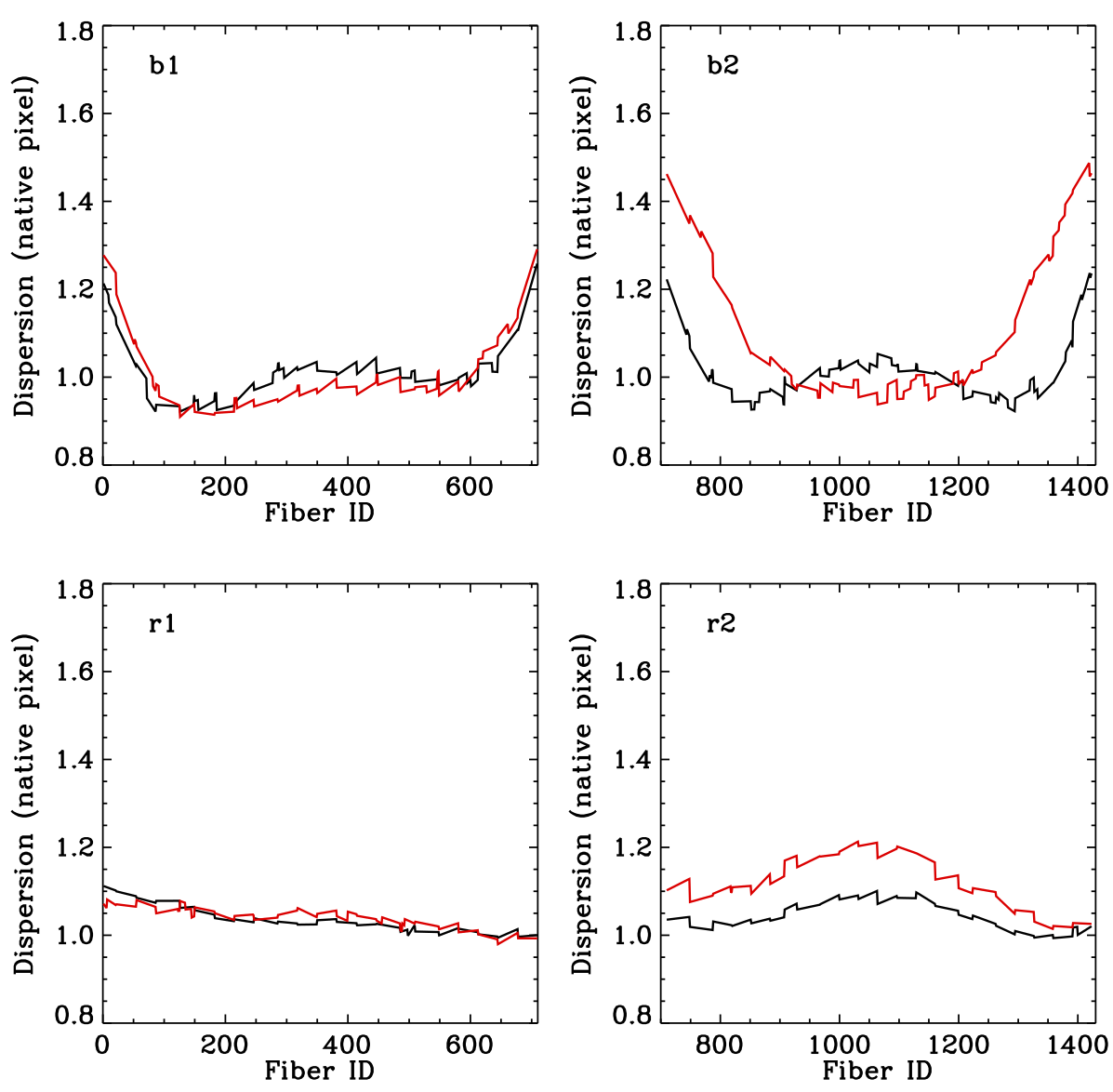}
\caption{Comparison of the instrumental line dispersion ($\sigma$ in native pixel units) in the 4 cameras as a function of fiber ID for Row 2000 (middle of detectors) for two arc frames (black vs. red lines) separated by 4.82 hours. There is significant focus drift in b2 and r2 causing the line dispersion to change significantly. The effect is strongest in b2 near the edges of the slit. This effect is taken into account in the delivered instrumental resolution by matching to the widths of sky lines.}
\label{fig:focusdrift}
\end{center}
\end{figure}

Figure~\ref{fig:specres} shows the distribution of the obtained spectral resolution as a function of wavelength for all 1390 galaxies observed in the first year. As discussed above, the focal plane is not flat at the CCD. Thus, there are variations in instrumental dispersion as a function of both slit position and wavelength. The variation is large in the blue camera because its focal plane is strongly curved in the spatial direction. There can also be large variations in resolution within a single IFU, especially if the IFU is placed close to the edges of the slit where the focal plane has the steepest slope relative to the CCD.
The arc frame provides the basic measurement of the instrumental resolution. Two other factors also change the instrumental line spread function (LSF) in the science exposures compared to the arc exposures. The science exposures are 15 minutes long. Over this 15 minute integration time, as the telescope tracks the field, there can be small focus drifts and small detector movements due to instrument flexure. These give a slight, additional broadening to the LSF. Since we often take flats and arcs both at the beginning and the end of the exposure sequence, we found that the cameras in spectrograph 2 (b2 and r2) can have significant focus drifts over a few hours. This is illustrated in Figure~\ref{fig:focusdrift} which compares the instrumental dispersion as a function of fiber ID for the four cameras between two arc exposures separated by 4.8 hours. Significant changes are apparent in the edge fibers in b2. Additionally, the detector can experience lateral movements in both the spatial and spectral directions at a small fraction of a pixel. Flexure tests show that at an altitude of 60 degrees, over 360 degrees of rotation, the min-to-max lateral shift in the spatial direction can be as large as 0.4-0.6 pixels and in the spectral direction can be as large as 2-3 pixels. When we track on the sky, over 15 minutes, the change in gravity vector is much smaller, but there still can be small sub-pixel shifts that will smear the LSF slightly. The overall shift between the arc frame and the science frame is taken out in the reduction pipeline by shifting the wavelength solution to match the sky lines. 
In the pipeline, we use the measured strong sky line widths to modify the LSF and report a final instrumental resolution that includes the above factors. The reported LSF in the data released in DR13 and shown in Figure~\ref{fig:specres} are these final sky-line-matched dispersion values \citep{Law16}.

The SDSS Legacy software assumes the LSF is a Gaussian before pixel integration, i.e, the flux in a pixel equals the integration of a Gaussian profile over the pixel width. This is different from assumptions made in many other analyses which assume the flux in a pixel equals the value of a Gaussian profile sampled at the center of the pixel. The LSF is critically sampled (${\rm FWHM} \geqslant 2 ~{\rm pixel~width}$) by the native pixel width in most parts of our detectors. In this regime, these differing assumptions can cause a 4\% width difference in the measured LSF. In addition, when we resample the spectra from the native pixel sampling to a regularly-spaced linear or logarithmic sampling, we effectively broaden the LSF. Combined together, these two factors cause a roughly 10\% increase in LSF or equivalently a 10\% decrease in spectral resolution. This factor is not included in the data released in DR13 but will be addressed in future data releases. See \cite{Law16} for more details.


\begin{figure*}
\begin{center}
\includegraphics[height=0.3\textheight]{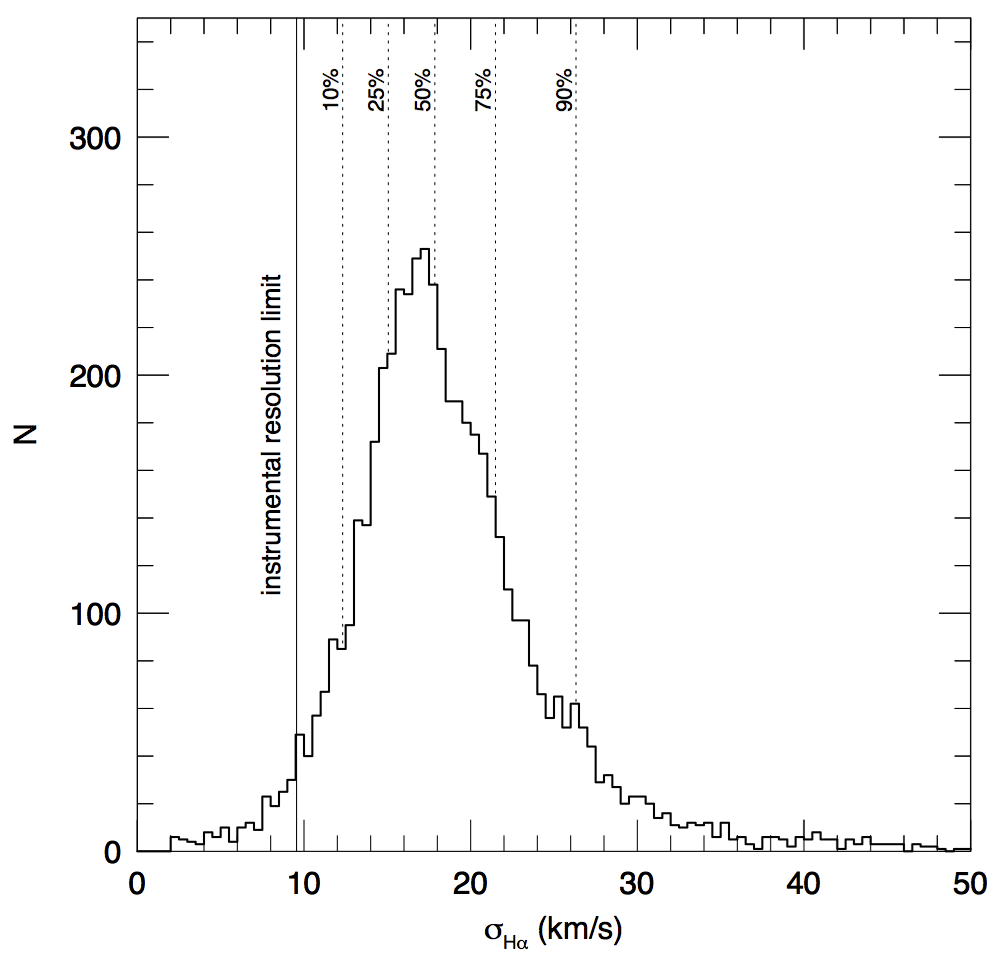}
\includegraphics[height=0.3\textheight]{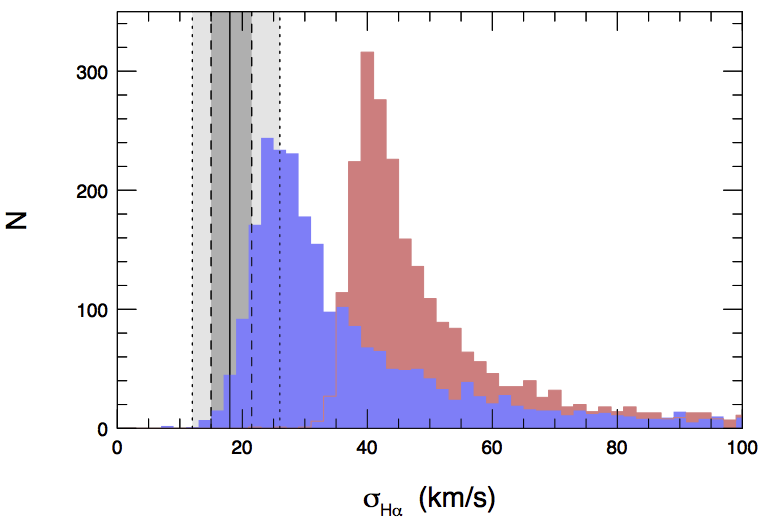}
\caption{Left: the distribution of intrinsic \hal\ line width among star-forming disk galaxies observed by \cite{Andersen06} with high resolution spectroscopy. Right: the distribution of intrinsic \hal\ line width measured in star-forming disk galaxies in MaNGA data before (pink histogram) and after (blue histogram) making a 10\% correction to the instrumental dispersion. The vertical lines indicate the 10-, 25-, 50-, 75-, and 90-percentiles of the distribution in the \cite{Andersen06} sample. The distribution in MaNGA galaxies after correction has a median that is much closer to, but still higher than, that measured by high resolution spectroscopy. This means we know our actual instrumental dispersion to better than 3\% accuracy. }
\label{fig:halinewidth}
\end{center}
\end{figure*}

Our ability to measure velocity dispersion for stars and gas below the intrumental resolution depends sensitively on how well we measure the effective intrumental resolution. We test this by measuring the distribution of \hal\ line width in star-forming galaxies observed in the first year, and compare it to those measured using high resolution spectroscopy. In the left panel of Figure~\ref{fig:halinewidth} we show the \hal\ line width ($\sigma$) distribution in local face-on star-forming galaxies measured by \cite{Andersen06} using high resolution spectra. The line widths are corrected for intrumental dispersion. The median value is around 18 km/s. In the right panel of Figure~\ref{fig:halinewidth}, we show the intrinsic \hal\ line width distribution measured in MaNGA spectra. The line widths are also corrected for instrumental dispersion. The two histograms shown use different estimates of the instrumental dispersion. The pink histogram uses the instrumental dispersion reported by the DR13 version of the pipeline directly, while the blue histogram broadens the instrumental dispersion by 10\% before using, as described above. The latter version yields an intrinsic \hal\ line width distribution with a median around 26 km/s, much closer to the value measured by high resolution spectroscopy. There is still a small difference, which means we may still be underestimating the instrumental dispersion ($\sim69{\rm km~s^{-1}}$) slightly, by about 3\%. This means we would have at most a 10\% bias on velocity dispersion around 40 km/s, which is 40\% lower than our instrumental resolution.

%



%

\subsection{Quality of Sky Subtraction}
One unique advantage of MaNGA is its very wide wavelength coverage, especially the coverage in the near-infrared which includes many spectral features important for ISM, stellar population, initial mass function, and kinematics diagnostics. However, the red part is full of the atmospheric emission lines. Reliable sky subtraction is critical to take advantage of this region of the spectra. \cite{Law16} give extensive details on how we perfrom sky subtraction and a detailed assessment of the quality. Here we briefly summarize the result. Using specially-built plates on which all fibers point at empty sky locations ("all-sky plates"), we tested our sky subtraction. We found the residual in our sky-subtracted sky spectra have a distribution that is very consistent with the expected uncertainty due to read noise and Poisson counting statistics. For individual wavelengths, the residuals are consistent with Poisson expectations in line-free regions of the spectra and is slightly above Poission expectation in strong line regions. 
Using all-sky plates, we have verified that the sky-subtracted sky fibers have no significant systematic residuals. This is done by stacking a large number of these residual spectra and verifying that the RMS of the stack decreases following the expectation of Poisson statistics.

\begin{figure}
\begin{center}
\includegraphics[width=0.5\textwidth]{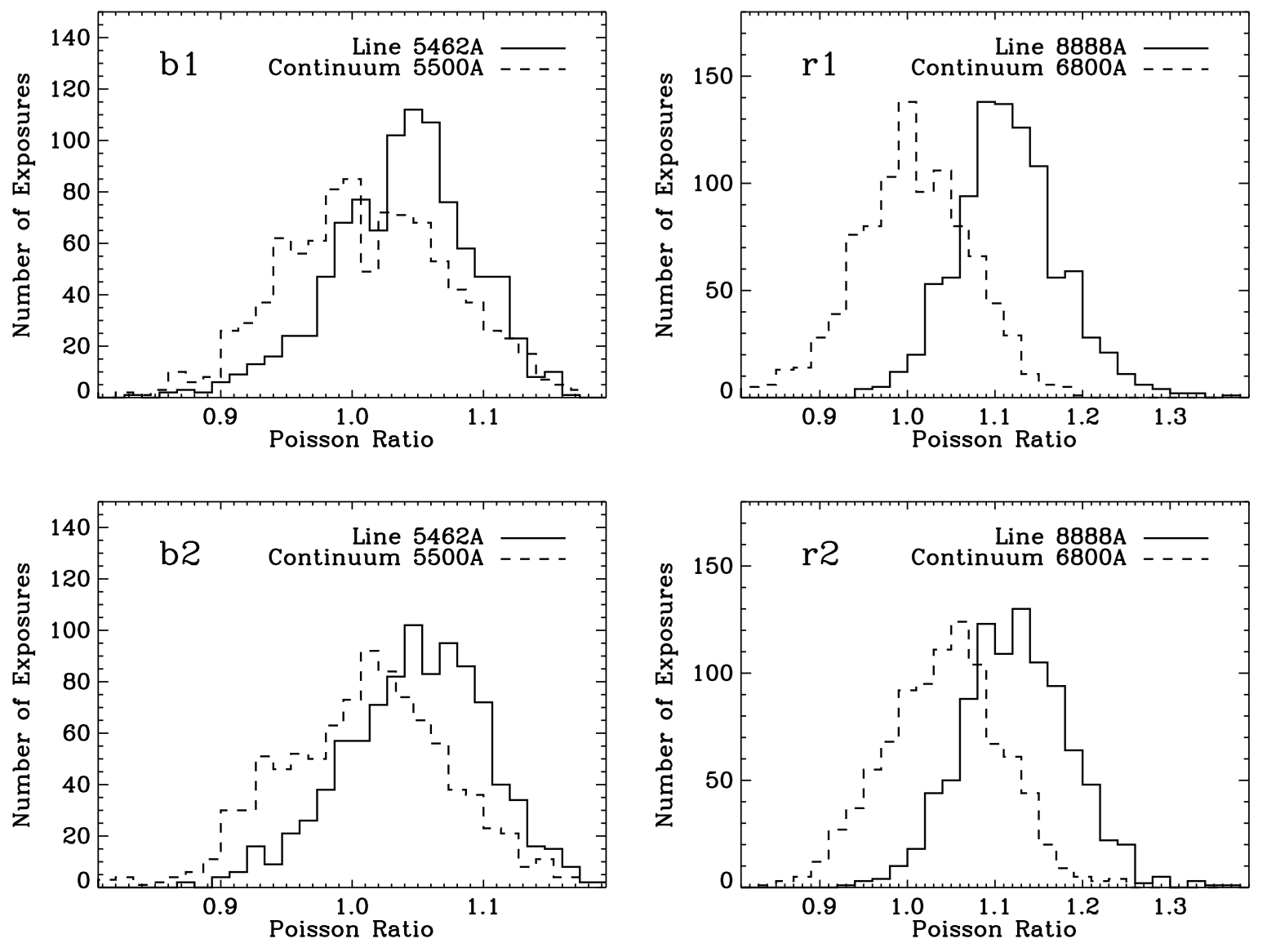}
\caption{The Poisson Ratio distribution among all exposures taken in the first year, for different wavelengths in the 4 cameras. The Poisson Ratio indicates how the distribution of the residual noise compare to the expected uncertainty given by read noise and the Poisson counting statistics. Each panel shows one camera. The solid histograms show the Poisson distribution around moderately strong sky emission lines; the dashed histogram show the Poisson Ratio distribution around line-free continuum regions. The wavelengths are indicated in the legend. The Sky subtraction is very close to Poisson in the continuum and slightly above Poisson around emission lines.}
\label{fig:skysubtraction}
\end{center}
\end{figure}

Here we present an evaluation of the sky subtraction accuracy in every science plate using the sky fibers. We take the standard deviation of the residual in the sky-subtracted sky spectra and then divide it by the expected uncertainty given by read noise and Poisson statistics, resulting in what we call `the Poisson ratio'. We evaluate this Poisson ratio for four wavelengths in the spectra: two are centered on moderately strong sky lines ($5462{\rm \AA}$ in the blue and $8888{\rm \AA}$ in the red), and two are on line-free continuum regions ($5500{\rm \AA}$ in the blue and $6800{\rm \AA}$ in the red) . Figure~\ref{fig:skysubtraction} shows the distribution of this Poisson ratio for all exposures taken in the first year. Because the sky model is built from the sky fibers, evaluating the Poisson ratio using the sky fibers underestimates the actual Poisson ratio. 
Using 14 exposures taken on multiple all-sky plates, we compare the Poisson ratio between science fibers and sky fibers for these four wavelengths. We found the science fibers show larger Poisson ratios than sky fibers by different factors on these different wavelengths: 2\% at 5500\AA, 7\% at 6800\AA, 12\% at 5462\AA, and 15\% at 8888\AA. 
Therefore, in Figure~\ref{fig:skysubtraction}, we have scaled up the Poisson ratio by these factors. The subtraction is very close to Poisson in the continuum and slightly above Poisson around sky emission lines. 


\subsection{Quality of Flux Calibration}

As described in detail by \cite{Yan16}, our flux calibration algorithm is different from single-fiber spectroscopy surveys because we are performing imaging spectroscopy. We would like to only correct for the flux lost due to the imperfect system response and atmosphere extinction, but not for any flux lost due to the limited aperture of each fiber. The separation of the two flux loss factors is achieved by modeling a star's flux as received by the 7 fibers in a mini-bundle. Given an initial guess of the PSF provided by the guider, we use the flux ratios among the 7 fibers to constrain the exact position of the star relative to the bundle, the size of the PSF, and the level of differentiatial atmosphere refraction. With this spatial model accounting for the aperture-induced flux loss, we can then estimate the flux loss due to the system response. We target 12 stars per plate with 6 per spectrograph. The average of the 6 stars per spectrograph provide the throughput correction that is applied to all galaxy fiber spectra. 

In \cite{Yan16}, we provided two assessments of the flux calibration accuracy. From comparison to broadband imaging of galaxies, we showed that the relative calibration in $g-r$, $r-i$, and $i-z$ colors are good to $\pm3\%$. From the comparison between completely independent measurements of the throughput curves, we showed that we achieve better than 5\% absolute calibration for 89\% of the wavelength range, and achieve a relative calibration RMS of 1.7\% between \hal\ and \hb, and 4.7\% between \niiw\ and \oiiw \citep{Yan16}. 

\begin{figure*}
\begin{center}
\includegraphics[width=0.9\textwidth]{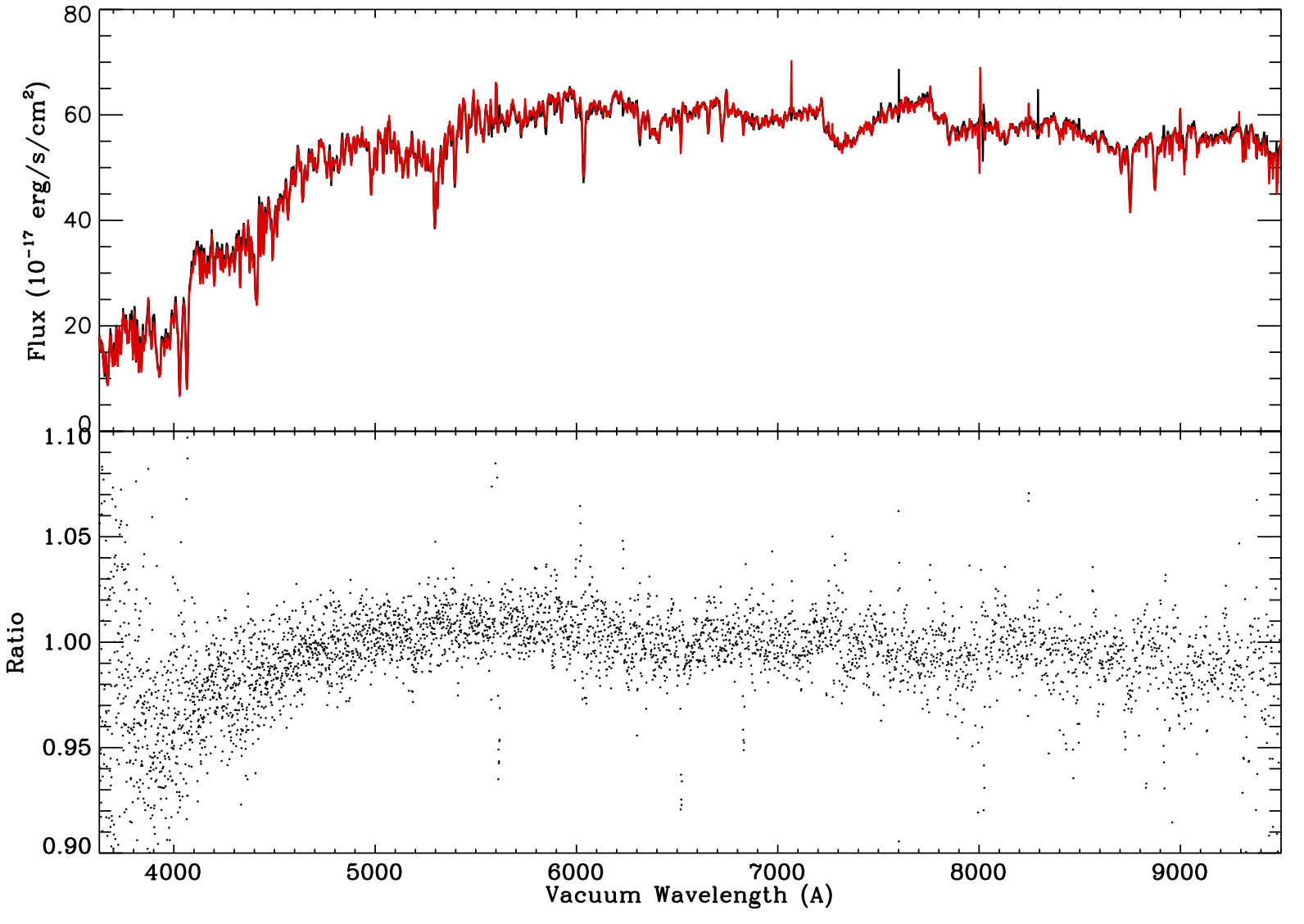}
\caption{Top: Comparison of the stacked spectra for the same galaxy observed on two different plates. Bottom: the ratio between the two stacked spectra illustrating a flux calibration uncertainty better than 5\%. }
\label{fig:repeatgals}
\end{center}
\end{figure*}

\begin{figure*}
\begin{center}
\includegraphics[width=1.0\textwidth]{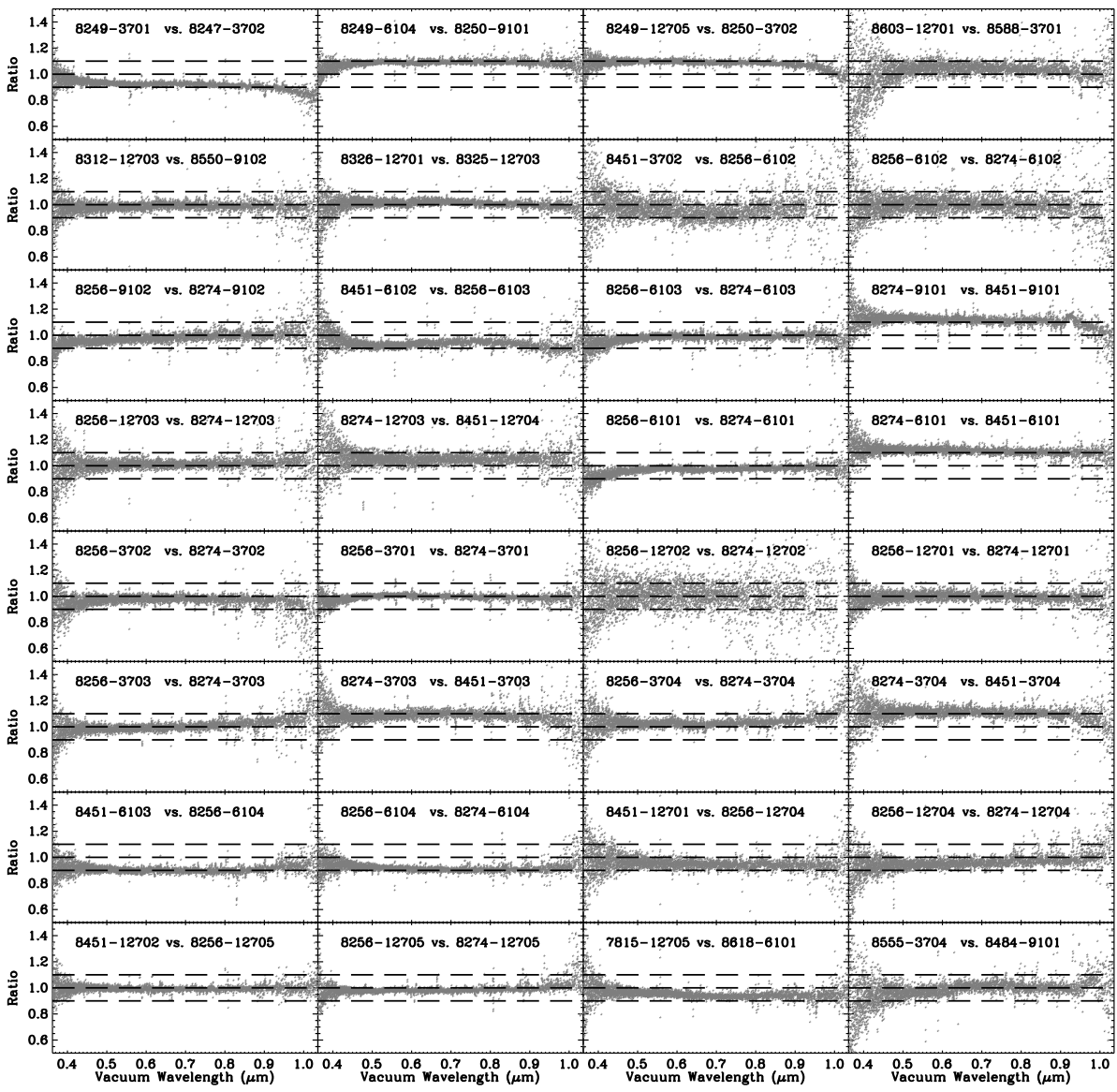}
\caption{The flux ratio as a function of wavelength between two independent observations of the same galaxy. The three dashed horizontal lines in each panel mark unity and $\pm10\%$. In the great majority of cases, the ratios are flat with wavelength, indicating excellent quality in relative calibration. In some cases, the absolute calibration between two observations can differ by about 10\%. } 
\label{fig:allresiduals}
\end{center}
\end{figure*}

Here, we provide yet another evaluation of the flux calibration accuracy based on comparison of repeated galaxies. These galaxies are observed on different plates with different standard stars. Therefore, the observations and calibrations are completely independent of each other. We stack the spectra from the data cubes in a 5\arcsec\ radius circle around the center of each object. In Figure~\ref{fig:repeatgals}, we show an example pair of these repeated observation of one galaxy. The bottom panel shows the ratio of the two spectra as a function of wavelength. The two spectra agree to within a few percent, and not more than 10\% at the wavelength extremes. In Figure~\ref{fig:allresiduals} we show the ratio plots for 32 pairs of repeated observations. In most cases, the ratio is very flat and is very close to 1. Sometimes, the absolute calibration of the two observations could differ by $\pm10\%$, but the relative calibration is mostly flat. 

\section{Verification of the Science Requirements} \label{sec:verification}

In this section, we verify that the science requirements set forth in Section~\ref{sec:requirements} are achieved with our first year data. 

\subsection{Star Formation Rate Surface Density}

We require the SFR surface density to be measured to better than 0.15 dex precision when the SFR density is above 0.01 ${\rm M_\odot~yr^{-1}~kpc^{-2}}$ and $E(B-V) < 0.5$. Given the derivation in Section~\ref{sec:requirements_deriv}, these limits correspond to a \hal\ SB of $6.58\times10^{-17} {\rm erg~s^{-1}~cm^{-2}~arcsec^{-2}}$ and a \hb\ SB of $1.36\times10^{-17} {\rm erg~s^{-1}~cm^{-2}~arcsec^{-2}}$. We evaluate the accuracy of these line measurements at this surface brightness using repeated observations. 

\begin{figure*}
\begin{center}
\includegraphics[width=0.8\textwidth]{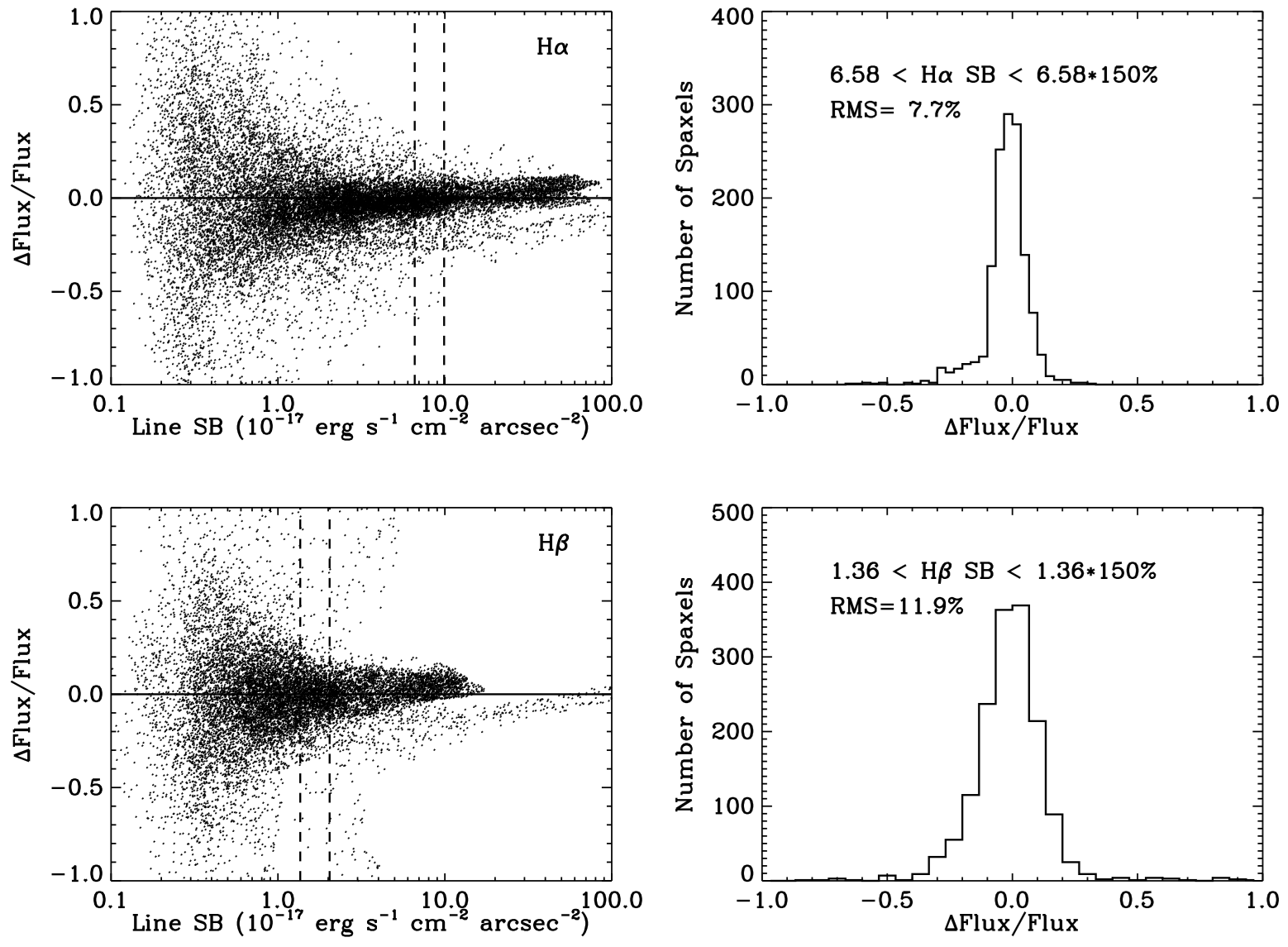}
\caption{Left panels: fractional flux difference as a function of average line flux between repeated observations of the same regions in same galaxies. Fractional uncertainty decreases with increasing flux. Right panels: distribution of the fractional difference for the surface brightnesses corresponding to the limits specified in our science requirements. The RMS of the distributions around 1.0 yield a fractional uncertainty of 5.4\% on \hal\ and 8.4\% on \hb\ for these threshold fluxes, which lead to 0.1 dex uncertainty on SFR surface density for 0.01 ${\rm M_\odot~yr^{-1}~kpc^{-2}}$ with an extinction $E(B-V) = 0.5$. }
\label{fig:lineaccuracy}
\end{center}
\end{figure*}

The reduced data cubes are processed by our dedicated data analysis pipeline (DAP, Westfall et al. in prep). Briefly, the emission lines are measured in the reduced data cube for each spaxel after the subtraction of the stellar continuum. 
We fit the emission-line only spectra with Gaussians around the lines, using multiple Gaussians when necessary (e.g. \hal+\nii\ triplet). 
For the comparison between repeated observations of the same galaxies, we first smoothed the emission line flux map by a $2.5\arcsec\times2.5\arcsec$ square kernel, which is equivalent to summing the flux in a resolution element. Then, we take the difference in line flux between the two independently measured flux maps of the same galaxy and then divide by their average. Figure~\ref{fig:lineaccuracy} shows the fractional difference in line flux vs. the average line flux. We can see the fractional differences decrease with increasing flux. At the threshold surface brightness, we found the fractional difference in \hal\ flux has a root-mean-square (RMS) of 7.7\% around 0 after one round of rejection of points more than $3\sigma$ away from zero, and \hb\ has an RMS of 11.9\%. Since this is the difference of two independent measurements, the actual uncertainty on the measurement is a factor of $\sqrt{2}$ smaller, at 5.4\% (\hal) and 8.4\% (\hb). According to Eqn. (12) in \cite{Yan16}, this would yield a final fractional uncertainty on SFR of 23.3\% or 0.1 dex. In this calculation, we have included the 1.7\% RMS relative calibration error between \hal\ and \hb\ and the 4\% RMS error in absolution calibration around \hal \citep{Yan16}, which do not dominate the uncertainty. We have met the science requirement on the SFR surface density. 

At lower SFR surface density, the uncertainty increases. At 0.003${\rm M_\odot~yr^{-1}~kpc^{-2}}$ and $E(B-V) < 0.5$, the fractional uncertainty is about 50\%, or 0.2 dex. At 0.001${\rm M_\odot~yr^{-1}~kpc^{-2}}$, the uncertainty is 75\%, or 0.3 dex.

There are a small fraction of spaxels with much larger fractional error. The reasons for these are still under investigation. There also appear to be systematic difference between \hal\ fluxes of repeated observations at high line fluxes, which is as large as 10\%. These cannot be caused by flux calibration error as the difference is not constant with changing flux. The actual cause is also to be investigated.

\subsection{Gas Metallicity Gradient}
\begin{figure}
\begin{center}
\includegraphics[width=0.45\textwidth]{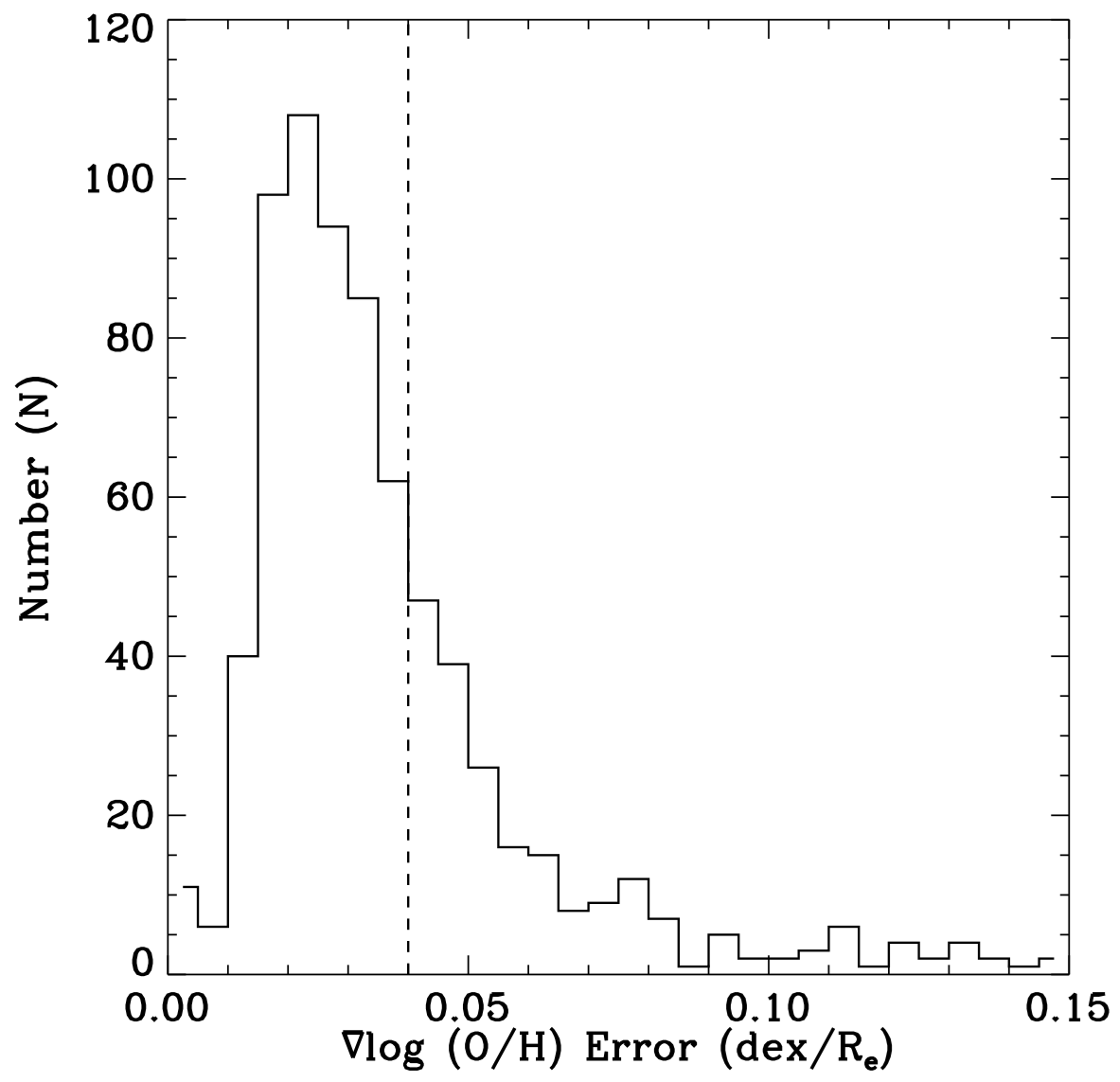}
\caption{Uncertainty distribution of the gas metallicity gradient in all star-forming galaxies observed in the first year. This is measured using the R23 metallicity indicator and the analytic formula given by \citep{TremontiHK04}. The vertical line marks the threshold of the science requirement. 
}
\label{fig:gas_metal_error}
\end{center}
\end{figure}

Our science requirement on gas metallicity is to measure the gradient to better than 0.04 dex per \Reff. For all galaxies observed in the first year, we subtracted the stellar continuum from the spectrum in each spaxel, perform Voronoi binning based on the \hal\ S/N, and then measured the emission fluxes in each bin, as done by \citep{Belfiore16}. We then classified the bins according to their positions on line ratio diagnostic diagrams. For all bins classified as star formation, we measure their gas phase metallicity, using the R23 metallicity indicator \citep{TremontiHK04}. For measuring the gradient, we group the bins into a set of elliptical annuli. In each annulus, we measure the mean metallicity and the error of the mean using the biweight estimator \citep{Beers90}. Then we fit a linear function through all the annuli. The distribution of the derived uncertainty is shown in Figure~\ref{fig:gas_metal_error}. We have experimented using different numbers of annuli and the results do not change statistically. For all star-forming galaxies, we found $\sim68\%$ of them have a gradient error that is smaller than 0.04 dex per \Reff. We met this science requirement for the majority of the sample.


Using a different metallicity indicator, such as O3N2 \citep{PettiniP04}, gives similar results in the error distribution, but different gradients. There are certainly systematic errors associated with the metallicity calibration adopted \citep{KewleyE08}. As stated in Section~\ref{sec:requirements}, our requirement is set only on the precision of the measurement. The systematic bias of different calibrations cannot be alleviated by getting deeper data. Comparison between different calibrators and more detailed theoretical modeling are needed to resolve their discrepancies.

\subsection{Stellar Population Gradients}

Our science requirement on stellar populations is to measure the age, metallicity, and abundance gradients in quiescent galaxies, and age gradients in star-forming galaxies, to better than 0.1 dex per decade in \Reff.

\begin{figure}
\begin{center}
\includegraphics[width=0.45\textwidth]{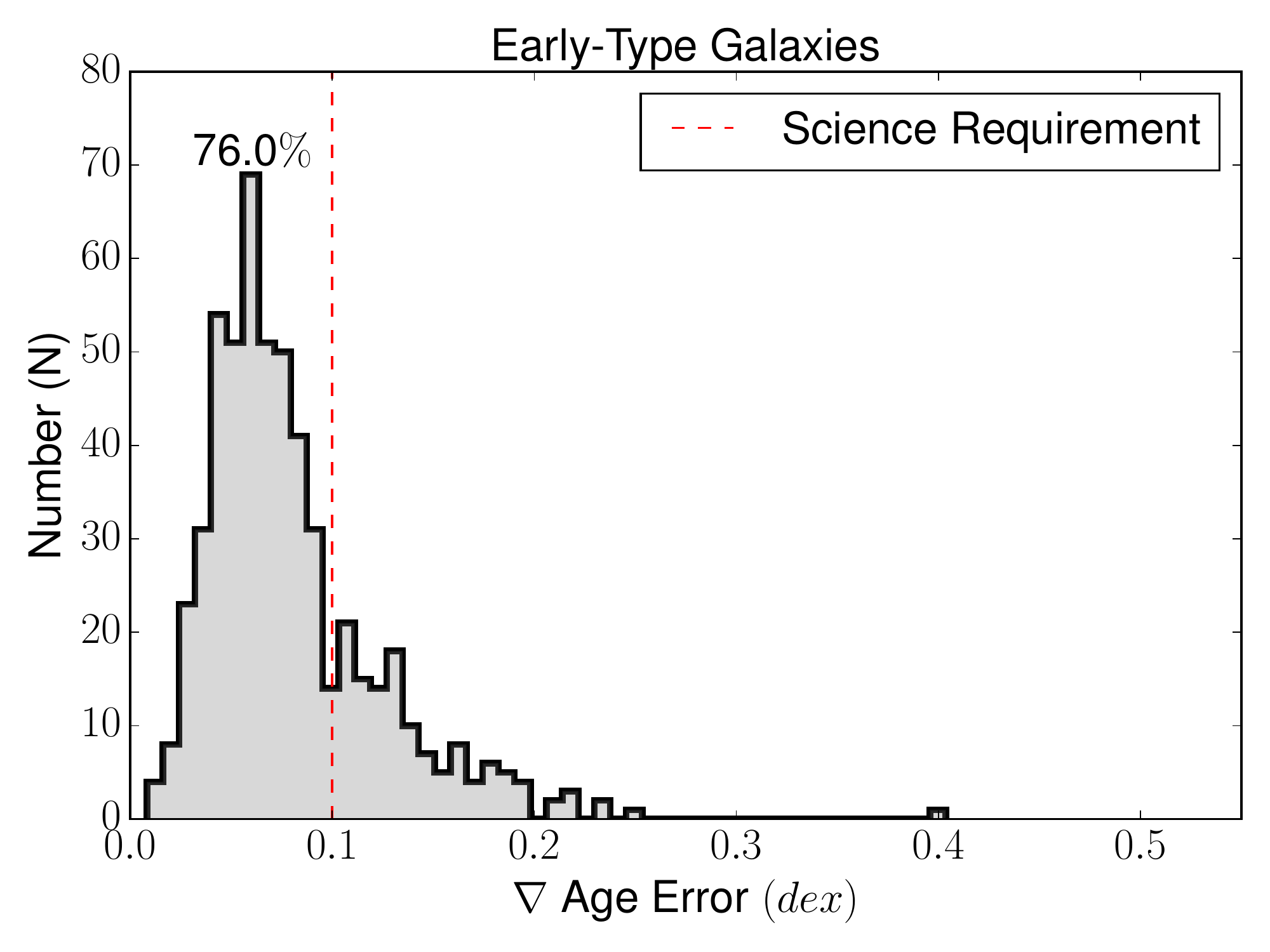}
\includegraphics[width=0.45\textwidth]{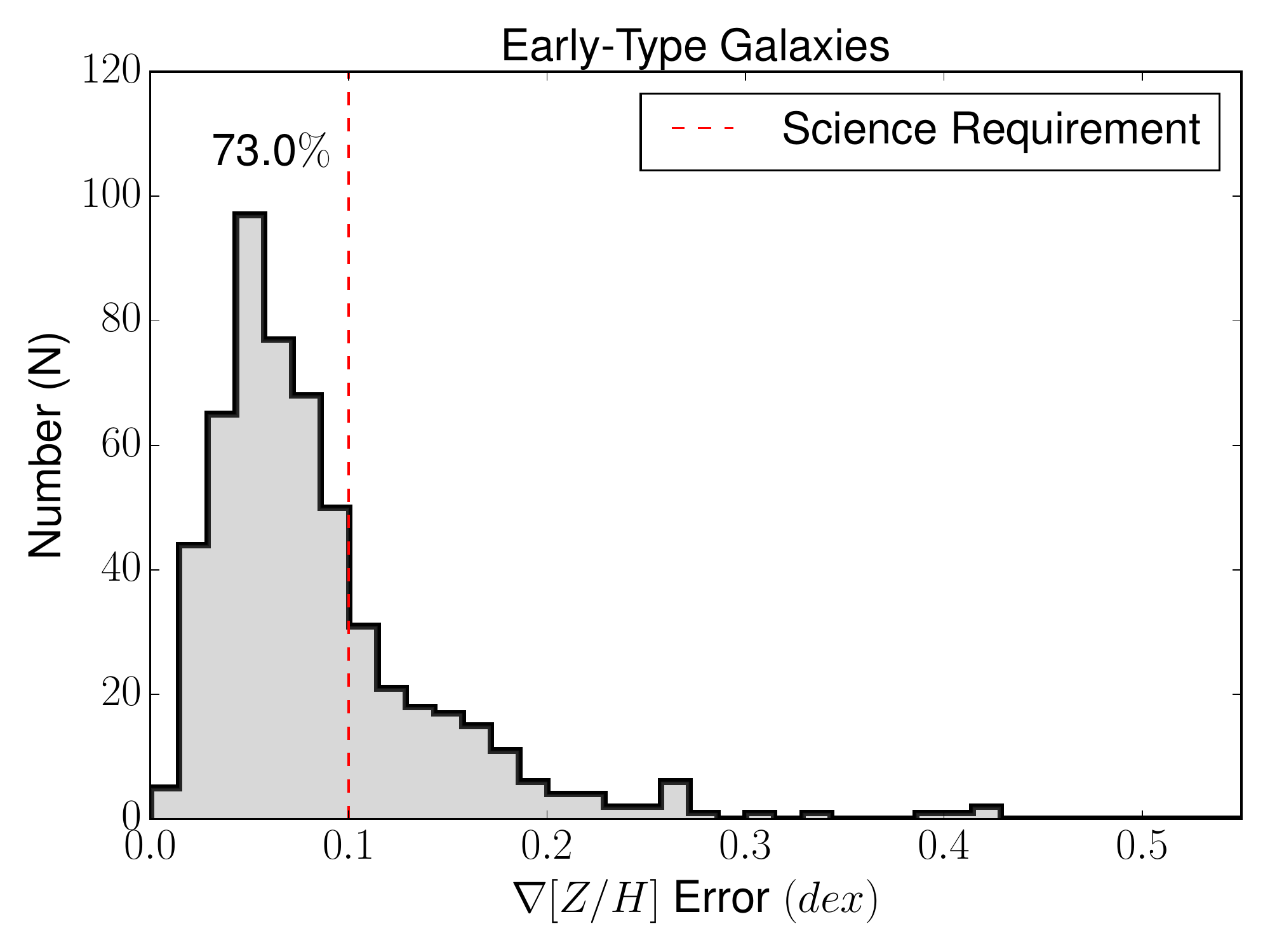}
\includegraphics[width=0.45\textwidth]{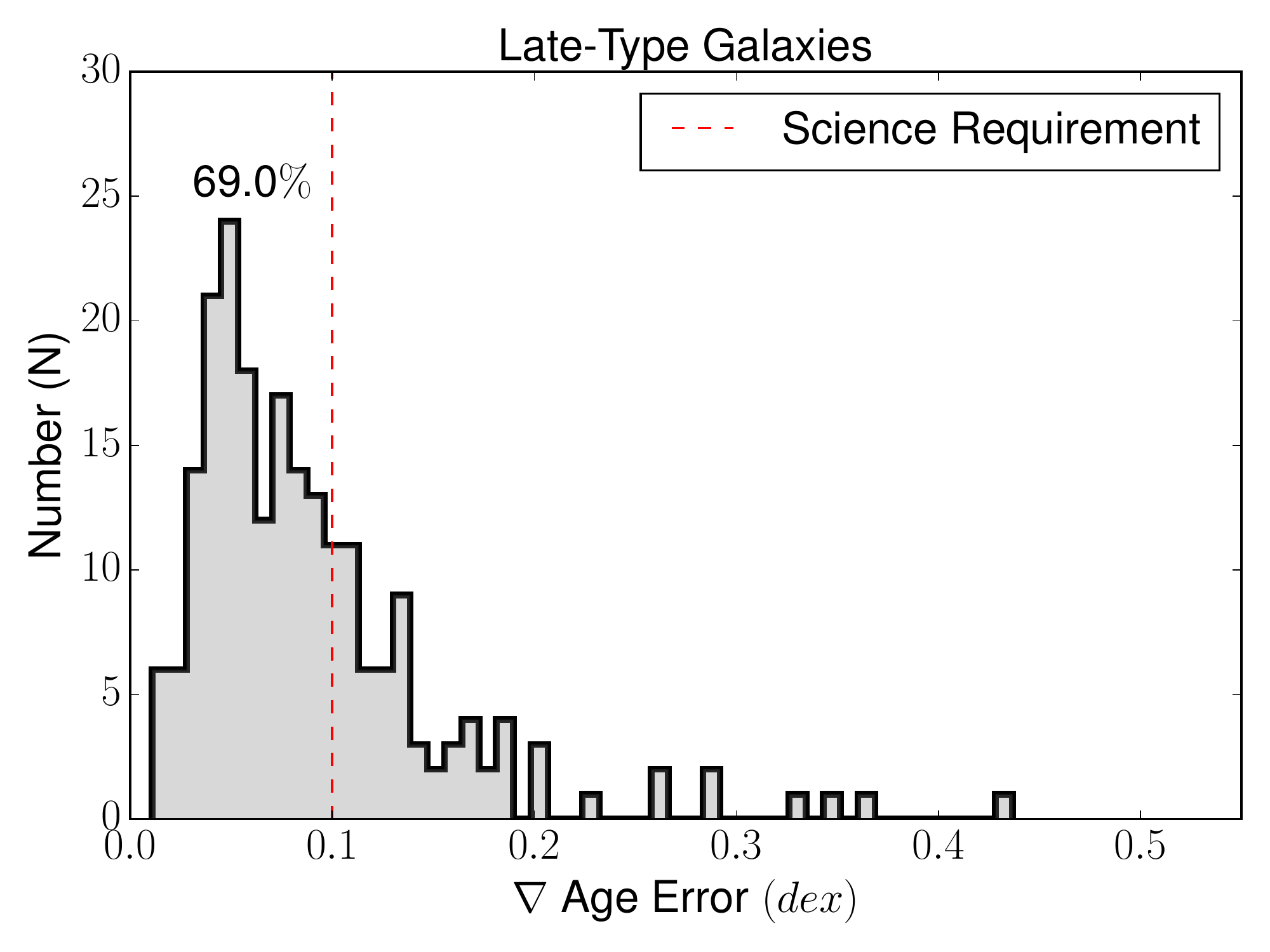}
\caption{The uncertainty distributions of the stellar age gradient (top panel) and the metallicity gradient (middle panel) among early-type galaxies observed in the first year. The bottom panel shows the uncertainty distribution for the age gradient in late-type galaxies. The units of the gradients are per dex per decade in \Reff. The vertical lines mark our science requirements, which are met by the great majority of galaxies.} 
\label{fig:ET_gradients_error}
\end{center}
\end{figure}

For each galaxy observed in the first year, using the data cube produce by the DRP, we Voronoi-bin the spaxels to have S/N greater than 5 per bin. We measured the stellar age and metallicity for each Voronoi bin, then fit the radial gradient with a linear function. We evaluate the uncertainty of the gradients using a Monte Carlo bootstrap resampling method. From a 1000 resampling of the original distribution for each galaxy, we measure the error on the slope. Figure~\ref{fig:ET_gradients_error} shows the uncertainty distribution of the stellar age and metallicity gradients for all early-type galaxies in the first-year data, and that of the age gradient for late-type galaxies. The measurement will be described in Goddard et al. in prep. We meet the stellar population gradient requirement for $\gtrsim 70\%$ of early-type and late-type galaxies.

In Section~\ref{sec:requirements} where we derived the S/N needed to meet the science requirement, we have been assuming that we will make the measurement by stacking all spaxels within an annulus and produce one measurement per annulus with an associated uncertainty. In reality, such an approach would likely underestimate the uncertainty. There are two reasons. First, it does not include any intrinsic physical variation within an annulus. Second, whatever algorithm we use to estimate error for one data point may not be robust. It is much more reliable to conduct measure the concerned quantity in many Voronoi bins within an annulus, then estimate the scatter among them. This scatter would include both the intrinsic scatter and the actual measurement uncertainty. The error on the final gradient derived from this would be much more robust. 

\subsection{Specific Angular Momentum}

We require the specific angular momentum within 1\Reff\ to be measured to better than 0.05 around $\lambda_R=0.1$ so that we could distinguish fast and slow rotators. 

We have measured $\lambda_{R_e}$ for all galaxies observed in the first year (Graham et al. in prep). We estimated the uncertainty on $\lambda_{R_e}$ by generating random normal distributons for both velocity and velocity dispersion according to the measurement errors on them. We generated 100 pairs of these random kinematic maps and computed $\lambda_{R_e}$ for each. The uncertainty on $\lambda_{R_e}$ is derived by taking the standard deviation among them. 

\begin{figure}
\begin{center}
\includegraphics[width=0.5\textwidth]{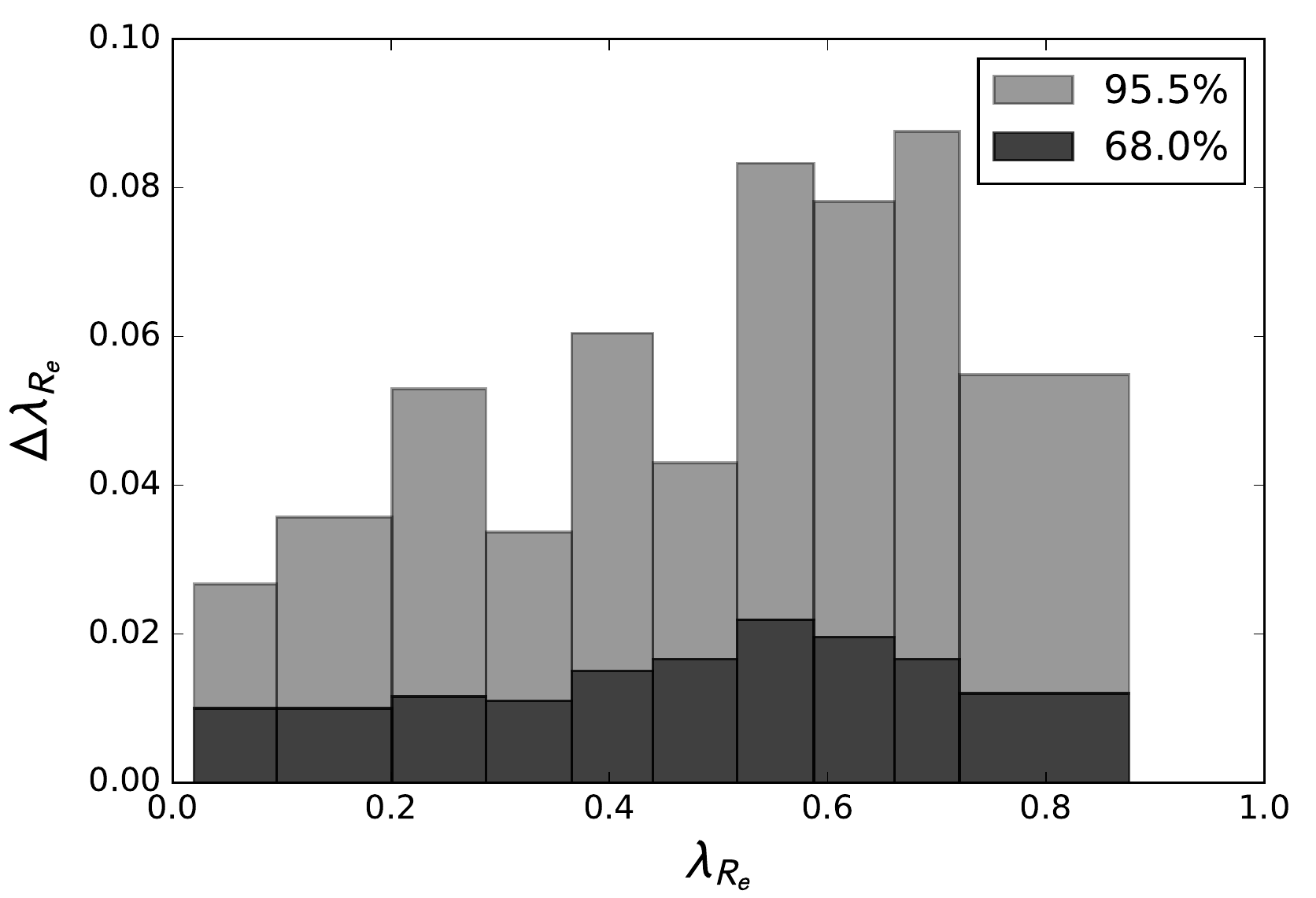}
\caption{Distribution of the uncertainty on the specific angular momenum ($\Delta\lambda_{R_e}$) as a function of the $\lambda_{R_e}$. The black histogram shows the 68-percentile in the $\Delta\lambda_{R_e}$ in each bin of $\lambda_{R_e}$. The grey histograms show the 95-percentiles.} 
\label{fig:lambdaR_err}
\end{center}
\end{figure}

Figure~\ref{fig:lambdaR_err} shows the uncertainty of $\lambda_{R_e}$ as a function of $\lambda_{R_e}$. Around $\lambda_{R_e}$ of 0.1, we can see nearly all galaxies have uncertainty better than 0.05. This meets our requirement. However, in this calculation, we have not considered systematics due to beam smearing effect. This would need to be assessed by simulations and will be addressed in future work. 

\subsection{Enclosed Gravitating Mass and Dark Matter Fraction}

In this section, we evaluate whether we meet the 10\% accuracy requirement on the enclosed gravitating mass for all axissymmetric galaxies, and the 10\% precision requirement the dark matter fraction within 1.5\Reff\ for early-type galaxies. We first address the gravitating mass requirement on rotation-dominated disk galaxies, then we address this and the dark matter requirement on early-type galaxies.

For rotation-dominated disk galaxies, we estimate the enclosed mass using the gas rotational velocity. The uncertainty is dominated by the error in the inclination. Compared to inclination error, the fractional error on the gas velocity is much smaller. We can measure the inclination from either photometry or kinematics. The difference between the photometric and kinematic inclinations can provide an indication of the uncertainty, which is typically much larger than the formal error provided by either measurement. To assess this, we select all rotation-dominated galaxies from the first year observations that have stellar line-of-sight velocity more than twice as large as the stellar line-of-sight velocity dispersion at 1 \Reff, and have kinematic inclination between 15 and 75 degree. This yields a subsample of 361 galaxies. Using the difference between photometric and kinematic inclination to derive the error on inclination ($1/\sqrt{2}$ of the difference), we obtain the fractional uncertainty on enclosed mass according to the following formula.

\begin{equation}
{\Delta M\over M} \simeq \sqrt{2} {\Delta \sin i \over \sin i} = 2 {(\sin i_{\rm phot}-\sin i_{\rm kin}) \over (\sin i_{\rm phot} + \sin i_{\rm kin})}
\end{equation}

\begin{figure}
\begin{center}
\includegraphics[width=0.5\textwidth]{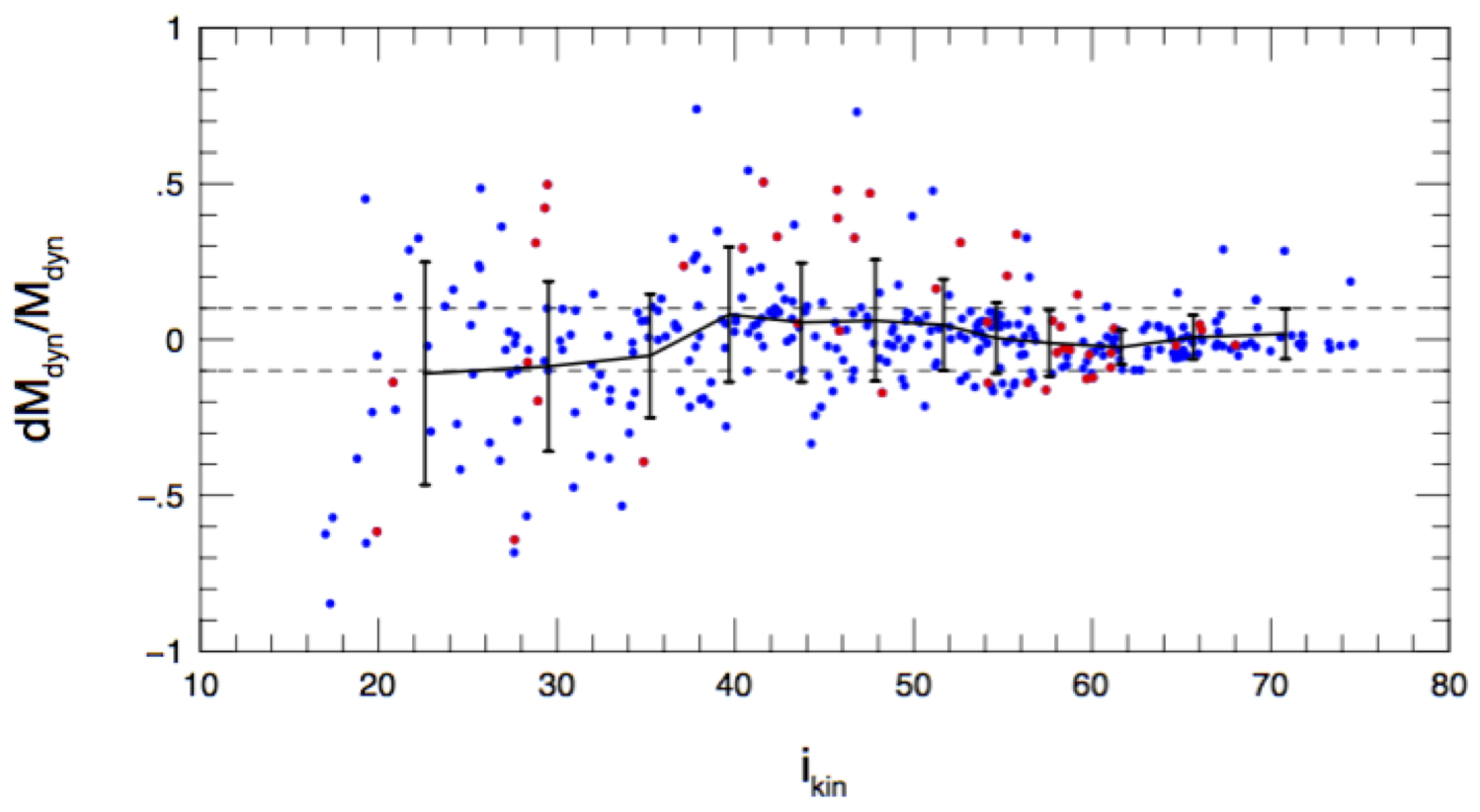}
\caption{Fractional uncertainty on enclosed gravitating mass within 1.5 \Reff\ as a function of the kinematic inclination, for all rotation-dominated galaxies observed by MaNGA in the first year. The blue and red points indicate blue and red galaxies, respectively. The solid line with error bars mark the mean values and standard deviation in bins of inclination. 
The error is dominated by uncertainty on derived inclination, which is estimated from the difference between photometric inclination and kinematic inclination. Adding the measurement errors of the gas or stellar velocity makes little difference to the results. The horizontal lines mark the science requirements. The majority of our sample satisfy this requirement.}
\label{fig:disk_mass_error}
\end{center}
\end{figure}

Figure~\ref{fig:disk_mass_error} shows the fractional uncertainty on mass as a function of kinematic inclination. The mean fractional errors (marked by the solid line) in bins of inclination indicate the systematic errors of the dynamical mass estimates; the standard deviations (marked by the error bar) indicate the random errors for individual galaxies. The systematic errors are better than 10\% in all bins and the random errors are better than 10\% at inclinations above 55 degree.  Overall, we expect 62\% of the sample to have a fractional error less than 10\%. 

\begin{figure}
\begin{center}
\includegraphics[width=0.45\textwidth]{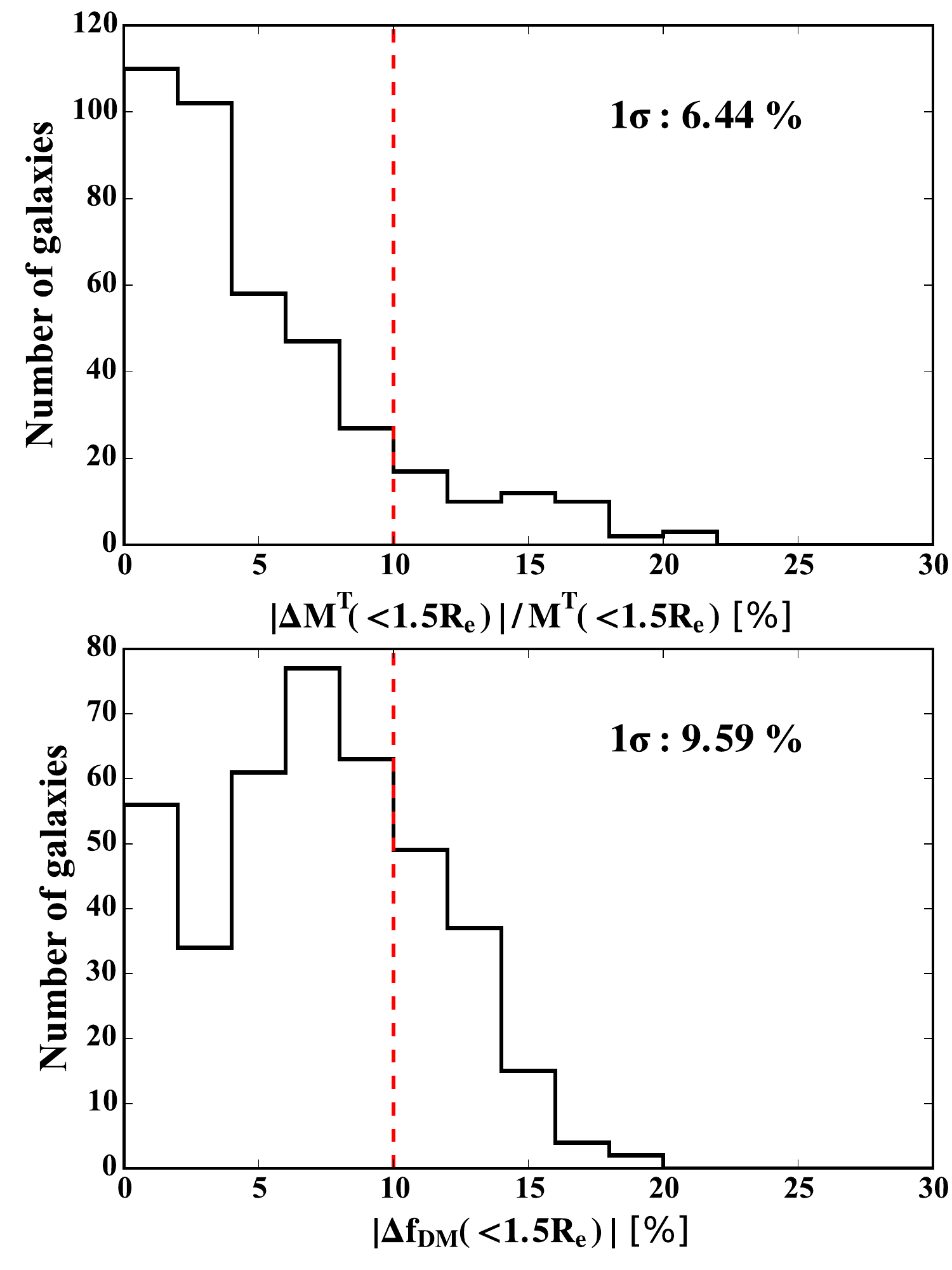}
\caption{Top: distribution of the fractional error on enclosed gravitating mass within 1.5\Reff\ for early-type galaxies. Bottom: distribution of the error on dark matter fraction within 1.5\Reff\ for early-type galaxies. The vertical lines mark the science requirements. These error are derived from the 1D marginalized MCMC distrubtion. They do not include the much larger systematic uncertainty associated with the JAM method. See text for detail.}
\label{fig:dynamics_error}
\end{center}
\end{figure}

For early-type galaxies, we estimate the enclosed gravitating mass and dark matter fraction in a different way.
With the first year data, Li et al. (in prep) applied JAM to derive the dynamical mass estimate for all elliptical galaxies. The sample is defined by Galaxy Zoo classification being 'elliptical', or by Sersic index greater than 2.5 and deVaucoulers fraction greater than 0.8 when Galaxy Zoo classification is 'uncertain'. Among 562 elliptical galaxies observed in the first year, 160 are rejected due to one of the four reasons: many pixels having unphysical velocity dispersion (38), having fewer than 20 Voronoi bins (51), having a foreground star (12), and being a merger or in a close pair (59). Among the remaining 402 galaxies, we run JAM within an MCMC framework, as described by \cite{LiH16}. From these, we estimated the statistical uncertainty on the enclosed gravitating mass and the dark matter fraction, using the 1D marginalized MCMC distributions. The distributions for these uncertainties are shown in Figure~\ref{fig:dynamics_error}. About 85\% of these 402 galaxies have a fractional error on total enclosed gravitating mass less than 10\% and 72\% have a dark matter fraction error less than 10\%. The overwheliming majority of the spaxels in most of these galaxies have velocity dispersion significantly above our instrumental resolution, thus the velocity and velocity dispersion are reliably measured.

However, these errors are only relevant for the precision of the estimates, which is what we defined in the science requirements. In light of the recent work by \cite{LiH16}, it is apparent that these random errors are dwarfed by the systematic error associated with the JAM method. \cite{LiH16} evaluated the accuracy of the JAM method using simulated galaxies from Illustrius project \citep{Genel14, Nelson15}. They found that, with a 0.5 kpc imaging resolution and 2 kpc velocity field resolution, the true fractional uncertainty is 11-16\% on the total mass, and $\sim33\%$ on the dark matter mass within 2.5\Reff, with relatively little bias (3\%) in the median value. If one degrade the imaging resolution to 2 kpc, there will also be a 10\% systematic bias in the median value of the measured dark matter mass. Going for smaller radius may also result in worse uncertainty.
This large systematic error is likely due to the simplified assumptions made in the JAM method, such as oblateness, constant mass-to-light ratio, constant anisotropy in the meridional plane, and a double power-law dark matter profile. These assumptions could fail for a significant fraction of galaxies. 

Therefore, although it appears that we have met the science requirements on these quantities, currently it is unclear whether we can reach our science goal of measuring the stellar mass-to-light ratio to better than 25\% in order to improve the constraints on the IMF. Further simulations done with the actual MaNGA resolution for both 1.5\Reff\ and 2.5\Reff\ spatial coverage will provide the answers. In addition, the accuracy of the measurements are significantly better when higher resolution imaging is available. Therefore, we can do significantly better in areas overlapping with HSC deep imaging fields. 

\section{Summary} \label{sec:summary}

MaNGA is an integral field spectroscopic survey of 10K nearby galaxies with wide wavelength coverage at medium resolution with uniform spatial coverage in units of \Reff. Up to the time of writing, we have already obtained observations for more than 2550 galaxies and are on track to finish $\sim10{\rm K}$ by summer of 2020.

In this paper, we have detailed the survey science requirements, both in terms of random and systematic errors, and how the high-level science requirements flow down in an interconnected way to the low level requirements on the hardware, sample selection, observations, and analysis. In this context we have described in detail how the sample selection is carried forward to generating a survey footprint on the sky, how this footprint is parsed into tiles, how these tiles are targeted with plates, and how these plates are designed, fabricated and scheduled for observation. The observing procedures are likewise detailed at a level necessary for a complete and reliable reconstruction of the survey execution. Finally, as proof of practice, we have given a complete demonstration of the data quality in both basic data products and high level derived science products across the full first year of data. 


The basic data quality of the survey is excellent. We have reached the S/N target while staying on track to finish observing 10K galaxies by 2020. We obtain a spatial resolution about 2.5\arcsec FWHM with a carefully characterized profile with uniform and near-critical sampling from multiple dithered observations. The sky subtraction is nearly Poisson even at near-infrared wavelengths. Both the absolute and relative flux calibrations are better than 5\%. The spectral resolution is a function of wavelength and is characterized for each fiber in each exposure. Exposure-to-exposure variations should be taken into account if the science case warrants it. 

The high level derived science products are also of high quality. We have met the majority of the science requirements set forth, such as the precision on the star formation rate surface density, the gas metallicity gradient, the stellar population age and metallicity gradient. On the several kinematics requirements, such as the specific angular momentum, the enclosed mass, and the dark matter fraction, the systematic errors due to simplified modeling assumptions dominate the precision of the measurements. The formal errors appear to meet the science requirements, but whether the scienctific goals on kinematics could be reached awaits further analysis facilitated by detailed simulations. 
The first year data will be released in SDSS Data Release 13\ in summer 2016.

\acknowledgements

We thank the referee for a very constructive report which helped us to improve the paper. KB is supported by World Premier International Research Center Initiative (WPI Initiative), MEXT, Japan.  MAB acknowledges support by grant NSF/AST 1517007. AW acknowledges support of a Leverhulme Trust Early Career Fellowship. AD acknowledges support from The Grainger Foundation. DB acknowledges support by grant RSF 14-50-00043. MC acknowledges support from a Royal Society University Research Fellowship. SM and HL acknowledge supported by the Strategic Priority Research Program “The Emergence of Cosmological Structures” of the Chinese Academy of Sciences Grant No. XDB09000000, and by the National Natural Science Foundation of China (NSFC) under grant number 11333003 and 11390372 (SM). KM acknowledges support by STFC.

Funding for the Sloan Digital Sky Survey IV has been provided by
the Alfred P. Sloan Foundation, the U.S. Department of Energy Office of
Science, and the Participating Institutions. SDSS-IV acknowledges
support and resources from the Center for High-Performance Computing at
the University of Utah. The SDSS web site is www.sdss.org.

SDSS-IV is managed by the Astrophysical Research Consortium for the 
Participating Institutions of the SDSS Collaboration including the 
Brazilian Participation Group, the Carnegie Institution for Science, 
Carnegie Mellon University, the Chilean Participation Group, the French Participation Group, Harvard-Smithsonian Center for Astrophysics, 
Instituto de Astrof\'isica de Canarias, The Johns Hopkins University, 
Kavli Institute for the Physics and Mathematics of the Universe (IPMU) / 
University of Tokyo, Lawrence Berkeley National Laboratory, 
Leibniz Institut f\"ur Astrophysik Potsdam (AIP),  
Max-Planck-Institut f\"ur Astronomie (MPIA Heidelberg), 
Max-Planck-Institut f\"ur Astrophysik (MPA Garching), 
Max-Planck-Institut f\"ur Extraterrestrische Physik (MPE), 
National Astronomical Observatory of China, New Mexico State University, 
New York University, University of Notre Dame, 
Observat\'ario Nacional / MCTI, The Ohio State University, 
Pennsylvania State University, Shanghai Astronomical Observatory, 
United Kingdom Participation Group,
Universidad Nacional Aut\'onoma de M\'exico, University of Arizona, 
University of Colorado Boulder, University of Oxford, University of Portsmouth, 
University of Utah, University of Virginia, University of Washington, University of Wisconsin, 
Vanderbilt University, and Yale University.

\bibliographystyle{apj}
\bibliography{astro_refs}

\end{document}